\documentclass[prd, onecolumn,superscriptaddress, nofootinbib, notitlepage, floatfix]{revtex4-1}
\usepackage{xcolor}
\usepackage{etoolbox}
\usepackage{graphicx}
\usepackage{wrapfig}
\usepackage{amsmath}
\usepackage{amsfonts}
\usepackage{amssymb}
\usepackage{multirow}
\usepackage{slashed}
\usepackage{physics}
\usepackage{ulem}
\usepackage[colorlinks=true, linkcolor=blue, citecolor=blue, urlcolor=blue]{hyperref}

\newcommand{\pvec}{\vec{p}}
\newcommand{\qvec}{\vec{q}}
\newcommand{\xvec}{\vec{x}}
\newcommand{\yvec}{\vec{y}}
\newcommand{\zvec}{\vec{z}}

\DeclareRobustCommand{\Eq}[1]{Eq.~\eqref{eq:#1}}

\DeclareRobustCommand{\fig}[1]{Fig.~\ref{fig:#1}}

\DeclareRobustCommand{\app}[1]{App.~\ref{app:#1}}
\DeclareRobustCommand{\sec}[1]{Sec.~\ref{sec:#1}}

\DeclareRobustCommand{\tb}[1]{Table~\ref{tb:#1}}

\DeclareRobustCommand{\refcite}[1]{Ref.~\cite{#1}}

\newcommand\bets{\begin{table*}}
\newcommand\eets[1]{\label{tb:#1}\end{table*}}
\begin{document}
\widetext

\title{Moments of proton GPDs from the OPE of nonlocal quark bilinears up to NNLO} 

\author{Shohini Bhattacharya}
\affiliation{RIKEN BNL Research Center, Brookhaven National Laboratory, Upton, NY 11973, USA}
\author{Krzysztof Cichy}
\affiliation{Faculty of Physics, Adam Mickiewicz University, ul.\ Uniwersytetu Pozna\'nskiego 2, 61-614 Pozna\'{n}, Poland}
\author{Martha Constantinou}
\affiliation{Department of Physics,  Temple University,  Philadelphia,  PA 19122 - 1801,  USA}
\author{Xiang Gao}
\email{gaox@anl.gov}
\affiliation{Physics Division, Argonne National Laboratory, Lemont, IL 60439, USA}
\author{Andreas Metz}
\affiliation{Department of Physics,  Temple University,  Philadelphia,  PA 19122 - 1801,  USA}
\author{Joshua Miller}
\affiliation{Department of Physics,  Temple University,  Philadelphia,  PA 19122 - 1801,  USA}
\author{Swagato Mukherjee}
\affiliation{Physics Department, Brookhaven National Laboratory, Upton, New York 11973, USA}
\author{Peter Petreczky}
\affiliation{Physics Department, Brookhaven National Laboratory, Upton, New York 11973, USA}
\author{Fernanda Steffens}
\affiliation{Institut f\"ur Strahlen- und Kernphysik, Rheinische Friedrich-Wilhelms-Universit\"at Bonn,\\ Nussallee 14-16, 53115 Bonn}
\author{Yong Zhao}
\affiliation{Physics Division, Argonne National Laboratory, Lemont, IL 60439, USA}

\date{\today}

\begin{abstract}
For the first time, we present a lattice QCD determination of Mellin moments of unpolarized generalized parton distributions (GPDs) of the proton from an analysis of the quasi-GPD matrix elements within the short-distance factorization framework. We perform our calculation on an $N_f$=2+1+1 twisted mass fermions ensemble with a clover improvement at lattice spacing $a=0.093$ fm and a pion mass of $m_\pi=260$ MeV. Focusing on the zero-skewness case, the iso-vector and iso-scalar quasi-GPDs are calculated from the $\gamma_0$ definition, as well as a recently proposed Lorentz-invariant definition. We utilize data on both symmetric and asymmetric kinematic frames, which allows us to obtain the Mellin moments for several values of the momentum transfer, $-t$, in the range 0.17 to $2.77~\rm{GeV}^2$. 
We use the ratio scheme for GPDs, i.e. renormalization group invariant ratios
with leading-twist factorization formula
and perturbatively calculated matching coefficients up to the next-next-to-leading order (NNLO)
to extract Mellin moments of GPDs, which are consistent with renormalization-group improved results. 
We compare our determination from quasi-GPDs with the results extracted using standard calculations of Mellin moments of local operators, specifically those related to the electromagnetic and gravitational form factors. 
We estimated the moments of GPDs up to the fifth ones for the first time. 
By extrapolating the Mellin moments to $-t=0$, we obtained the quark charges, momentum fraction, as well as the angular momentum contributions to the proton spin. The impact parameter space interpretation of the GPD moments is discussed, which provides insights into the spatial distribution of unpolarized quarks and their correlations in the transverse plane of an unpolarized or transversely polarized proton.
\end{abstract}

\maketitle

\section{Introduction}\label{sec:intro}

The introduction of generalized parton distributions (GPDs)~\cite{Mueller:1998fv, Ji:1996ek, Radyushkin:1996nd} has opened up a new perspective into the three-dimensional imaging of the nucleon. These pioneering quantities reveal information on hadron structure far beyond the traditional one-dimensional parton distribution functions (PDFs), typically investigated in deep-inelastic scattering (DIS) experiments, and the transverse structure encoded in the various form factors. In particular, GPDs provide a more comprehensive and nuanced view of the internal structure of the nucleon, offering insights into the spatial distributions of quarks and gluons~\cite{Burkardt:2000za, Ralston:2001xs, Diehl:2002he, Burkardt:2002hr}. What is more, the moments of GPDs are related to the matrix elements of the energy-momentum tensor (EMT), from which we can gain valuable insights into the distribution of the hadron's internal energy, momentum, and pressure~\cite{Polyakov:2002wz, Polyakov:2002yz, Polyakov:2018zvc}, as well as the coupling of hadrons to gravity. Extracting the moments at zero momentum transfer allows the momentum fraction, spin, and angular momentum carried by the quarks and gluons inside the hadron to be determined. These information have the potential to improve our understanding of fundamental physics and could lead to significant advancements in various fields within nuclear and particle physics. 

There have been efforts to parameterize the GPDs and fit them from global experiments~\cite{Polyakov:2002wz,Guidal:2004nd,Goloskokov:2005sd,Mueller:2005ed,Kumericki:2009uq,Goldstein:2010gu,Gonzalez-hernandez:2012xap,Kriesten:2021sqc,Hashamipour:2021kes,Guo:2022upw,Guo:2023ahv}, but this field is still in its infancy, as it remains a challenge to extract the $x$-dependence of GPDs from 
{processes such as deeply-virtual Compton scattering (DVCS)}~\cite{Kumericki:2016ehc,Mueller:2014hsa,Bertone:2021yyz,Moffat:2023svr}. 
Other exclusive reactions, which could in principle alleviate this problem, include double DVCS~\cite{Guidal:2002kt, Belitsky:2002tf} and further processes where two particles (with one or both of them being photons) are detected in addition to the final-state nucleon~\cite{Pedrak:2017cpp,Duplancic:2018bum, Qiu:2022bpq, Qiu:2022pla, Duplancic:2023kwe}.
However, those are typically very difficult to measure.
Given the challenges in extracting GPDs from experiments, it is highly desirable to have lattice QCD results that provide complementary knowledge and potential guidance to experiments.

In this work, we focus on the GPDs with unpolarized quarks, $H$ and $E$, which are obtained from the Fourier transform of light-cone correlators
\begin{align}
F^{\mu}(z, P, \Delta) & = \langle p_f | \bar{q} (-\tfrac{z}{2}) \gamma^\mu \, {\cal W}(-\tfrac{z}{2}, \tfrac{z}{2})  q (\tfrac{z}{2}) | p_i\rangle \, ,
\label{eq:GPD_mx}
\end{align}
with $\gamma^\mu=\gamma^+$. The quark and anti-quark are separated along the light-cone direction $z=l n_-$ and are connected by the Wilson-line $\mathcal{W}(-\frac{z}{2},\frac{z}{2})=\mathcal{P}\exp(i\int_{-l n_-/2}^{l n_-/2}dl'A^+)$. Because of Lorentz invariance, the GPDs are functions of $x$ and two Lorentz-invariant products of the vectors $p_f,\,p_i$~\footnote{Alternatively, $P=(p_f+p_i)/2$, $\Delta=p_f-p_i$.} and $n_-$, which is conventionally chosen as $\xi=-(\Delta n_-)/(2Pn_-)$ and $t=\Delta^2$. Though lattice QCD can compute non-perturbative matrix elements, these are time-dependent quantities and cannot be computed for a lattice defined in Euclidean space. It is instead the first few Mellin moments of GPDs that are traditionally extracted from the lattice~\cite{Hagler:2003jd, QCDSF-UKQCD:2007gdl, Alexandrou:2011nr, Alexandrou:2013joa,Constantinou:2014tga,Green:2014xba,Alexandrou:2017ypw,Alexandrou:2017hac,Hasan:2017wwt,Gupta:2017dwj,Capitani:2017qpc,Alexandrou:2018sjm,Shintani:2018ozy,Jang:2018djx,Bali:2018qus,Bali:2018zgl,Alexandrou:2019ali,Constantinou:2020hdm,Alexandrou:2022dtc,LHPC:2007blg}, as matrix elements of local operators. However, due to signal decay and power-divergent mixing under renormalization, there are no moments beyond the third that exist.

Breakthrough was made about a decade ago when the quasi-PDF method was proposed~\cite{Ji:2013dva,Ji:2014gla} that utilizes operators holding the same form as \Eq{GPD_mx} but with equal-time quark fields that are separated along a spatial direction. Without loss of generality, the direction is chosen to be $z=ln_3$. This approach paves a way to relate the Euclidean matrix elements to the light-cone PDFs through factorization by an expansion in powers of $1/P_3$, with $P_3$ the hadron momentum, and makes it possible to compute $x$-dependent PDFs from lattice QCD. The quasi-PDF approach, often referred to as large momentum effective theory (LaMET)~\cite{Ji:2020ect}, was then extended to the GPDs and other light-cone quantities. Starting from the same quasi-PDF operator, the so-called Ioffe-time pseudo-distributions or  pseudo-PDFs (pPDFs)~\cite{Radyushkin:2017cyf, Orginos:2017kos} were proposed to extract either the Mellin moments or $x$-dependent PDFs by expanding in $z^2$. Several other approaches~\cite{Liu:1993cv,Braun:1994jq,Aglietti:1998ur,Detmold:2005gg,Chambers:2017dov,Ma:2014jla,Detmold:2021uru} also became available in the past few years.

Soon after the theoretical breakthrough, significant progress has been made for the calculation of PDFs~\cite{Lin:2014zya,Alexandrou:2015rja,Chen:2016utp,Alexandrou:2016jqi,Alexandrou:2017huk,Chen:2017mzz,Orginos:2017kos,Lin:2017ani,Alexandrou:2018pbm,Lin:2018pvv,Alexandrou:2018eet,Liu:2018uuj,Zhang:2018nsy,Sufian:2019bol,Alexandrou:2019lfo,Izubuchi:2019lyk,Joo:2019jct,Joo:2019bzr,Chai:2020nxw,Joo:2020spy,Bhat:2020ktg,Alexandrou:2020uyt,Alexandrou:2020qtt,Lin:2020ssv,Fan:2020nzz,Gao:2020ito,Lin:2020fsj,Karpie:2021pap,Alexandrou:2021oih,Egerer:2021ymv,HadStruc:2021qdf,Gao:2021dbh,Gao:2022iex,Gao:2022ytj,LatticeParton:2022xsd}, including higher-twist distributions~\cite{Bhattacharya:2020xlt,Bhattacharya:2020jfj,Bhattacharya:2021boh,Bhattacharya:2021moj}, parton distribution amplitudes~\cite{Zhang:2017zfe,Zhang:2017bzy,Bali:2018spj,Zhang:2020gaj,Hua:2020gnw,Detmold:2021qln,Hua:2022kcm,Gao:2022vyh}, GPDs~\cite{Chen:2019lcm,Alexandrou:2020zbe,Lin:2020rxa,Alexandrou:2021bbo,CSSMQCDSFUKQCD:2021lkf,Bhattacharya:2021oyr,Lin:2021brq,Bhattacharya:2022aob,Constantinou:2022fqt,Bhattacharya:2023tik} as well as transverse-momentum dependent parton distributions \cite{Shanahan:2020zxr,Zhang:2020dbb,Li:2021wvl,Shanahan:2021tst,Schlemmer:2021aij,LPC:2022ibr}. More information can be found in several recent reviews~\cite{Cichy:2018mum,Ji:2020ect,Constantinou:2020pek,Cichy:2021lih,Cichy:2021ewm}. 

For GPDs in particular, the Dirac structure $\gamma^\mu=\gamma^0$ was usually taken for the quasi-GPD matrix elements as proposed for quasi-PDFs and inspired by the $\gamma^{+}$ definition in the light cone. This definition is, however, frame-dependent for GPDs, requiring that calculations are defined in the symmetric kinematic frame: the momentum transfer is symmetrically distributed as $\vec{p}_f=\vec{P}+\vec{\Delta}/2$ and $\vec{p}_i=\vec{P}-\vec{\Delta}/2$. Encouraging results were reported, with, however, a heavy computational cost for every value of the momentum transfer~\cite{Chen:2019lcm,Alexandrou:2020zbe,Lin:2020rxa,Alexandrou:2021bbo,CSSMQCDSFUKQCD:2021lkf,Bhattacharya:2021oyr,Lin:2021brq}.

It was recently proposed and numerically proven that one can equivalently relate the matrix elements in any kinematic frame (e.g., $\vec{p}_f=\vec{P}$ and $\vec{p}_i=\vec{P}-\vec{\Delta}$) to the symmetric one, through Lorentz-invariant amplitudes based on the Lorentz-covariant parameterization of the matrix elements~\cite{Bhattacharya:2022aob,Constantinou:2022fqt,Bhattacharya:2023tik,Cichy:2023dgk}. This finding established a basis for faster and more efficient computation of GPDs using lattice QCD in asymmetric frames, allowing for flexibility in the distribution of transferred momentum between the initial and final states. Additionally, a novel Lorentz-invariant definition of quasi-GPDs matrix elements was proposed in the same work, which may lead to smaller power corrections in matching to the light-cone GPDs. Based on that, exploratory results using $x$-space matching and RI-MOM renormalization were established. In this work, we instead will apply the leading-twist short-distance factorization in coordinate space with a ratio-scheme renormalization to extract the first moments of GPDs at a broad range of values for the momentum transfer. By comparing our results with traditional moment calculations, we will be able to assess the efficacy of different definitions for quasi-GPDs, as well as access higher-order moments that are difficult or even impossible to calculate through traditional methods. 

The plan of the paper is as follows. In \sec{HandE}, we review the definition of quasi $H$ and $E$ GPDs. In \sec{bm+ratio}, we present our bare quasi-GPD matrix elements and the ratio scheme renormalization. In \sec{SDF}, we discuss the determination of Mellin moments of the proton GPDs using short-distance factorization with perturbative matching. In \sec{tdependence}, we discuss the $t$-dependence of GPD moments to determine the quark charges, momentum fraction, and total spin contribution to the proton. Additionally, we provide an interpretation of the results in the impact parameter space. Finally, \sec{conclusion} contains our conclusions. Some supplementary material is presented in the Appendix.

\section{Quasi $H$ and $E$ GPD matrix elements}\label{sec:HandE}

As discussed in \refcite{Bhattacharya:2022aob}, the quasi-GPD matrix elements can be parametrized in terms of Lorentz-invariant amplitudes ${\cal A}_i$, with certain kinematic factors. 
Therefore, quasi-GPD matrix elements in different frames can be related to each other, which largely reduces the computational cost for the study of GPDs at various values for the momentum transfer. This can be achieved via a calculation in a convenient asymmetric frame, in which the initial or final states do not carry any momentum transfer. Here, we choose $\vec{p}_f=\vec{P}$ and $\vec{p}_i=\vec{P}-\vec{\Delta}$. In this section, we will review some of the key aspects of \refcite{Bhattacharya:2022aob}, on which this work relies.

For spin-1/2 particles like the proton, the matrix elements defined in \Eq{GPD_mx} can be parametrized in terms of eight linearly-independent Dirac structures multiplied by eight Lorentz-invariant (frame-independent) amplitudes,
\begin{align}
\label{eq:parametrization_general}
F^{\mu} (z,P,\Delta) & = \bar{u}(p_f,\lambda') \bigg [ \dfrac{P^{\mu}}{m} {\cal A}_1 + m z^{\mu} {\cal A}_2 + \dfrac{\Delta^{\mu}}{m} {\cal A}_3 + i m \sigma^{\mu z} {\cal A}_4 + \dfrac{i\sigma^{\mu \Delta}}{m} {\cal A}_5 \nonumber \\[1ex]
& \hspace{5cm} + \dfrac{P^{\mu} i\sigma^{z \Delta}}{m} {\cal A}_6 + m z^{\mu} i\sigma^{z \Delta} {\cal A}_7 + \dfrac{\Delta^{\mu} i\sigma^{z \Delta}}{m} {\cal A}_8  \bigg ] u(p_i, \lambda) \, ,
\end{align}
where
$\sigma^{\mu \nu} \equiv \tfrac{i}{2} (\gamma^\mu \gamma^\nu - \gamma^\nu \gamma^\mu)$,  
$\sigma^{\mu z} \equiv \sigma^{\mu \rho} z_\rho$, 
$\sigma^{\mu \Delta} \equiv \sigma^{\mu \rho} \Delta_\rho$, $\sigma^{z \Delta} \equiv \sigma^{\rho \tau} z_\rho \Delta_\tau$, ${\cal A}_i \equiv {\cal A}_i (z\cdot P, z \cdot \Delta, \Delta^2, z^2)$. For convenience, we use the compact notation ${\cal A}_i \equiv {\cal A}_i (z\cdot P, z \cdot \Delta, \Delta^2, z^2)$. For the case of unpolarized quarks, there are two (vector) light-cone GPDs $H$ and $E$  defined through~\cite{Diehl:2002he},
\begin{align}
F^{+} (z, P, \Delta) & = \bar{u}(p_f, \lambda ') \bigg [\gamma^{+} H(z, P, \Delta)  + \frac{i\sigma^{+\mu}\Delta_{\mu}}{2m} E(z, P, \Delta) \bigg ] u(p_i, \lambda)\,.\label{eq:GPD_para_spin1/2}
\end{align}
Combining \Eq{parametrization_general} and \Eq{GPD_para_spin1/2}, one can derive,
\begin{align}
\label{eq:H}
H (z,P,\Delta) & = {\cal A}_1 + \dfrac{\Delta^{+}}{P^{+}} {\cal A}_3 \, ,\\
\label{eq:E}
E (z,P,\Delta) & = - {\cal A}_1 - \dfrac{\Delta^{+}}{P^{+}} {\cal A}_3 + 2 {\cal A}_5 + 2P^{+}z^- {\cal A}_6 + 2 \Delta^{+} z^- {\cal A}_8 \, ,
\end{align}
where $z^2=0$ is always ensured, since $z=ln_-$ is along the light-cone direction. Meanwhile, it is worth noting that the light-cone GPDs are solely dependent on Lorentz scalars, rendering them frame-independent. 

A similar approached can be followed in a Euclidean lattice calculation. Thus, historically, the quasi-GPD are defined from the $\gamma^\mu=\gamma^0$ matrix elements with spatial separation $z=(0,0,0,z_3)$, inspired by the success of quasi-PDFs and the lack of finite mixing~\cite{Constantinou:2017sej},
\begin{align}
F^{0} (z, P,\Delta) & = \langle p_f, \lambda' | \bar{q} (-\tfrac{z}{2}) \gamma^0 q (\tfrac{z}{2}) | p_i, \lambda\rangle 
\,=\, \bar{u}(p_f, \lambda ') \bigg [\gamma^{0} {\cal{H}}_0 (z,P,\Delta) + \frac{i\sigma^{0\mu}\Delta_{\mu}}{2m} {\cal{E}}_0 (z,P,\Delta) \bigg ] u(p_i, \lambda) \, .
\label{e:historic}
\end{align}
We can again express them in terms of the Lorentz-invariant amplitudes, ${\cal A}_i$. For example, in the symmetric frame with $\vec{p}_f=\vec{P}^s+\vec{\Delta}^s/2$ and $\vec{p}_i=\vec{P}^s-\vec{\Delta}^s/2$, one obtains
\begin{align}
\label{eq:quasiH_symm}
{\cal{H}}^{s}_0 (z,P^{s},\Delta^{s}) & = {\cal A}_1 + \dfrac{\Delta^{0,s}}{P^{0,s}} {\cal A}_3 - \dfrac{m^{2} \Delta^{0,s} z^3}{2P^{0,s} P^{3,s}} {\cal A}_4 + \bigg [ \dfrac{(\Delta^{0,s})^{2} z^{3}}{2P^{3,s}}  - \dfrac{\Delta^{0,s} \Delta^{3,s} z^3 P^{0,s}}{2(P^{3,s})^2} - \dfrac{z^{3} (\Delta^{s}_\perp)^2}{2P^{3,s}} \bigg ] {\cal A}_6 \nonumber \\[1ex]
& +  \bigg [ \dfrac{(\Delta^{0,s})^{3} z^{3}}{2P^{0,s} P^{3,s}}  - \dfrac{(\Delta^{0,s})^2 \Delta^{3,s} z^3}{2(P^{3,s})^2} - \dfrac{\Delta^{0,s} z^{3} (\Delta^{s}_\perp)^2}{2P^{0,s} P^{3,s}} \bigg ] {\cal A}_8 \, ,\\[3ex]
\label{eq:quasiE_symm}
{\cal{E}}^{s}_0 (z,P^{s},\Delta^{s}) & = - {\cal A}_1 - \dfrac{\Delta^{0,s}}{P^{0,s}} {\cal A}_3 + \dfrac{m^2 \Delta^{0,s} z^{3}}{2P^{0,s} P^{3,s}} {\cal A}_4 + 2 {\cal A}_5 + \bigg [ - \dfrac{(\Delta^{0,s})^{2}z^{3}}{2P^{3,s}}  + \dfrac{P^{0,s} \Delta^{0,s} \Delta^{3,s} z^{3}}{2 (P^{3,s})^{2}} + \dfrac{z^{3} (\Delta^{s}_\perp)^2}{2 P^{3,s}} - \dfrac{2z^{3}(P^{0,s})^{2}}{P^{3,s}} \bigg ] {\cal A}_6 \nonumber \\[1ex]
& + \bigg [ - \dfrac{(\Delta^{0,s})^{3}z^{3}}{2P^{0,s} P^{3,s}}  + \dfrac{(\Delta^{0,s})^2 \Delta^{3,s} z^{3}}{2(P^{3,s})^{2}} + \dfrac{\Delta^{0,s} z^{3}(\Delta^{s}_\perp)^2}{2P^{0,s} P^{3,s}} - \dfrac{ 2z^{3}P^{0,s} \Delta^{0,s}}{P^{3,s}} \bigg ] {\cal A}_8 \, . \\ \nonumber
\end{align}
while for the asymmetric frame with $\vec{p}_f=\vec{P}^a$ and $\vec{p}_i=\vec{P}^a-\vec{\Delta}^a$, the quasi-GPDs read
\begin{align}
\label{eq:quasiH_nonsymm}
{\cal{H}}^{a}_0 (z,P^{a},\Delta^{a}) & = {\cal A}_1 + \dfrac{\Delta^{0,a}}{{P}^{0,a}} {\cal A}_3 - \bigg [ \dfrac{m^2 \Delta^{0,a} z^3}{2{P}^{0,a} {P}^{3,a}} - \dfrac{1}{(1+\tfrac{\Delta^{3,a}}{2{P}^{3,a}})} \dfrac{m^2 \Delta^{0,a} \Delta^{3,a} z^3}{4 {P}^{0,a} ({P}^{3,a})^2} \bigg ] {\cal A}_4 \nonumber \\
& + \bigg [ \dfrac{(\Delta^{0,a})^2 z^3}{2{P}^{3,a}} - \dfrac{1}{(1+\tfrac{\Delta^{3,a}}{2{P}^{3,a}})} \dfrac{(\Delta^{0,a})^2 \Delta^{3,a} z^3}{4 ({P}^{3,a})^2} - \dfrac{1}{(1+\tfrac{\Delta^{3,a}}{2{P}^{3,a}})} \dfrac{{P}^{0,a} \Delta^{0,a} \Delta^{3,a} z^3}{2 ({P}^{3,a})^2} - \dfrac{z^3 (\Delta^{a}_\perp)^2}{2 {P}^{3,a}} \bigg ] {\cal A}_6
\nonumber\\
& + \bigg [ \dfrac{(\Delta^{0,a})^3 z^3}{2{P}^{0,a} {P}^{3,a}} - \dfrac{1}{(1+\tfrac{\Delta^{3,a}}{2{P}^{3,a}})} \dfrac{(\Delta^{0,a})^3 \Delta^{3,a} z^3}{4 {P}^{0,a} ({P}^{3,a})^2} - \dfrac{1}{(1+\tfrac{\Delta^{3,a}}{2{P}^{3,a}})} \dfrac{(\Delta^{0,a})^2 \Delta^{3,a} z^3}{2({P}^{3,a})^2} - \dfrac{z^3 (\Delta^{a}_\perp)^2 \Delta^{0,a}}{2 {P}^{0,a} {P}^{3,a}} \bigg ] {\cal A}_8 \, , \\[3ex]
\label{eq:quasiE_nonsymm}
{\cal{E}}^{a}_0 (z,P^{a},\Delta^{a}) & = - {\cal A}_1 - \dfrac{\Delta^{0,a}}{{P}^{0,a}} {\cal A}_3 -  \bigg [ - \dfrac{m^2 \Delta^{0,a} z^3}{2 {P}^{0,a} {P}^{3,a}} - \dfrac{1}{(1 + \tfrac{\Delta^{3,a}}{2 {P}^{3,a}})} \bigg ( \dfrac{m^2 z^3}{{P}^{3,a}} - \dfrac{m^2 \Delta^{0,a} \Delta^{3,a} z^3}{4 {P}^{0,a} ({P}^{3,a})^2} \bigg ) \bigg ] {\cal A}_4 + 2{\cal A}_5 \nonumber \\
& + \bigg [ - \dfrac{(\Delta^{0,a})^2 z^{3}}{2 {P}^{3,a}} - \dfrac{1}{(1 +\tfrac{\Delta^{3,a}}{2{P}^{3,a}})} \bigg ( \dfrac{{P}^{0,a} \Delta^{0,a} z^3}{ {P}^{3,a}} - \dfrac{(\Delta^{0,a})^2 \Delta^{3,a} z^3}{4 ({P}^{3,a})^2} \bigg ) - \dfrac{1}{(1+\tfrac{\Delta^{3,a}}{2 {P}^{3,a}})} \bigg ( \dfrac{2z^3 ({P}^{0,a})^2}{{P}^{3,a}} 
\nonumber\\
& - \dfrac{{P}^{0,a} \Delta^{0,a} \Delta^{3,a} z^3}{2 ({P}^{3,a})^2} \bigg ) + \dfrac{z^3 (\Delta^{a}_\perp)^2}{2{P}^{3,a}} \bigg ] {\cal A}_6 + \bigg [ - \dfrac{(\Delta^{0,a})^3 z^{3}}{2 {P}^{0,a}{P}^{3,a}} - \dfrac{1}{(1 +\tfrac{\Delta^{3,a}}{2{P}^{3,a}})} \bigg ( \dfrac{ (\Delta^{0,a})^2 z^3}{{P}^{3,a}} - \dfrac{(\Delta^{0,a})^3 \Delta^{3,a} z^3}{4 \overline{P}^{0,a} ({P}^{3,a})^2} \bigg ) \nonumber \\
& - \dfrac{1}{(1+\tfrac{\Delta^{3,a}}{2 {P}^{3,a}})} \bigg ( \dfrac{2z^3 {P}^{0,a} \Delta^{0,a}}{{P}^{3,a}} - \dfrac{(\Delta^{0,a})^2 \Delta^{3,a} z^3}{2 ({P}^{3,a})^2} \bigg ) + \dfrac{z^3 (\Delta^{a}_\perp)^2 \Delta^{0,a}}{2 {P}^{0,a} {P}^{3,a}} \bigg ] {\cal A}_8 \, .
\end{align}
It is evident that when the hadron momentum $\vec{P}=(0,0,P_3)$ is finite, the definition of $\gamma^0$ varies across different frames, though such discrepancies vanish in the infinite momentum limit. What follows is the question of which definition could be an appropriate choice that can provide good convergence to the light-cone quantities, and can be described by the perturbative matching. To explore this possibility, inspired by \Eq{H} and \Eq{E}, it is natural to define the quasi-GPD matrix elements also in a Lorentz-invariant form as
\begin{align}
\label{eq:Hq_improved}
{\cal H}^{s/a}_{\rm LI}  (z\cdot P, z \cdot \Delta, (\Delta)^2, z^{2})& = {\cal A}_1 + \dfrac{\Delta \cdot z}{P \cdot z} {\cal A}_3 \, , \\[3ex]
\label{eq:Eq_improved}
{\cal E}^{s/a}_{\rm LI}(z\cdot P, z \cdot \Delta, (\Delta)^2 ,z^2) & = - {\cal A}_1 - \dfrac{\Delta \cdot z}{P \cdot z} {\cal A}_3 + 2 {\cal A}_5 + 2 P \cdot z {\cal A}_6 + 2 \Delta \cdot z  {\cal A}_8 \, ,
\end{align}
with the only difference with their light-cone counterpart $H$ and $E$ being $z^2\neq0$ originated from the spatial separation such as $z=(0,0,0,z_3)$. All the amplitudes ${\cal A}^{s/a}_i$ can be extracted from either the symmetric or an asymmetric frame through linear combinations of the various $F^\mu$ ($\mu=0,1,2,3$), and any choice can be used in Eqs.~\eqref{eq:quasiH_symm} - \eqref{eq:Eq_improved} as long as a Lorentz transformation is applied to the kinematic factors to match the values of $t$. Consequentially, we can construct the quasi-GPDs in any frame, including the Lorentz-invariant ones, as has been proven in ~\refcite{Bhattacharya:2022aob}. In addition, with a reduced number of additional terms, the proposed Lorentz-invariant definition could potentially converge to the light-cone GPDs faster than the $\gamma^0$ definition. To explore this possibility, we will analyze results from different definitions and compare them with the traditional Mellin-moments calculations in this work.

\section{Bare matrix elements and ratio scheme renormalization}\label{sec:bm+ratio}

\subsection{Lattice setup}

In this study, we use a gauge ensemble of $N_f=2+1+1$ twisted-mass fermions with a clover term and Iwasaki-improved gluons~\cite{Alexandrou:2018egz}. The lattice size and spacing of the ensemble are $N_s\times N_t=32^3 \times 64$ and $a$ = 0.093 fm, respectively. The quark masses are tuned to produce a pion mass of 260 MeV. To extract the bare matrix elements of quasi-GPDs, we need to compute the two-point and three-point functions, namely
\begin{align}
\begin{split}
    C^{\rm 2pt}(\Gamma_0, p;t_s) = \sum_{\yvec} e^{-i \pvec \cdot (\yvec - \xvec)} \Gamma^0_{\alpha \beta} \langle {N^{(s)}_\alpha (\yvec, t_s ) \overline{N}^{(s^\prime)}_\beta (\xvec, 0)} \rangle,
\end{split}
\end{align}
and,
\begin{align}
\begin{split}
    C^{\rm 3pt}_\mu (\Gamma_\kappa, p_f, p_i; t_s, \tau) = \sum_{\yvec, \zvec_0} e^{-i \pvec_f \cdot (\yvec - \xvec)} e^{-i \qvec \cdot (\xvec - \zvec_0)} \Gamma^\kappa_{\alpha \beta}\langle{N_\alpha (\yvec, t_s) \mathcal{O}_\mu (\zvec_0 + z \hat{z}, \tau) \overline{N}_\beta (\xvec, 0)} \rangle,
\end{split}
\end{align}
where $N^{(s)}$ is the standard nucleon source under momentum smearing~\cite{Bali:2016lva} to improve the overlap with the proton ground state and suppress gauge noise. 
It was found that the statistical noise is $z$-dependent and reduces by a factor of 4-5 in the real part, and 2-3 in the imaginary part of the unpolarized GPDs~\cite{Alexandrou:2020zbe}. 
$\mathcal{O}_\mu=\bar\psi\left(z\right) \gamma_\mu {\cal W}(0,z)\psi\left(0\right)$ is the equal-time non-local operator with the quark fields separated along the $z$ direction. In this work, we compute the iso-vector ($u-d$) and iso-scalar ($u+d$) flavor combination. For the iso-scalar case, we ignore the disconnected diagrams, which were found to be negligible for the unpolarized case~\cite{Alexandrou:2021oih}. The unpolarized and polarized parity projectors $\Gamma_0$ and $\Gamma_\kappa$ are defined as
\begin{align}
	\Gamma_0 &= \frac{1}{4} \left(1 + \gamma_0\right)\,, \\
	\Gamma_\kappa &= \frac{1}{4} \left(1 + \gamma_0\right) i \gamma_5 \gamma_\kappa\,, \quad \kappa=1,2,3\,.
\end{align}
Taking advantage of the correlations between two- and three-point functions, we construct the ratio,
\begin{align}
R^\mu_\kappa (\Gamma_\kappa, p_f, p_i; t_s, \tau) = \frac{C^{\rm 3pt}_\mu (\Gamma_\kappa, p_f, p_i; t_s, \tau)}{C^{\rm 2pt}(\Gamma_0, p_f;t_s)} \sqrt{\frac{C^{\rm 2pt}(\Gamma_0, p_i, t_s-\tau)C^{\rm 2pt}(\Gamma_0, p_f, \tau)C^{\rm 2pt}(\Gamma_0, p_f, t_s)}{C^{\rm 2pt}(\Gamma_0, p_f, t_s-\tau)C^{\rm 2pt}(\Gamma_0, p_i, \tau)C^{\rm 2pt}(\Gamma_0, p_i, t_s)}}\,,
\end{align}
which, in the $t_s\rightarrow\infty$ limit, corresponds to the bare matrix elements of proton ground state matrix elements $\lim\limits_{t_s\rightarrow\infty} R^\mu_\kappa=\Pi_\mu(\Gamma_\kappa)$. To keep the statistical noise under control, we use a source-sink separation of $t_s = 10a = 0.93$ fm and take a plateau fit with respect to $\tau$ in a region of convergence. 
We postpone the study of excited states for future work targeting precision control. In \tb{stat}, we show the momenta $\vec{P}=(0,0,P_3)$ and $\vec{\Delta}$ as well as the statistics used in this work, where the notation for the symmetric frame is,
\begin{equation}
\label{eq:pf_symm}
\vec{p}^{\,s}_f=\vec{P} + \frac{\vec{\Delta}}{2} = \left(+\frac{\Delta_1}{2},+\frac{\Delta_2}{2},P_3\right)\,,\qquad
\vec{p}^{\,s}_i=\vec{P} - \frac{\vec{\Delta}}{2}= \left(-\frac{\Delta_1}{2},-\frac{\Delta_2}{2},P_3\right)\,,
\end{equation}
and for the asymmetric frame, in which all the momentum transfer is assigned to the initial state, is
\begin{equation}
\vec{p}^{\,a}_f=\vec{P} =  \left(0,0,P_3\right) \,,\qquad
\label{eq:pi_nonsymm}
\vec{p}^{\,a}_i=\vec{P} - \vec{\Delta} =  \left(-\Delta_1,-\Delta_2,P_3\right)\,.
\end{equation}
While $\vec{P}$ and $\vec{\Delta}$ are the same for both frames, they lead to different values of $-t$ due to the different distribution of the momentum transfer, that is
\begin{eqnarray}
-t^s = \vec{\Delta}^2\,,\qquad
-t^a = \vec{\Delta}^2 - (E(p')-E(p))^2    \,.
\label{eq:Qboosted}
\end{eqnarray}
We note that this work focuses on zero skewness, namely $\Delta_3=0$, and most of the hadron momentum $P$ is fixed at 1.25 GeV throughout the calculation. We combine all data contributing to the same value of momentum transfer $t=-\Delta^2$ with definite symmetry with respect to $P_3 \to -P_3$, $z_3\to -z_3$, and $\vec{\Delta} \to -\vec{\Delta}$.

\begin{table}[h!]
\begin{center}
\renewcommand{\arraystretch}{1.9}
\begin{tabular}{lcccc|cccc}
\hline
frame & $P_3$ [GeV] & $\quad \mathbf{\Delta}$ $[\frac{2\pi}{L}]\quad$ & $-t$ [GeV$^2$] & $\quad \xi \quad $ & $N_{\rm ME}$ & $N_{\rm confs}$ & $N_{\rm src}$ & $N_{\rm tot}$\\
\hline
N/A       & $\pm$1.25 &(0,0,0)  &0   &0   &2   &329  &16  &10528 \\
\hline
symm      & $\pm$0.83 &($\pm$2,0,0), (0,$\pm$2,0)  &0.69   &0   &8   &67 &8  &4288 \\
symm      & $\pm$1.25 &($\pm$2,0,0), (0,$\pm$2,0)  &0.69   &0   &8   &249 &8  &15936 \\
symm      & $\pm$1.67 &($\pm$2,0,0), (0,$\pm$2,0)  &0.69   &0   &8   &294 &32  &75264 \\
symm      & $\pm$1.25 &$(\pm 2,\pm 2,0)$           &1.38   &0   &16   &224 &8  &28672 \\
symm      & $\pm$1.25 &($\pm$4,0,0), (0,$\pm$4,0)  &2.77   &0   &8   &329 &32  &84224 \\
\hline
asymm  & $\pm$1.25 &($\pm$1,0,0), (0,$\pm$1,0)  &0.17   &0   &8   &269 &8  &17216\\
asymm      & $\pm$1.25 &$(\pm 1,\pm 1,0)$       &0.34   &0   &16   &195 &8  &24960 \\
asymm  & $\pm$1.25 &($\pm$2,0,0), (0,$\pm$2,0)  &0.65   &0   &8   &269 &8  &17216\\
asymm      & $\pm$1.25 &($\pm$1,$\pm$2,0), ($\pm$2,$\pm$1,0) &0.81   &0   &16   &195 &8  &24960 \\
asymm  & $\pm$1.25 &($\pm$2,$\pm$2,0)          &1.24    &0   &16  &195 &8   &24960\\
asymm  & $\pm$1.25 &($\pm$3,0,0), (0,$\pm$3,0)  &1.38   &0   &8   &269 &8  &17216\\
asymm      & $\pm$1.25 &($\pm$1,$\pm$3,0), ($\pm$3,$\pm$1,0)  &1.52   &0   &16   &195 &8  &24960 \\
asymm  & $\pm$1.25 &($\pm$4,0,0), (0,$\pm$4,0)  &2.29   &0   &8   &269 &8  &17216\\
\hline
\end{tabular}
\caption{Statistics for the symmetric and asymmetric frame matrix elements are shown. $N_{\rm ME}$, $N_{\rm confs}$, $N_{\rm src}$ and $N_{\rm total}$ are the number of matrix elements, configurations, source positions per configuration and total statistics, respectively.}
\label{tb:stat}
\end{center}
\end{table}

\begin{figure}[h!]
    \centering
    \includegraphics[width=0.4\textwidth]{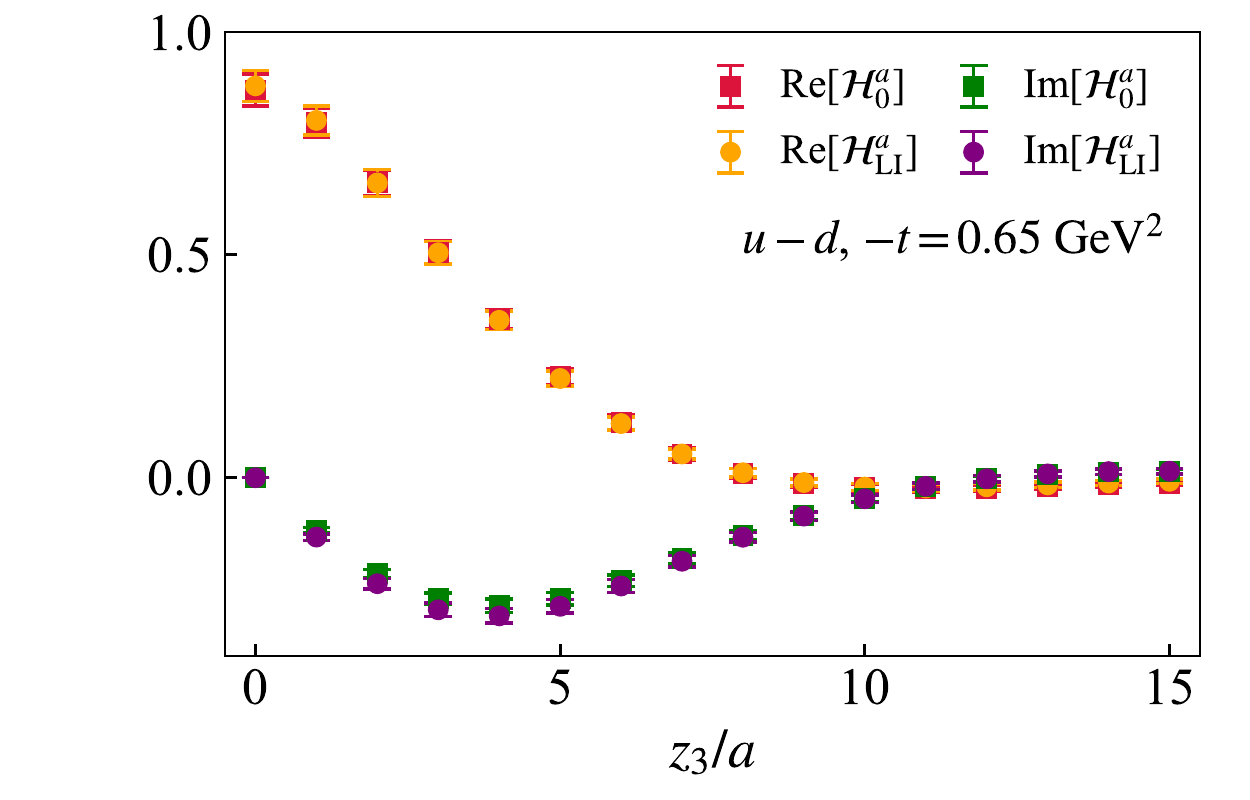}
    \includegraphics[width=0.4\textwidth]{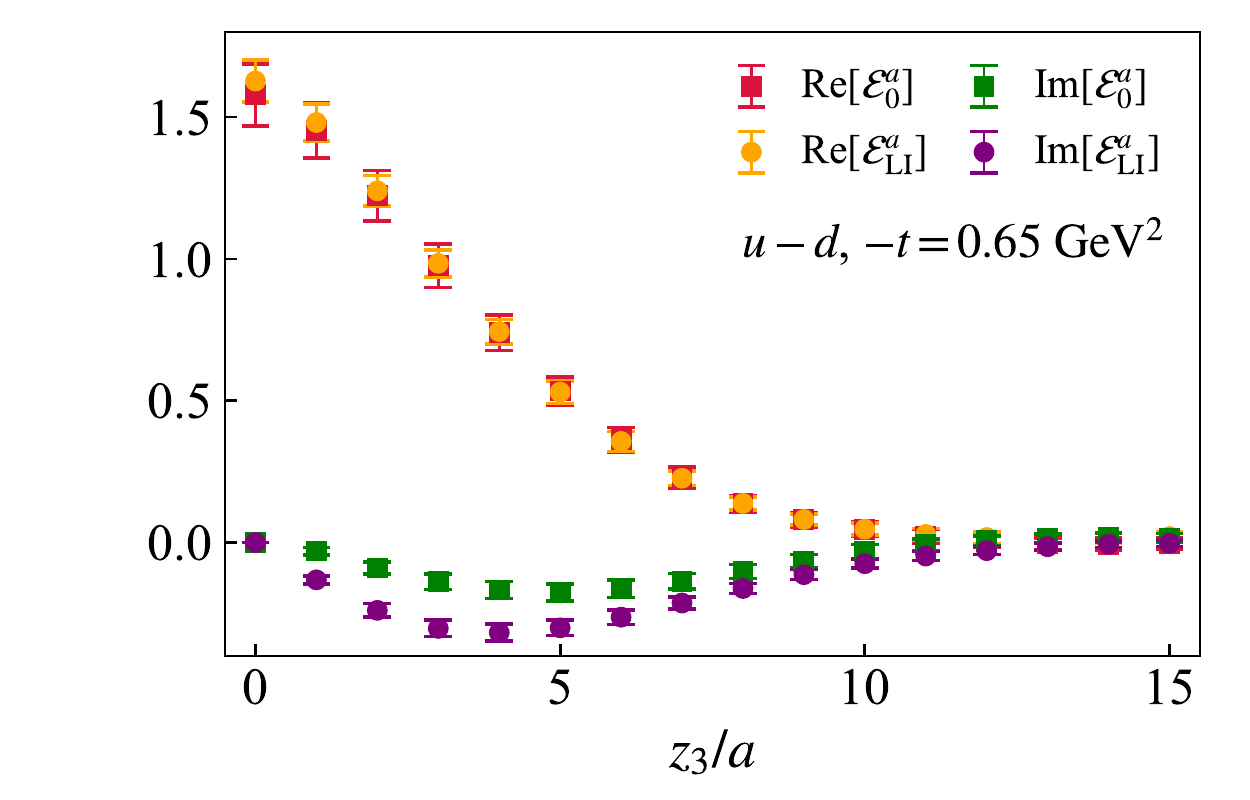}
	\caption{Bare matrix elements of the iso-vector quasi-GPDs from the asymmetric frame with momentum transfer $-t=0.65~\rm{GeV}^2$. The left panel shows $\mathcal{H}^a_0$ from the $\gamma_0$ definition and $\mathcal{H}^{a}_{\rm{LI}}$ from the Lorentz-invariant definition, while the right panel shows $\mathcal{E}^a_0$ from the $\gamma_0$ definition and $\mathcal{E}^{a}_{\rm{LI}}$ from the Lorentz-invariant definition.\label{fig:bm_t066}}
\end{figure}

\begin{figure}[h!]
    \centering
    \includegraphics[width=0.4\textwidth]{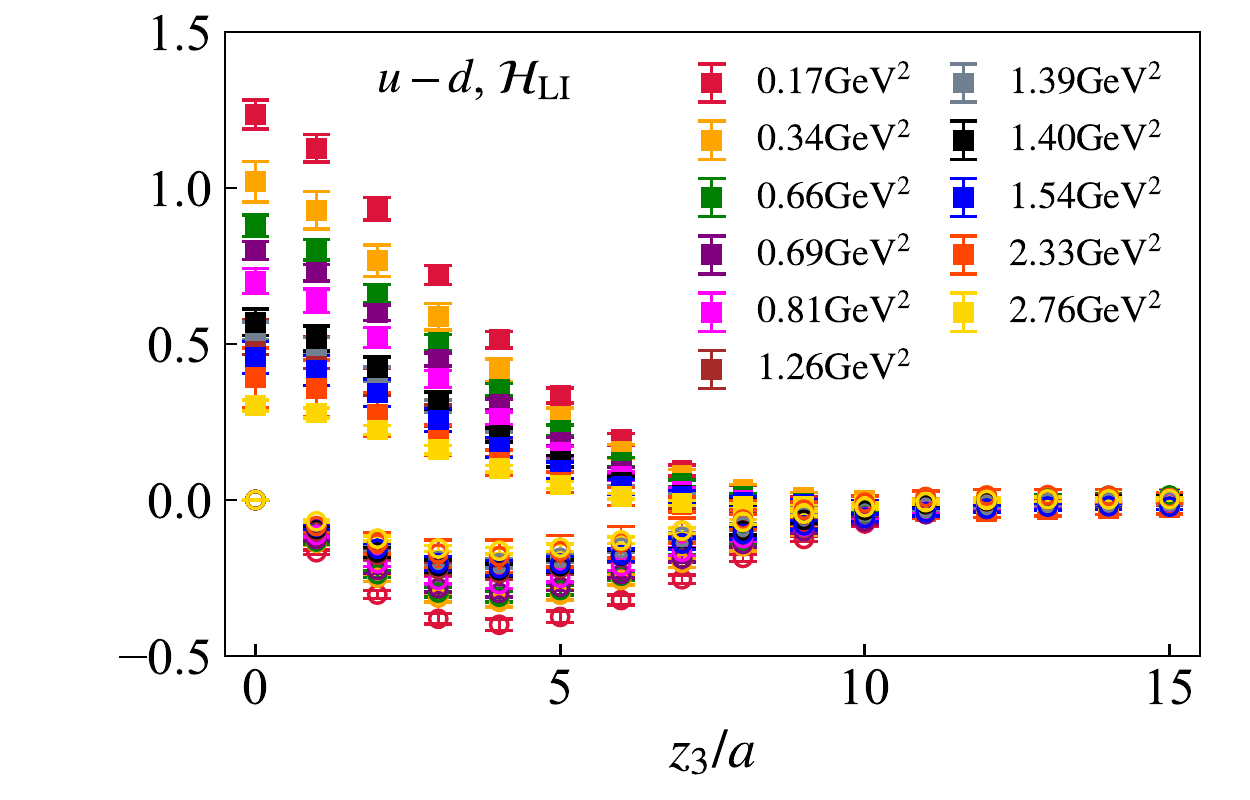}
    \includegraphics[width=0.4\textwidth]{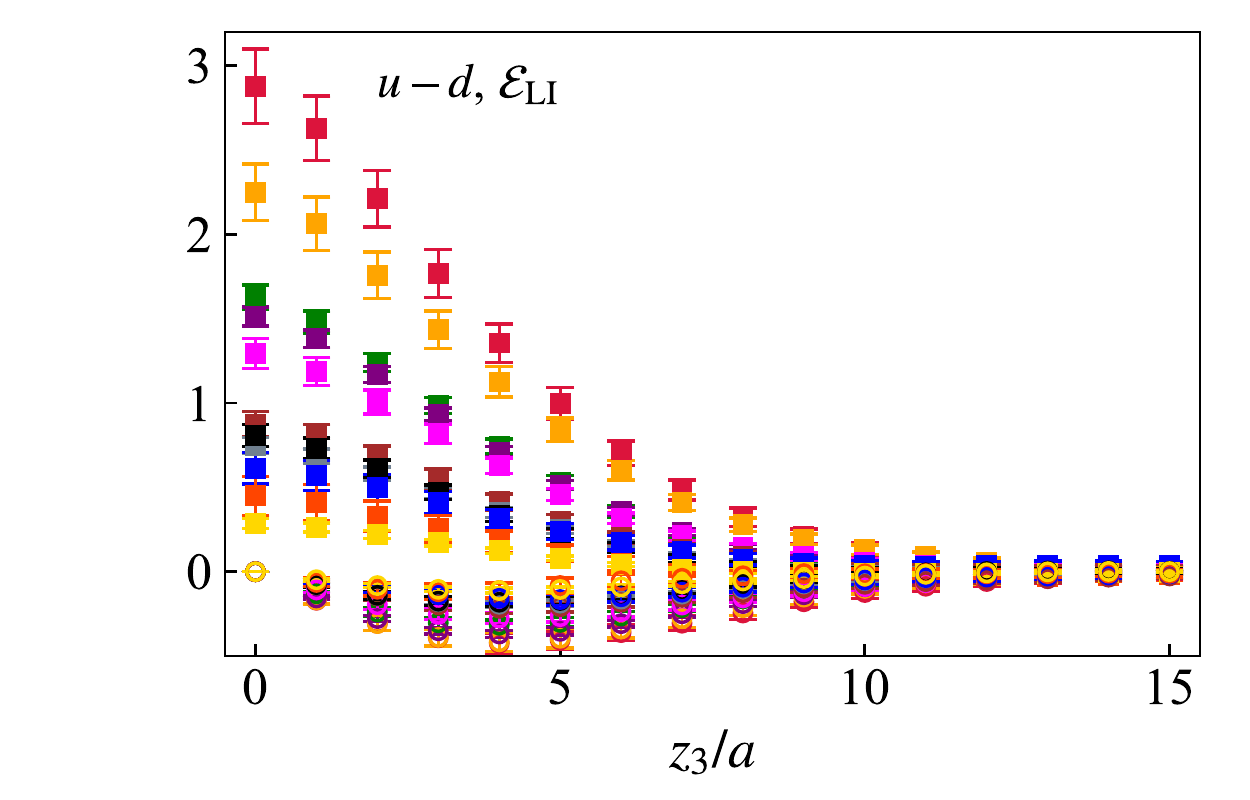}
    \includegraphics[width=0.4\textwidth]{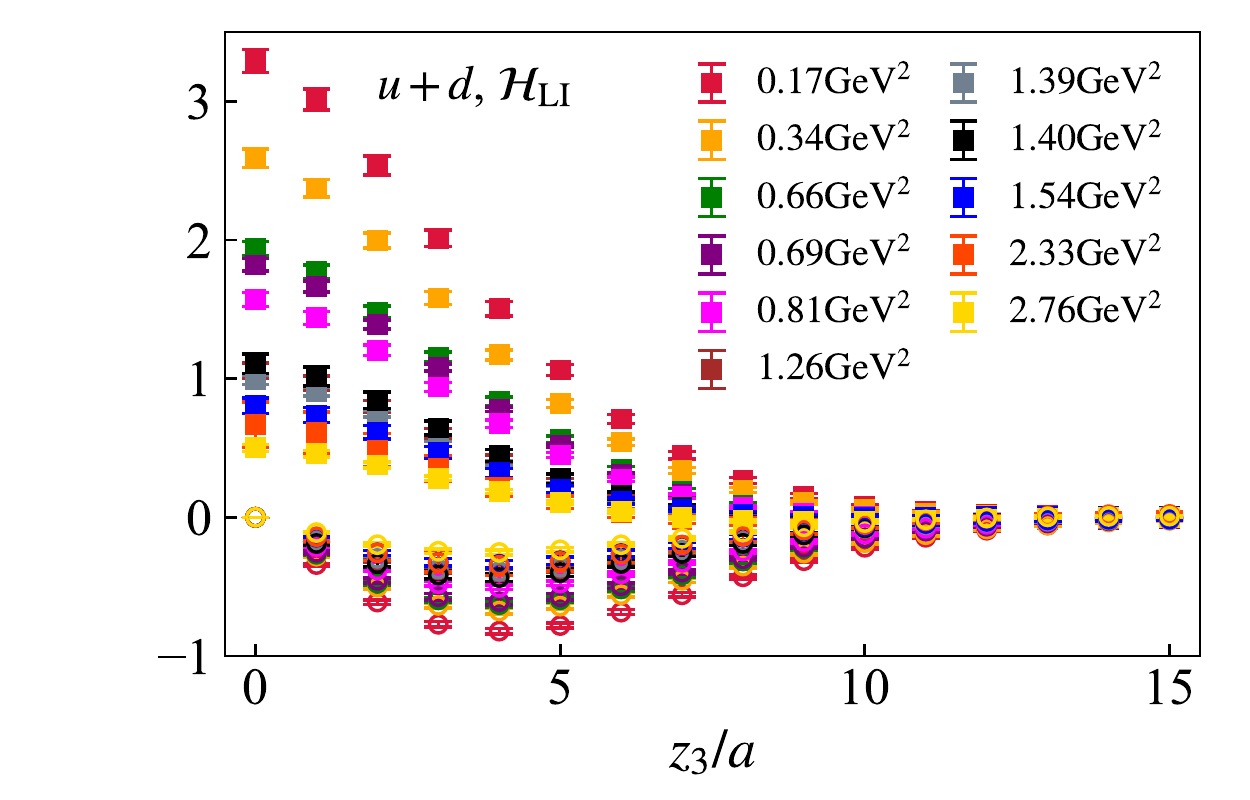}
    \includegraphics[width=0.4\textwidth]{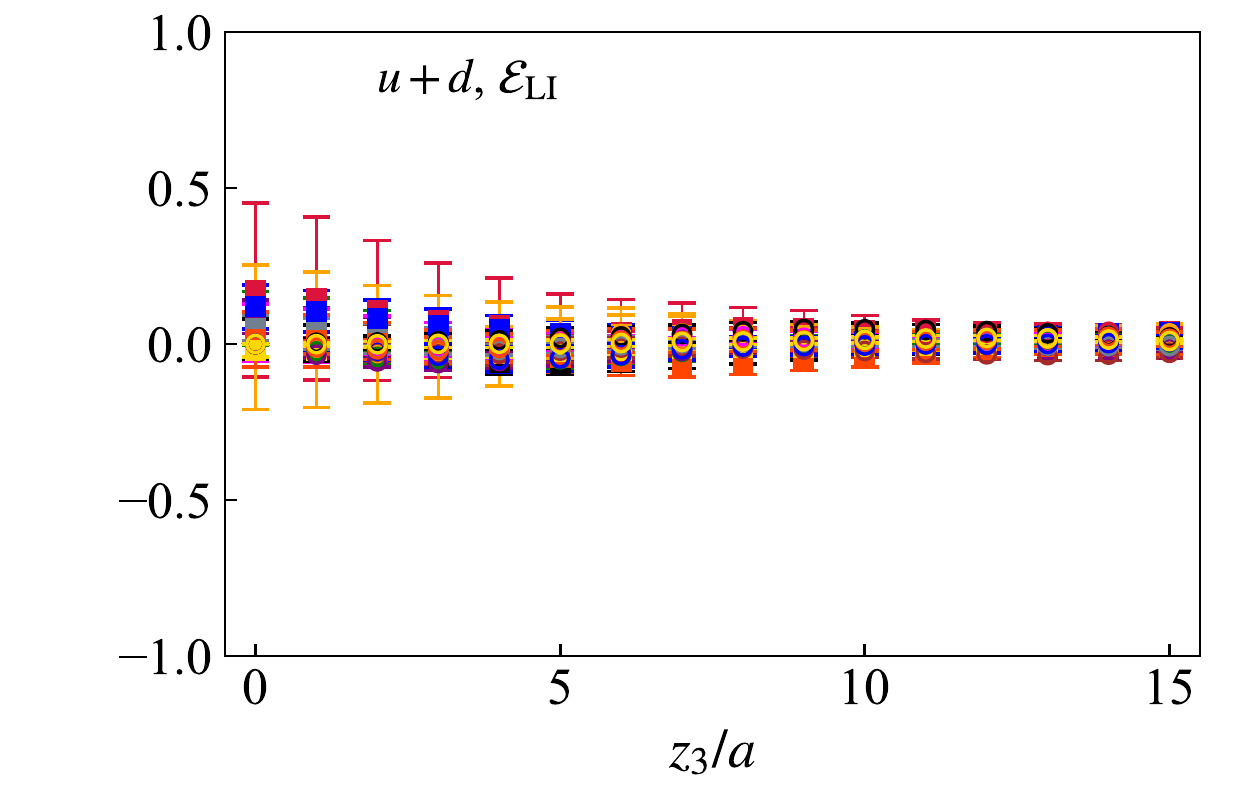}
	\caption{Bare matrix elements under Lorentz-invariant definition constructed from amplitudes ${\cal A}^{s/a}_i$ from both asymmetric and symmetric frame. The upper and lower panels are for the iso-vector and iso-scalar cases, with squared points for the real part and circled points for the imaginary part. The data shown are from hadron momentum $P_3$ = 1.25 GeV with all values of $-t$ indicated in Table~\ref{tb:stat}.\label{fig:bmLI}}
\end{figure}

\begin{figure}[h!]
    \centering
    \includegraphics[width=0.4\textwidth]{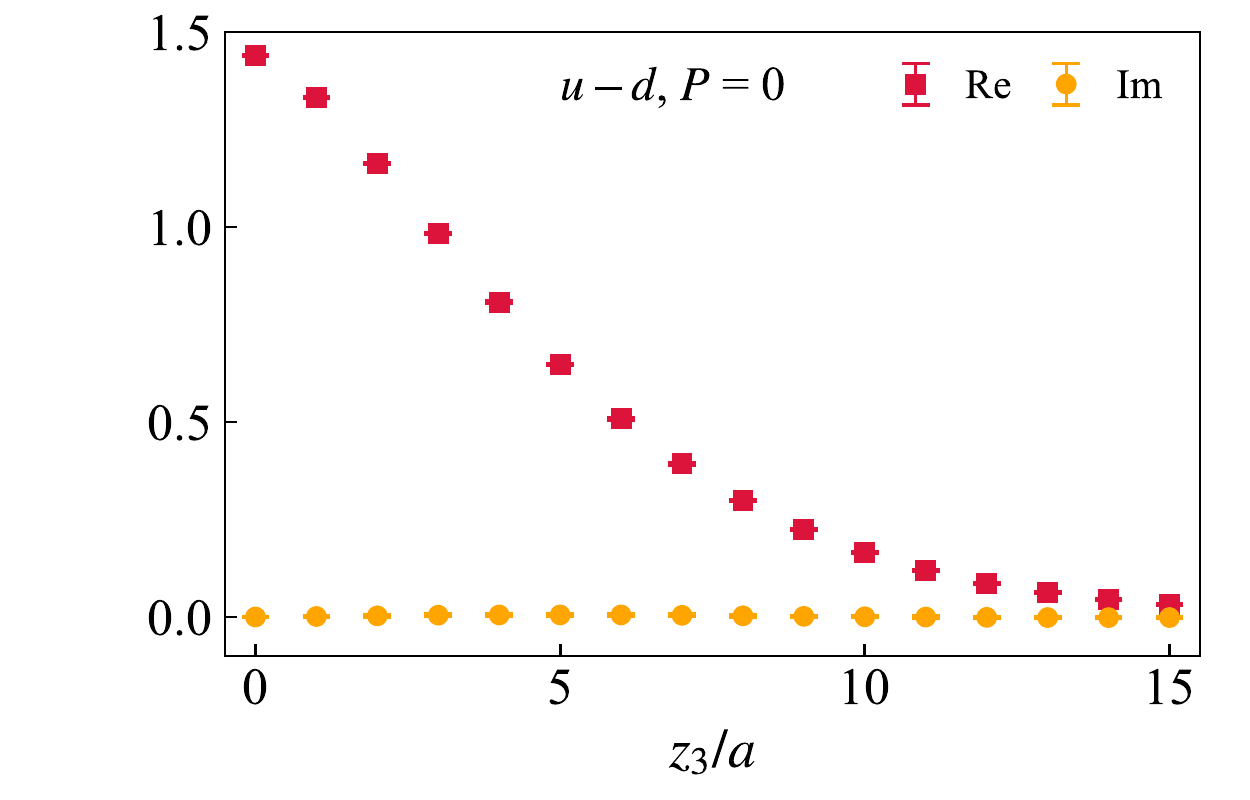}
	\caption{Bare matrix elements of iso-vector zero-momentum quasi-PDF matrix elements $\mathcal{F}(z,P=0,\Delta=0)$.\label{fig:bmqPDFP0}}
\end{figure}

In \fig{bm_t066}, we compare the iso-vector bare matrix elements under the $\gamma^0$ definition with the Lorentz-invariant definition, as extracted from the asymmetric frame data with momentum transfer $-t=0.65~\rm{GeV}^2$. As one can see, the difference between the two definitions for the $\mathcal{H}$ GPD is almost negligible, whereas, for the $\mathcal{E}$ GPD, particularly its imaginary part, the difference is considerable. This effect is similarly observed at other values of momentum transfer and will be reflected in the moments obtained through the subsequent analysis. 
In \fig{bmLI}, we summarize all the matrix elements at various values of $-t$ using the Lorentz-invariant definition derived from amplitudes ${\cal A}^{s/a}_i$ in both asymmetric and symmetric frames. The matrix elements under the $\gamma_0$ definition are shown in \app{g0bm}. We clarify that ${\cal H}^s_0$ and ${\cal E}^s_0$ use ${\cal A}^{s}_i$, while ${\cal H}^a_0$ and ${\cal E}^a_0$ use ${\cal A}^{a}_i$. We remind the reader that one may use the ${\cal A}_i$ from any frame along with the appropriate Lorentz transformation to the kinematic coefficients. The iso-vector and iso-scalar cases are both shown in the upper and lower panels, with squared points for the real part and circled points for the imaginary part. As one can observe, the matrix elements exhibit a decreasing trend as the momentum transfer $-t$ increases. In the iso-scalar case, the $\mathcal{E}_{\rm LI}$ are mostly consistent with zero within the errors. For completeness, we show in Fig.~\ref{fig:bmqPDFP0} the bare matrix element for the iso-vector quasi-PDF that is used in the ratio renormalization scheme, which are purely real.

\newpage
\subsection{Ratio scheme renormalization}\label{sec:ratio}

The bare matrix elements contain UV divergences related to the operator and the Wilson line. Thus, we have to renormalize the matrix elements prior to extracting physical quantities. It is known that the UV divergence of the quark bilinear operator is multiplicative and independent of the Dirac structure $\Gamma_\mu$ matrix as well as the hadron state~\cite{Ji:2017oey,Green:2017xeu,Ishikawa:2017faj}. One can therefore remove the UV divergence by constructing the renormalization group (RG) invariant ratios using the bare matrix elements with different hadron states but the same operators~\cite{Fan:2020nzz}. In this work, we construct the ratios,
\begin{align}
	\mathcal{M}(z,P,\Delta)=\frac{\mathcal{F}(\vec{z},\vec{P},\vec{\Delta})}{\mathcal{F}(\vec{z},\vec{P}=0,\vec{\Delta}=0)}=\frac{\mathcal{F}^R(\vec{z},\vec{P},\vec{\Delta})}{\mathcal{F}^R(\vec{z},\vec{P}=0,\vec{\Delta}=0)}\,,
\end{align}
where the ratio of bare matrix elements $\mathcal{F}$ is RG invariant, so that is equivalent to the ratio of renormalized matrix elements $\mathcal{F}^R$. $\mathcal{F}(\vec{z},\vec{P},\vec{\Delta})$ in the numerator can be either $\mathcal{H}$ or $\mathcal{E}$ quasi-GPDs for the iso-vector ($u-d$) and iso-scalar ($u+d$) cases. For the denominator, we chose to always take the iso-vector $\vec{P}=0$ quasi-PDF matrix elements as shown in \fig{bmqPDFP0}, since the renormalization factors are identical. A discussion of iso-scalar $\vec{P}=0$ quasi-PDF matrix elements can be found in \app{pz0}. At $z=0$, the iso-vector matrix element $\mathcal{F}(\vec{z}=0,\vec{P}=0,\vec{\Delta}=0)$ gives the renormalization constant $Z_V$. 

\begin{figure}[h!]
    \centering
    \includegraphics[width=0.4\textwidth]{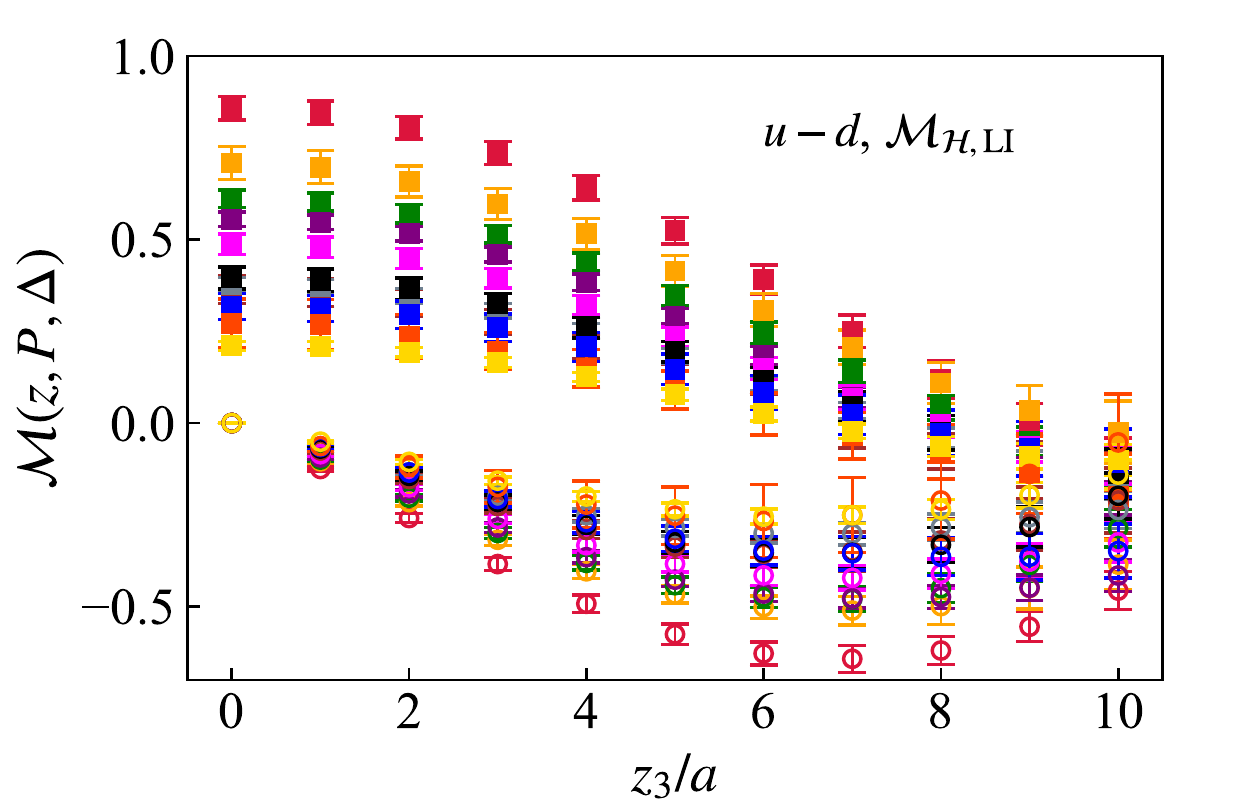}
    \includegraphics[width=0.4\textwidth]{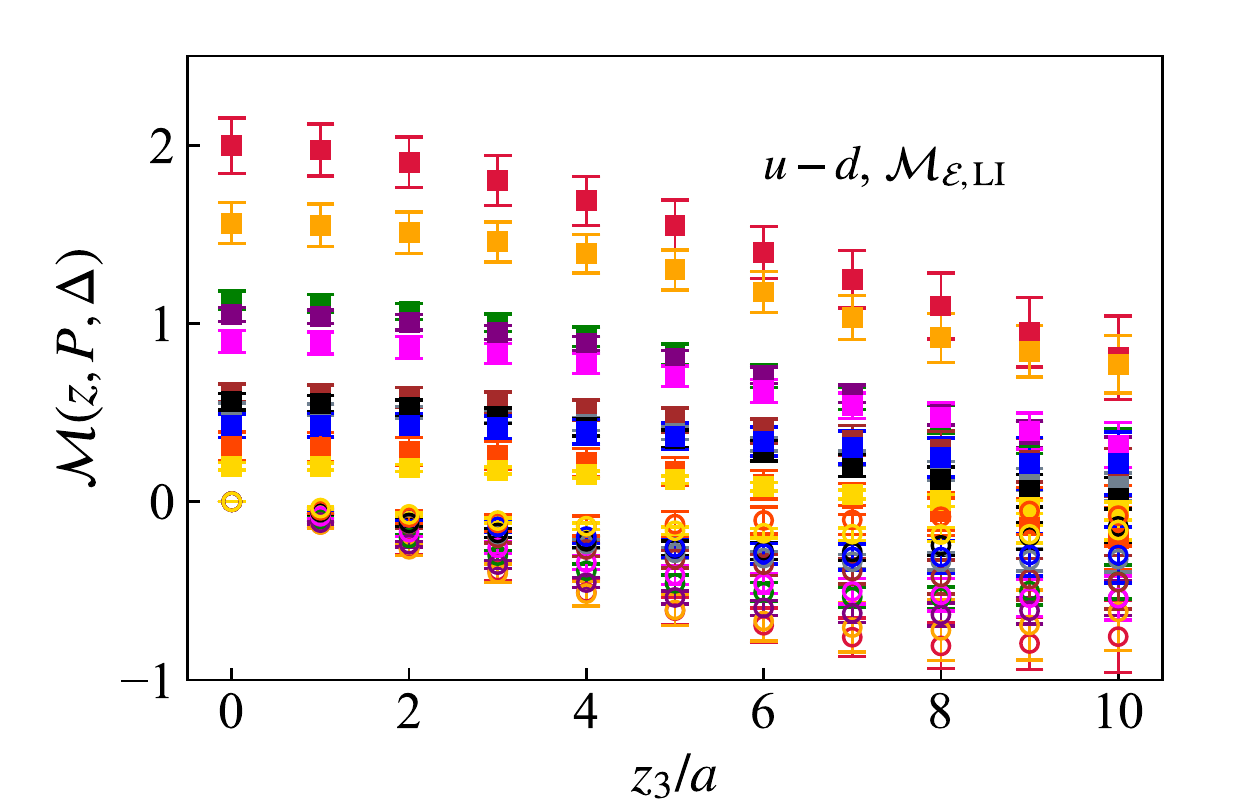}
    \includegraphics[width=0.4\textwidth]{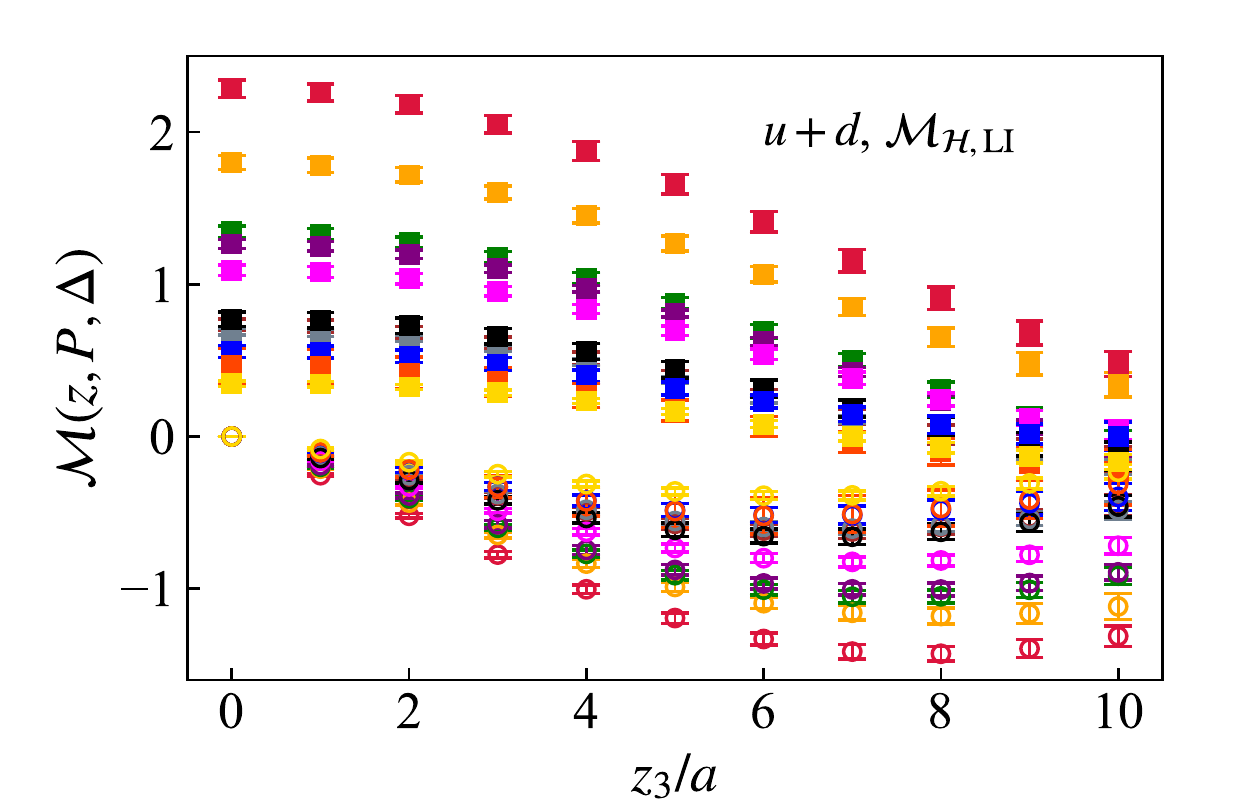}
    \includegraphics[width=0.4\textwidth]{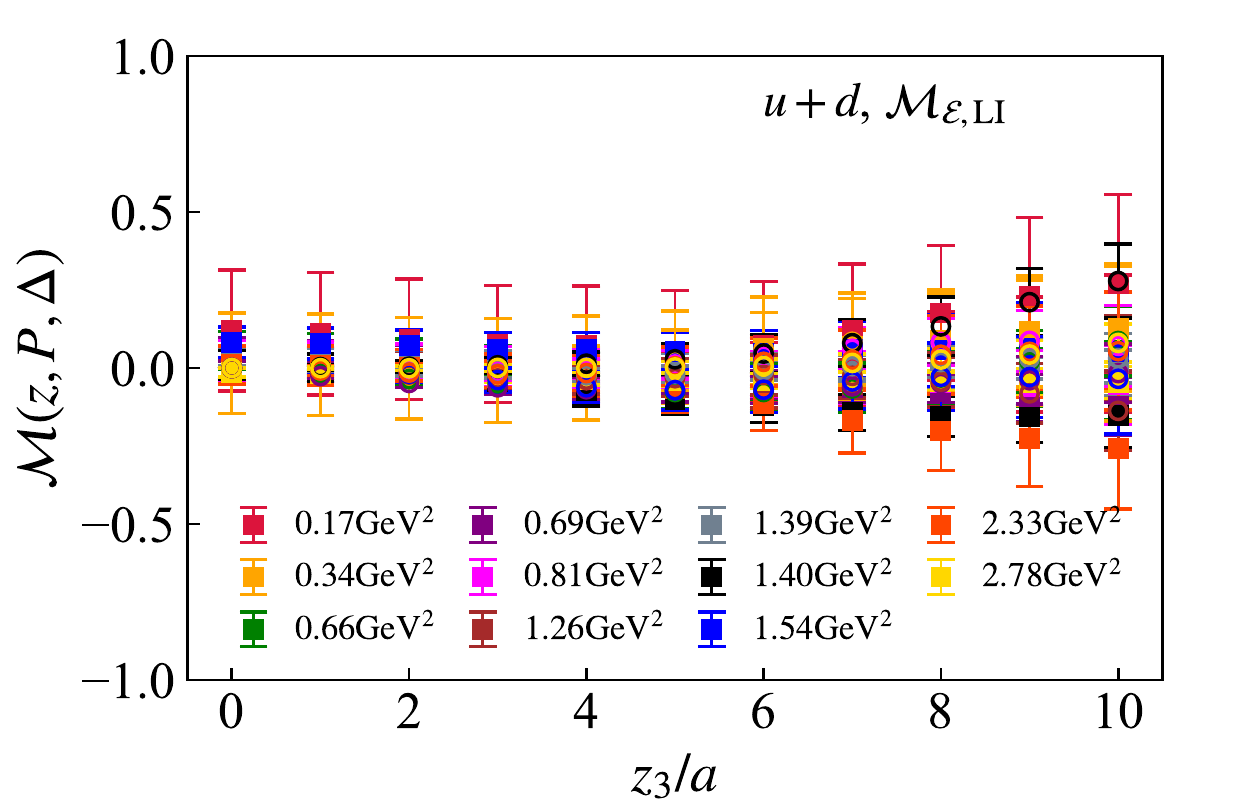}
    	\caption{Ratio scheme  renormalized matrix elements $\mathcal{M}(z,P_z,P_z^0)$ for isovector $(u-d)$ case (upper panels) and iso-scalar $(u+d)$ case (lower panels). The filled squared symbols are for the real part, while the circled open symbols are for the imaginary part.\label{fig:MisoVS}}
\end{figure}

The RG invariant ratios, $\mathcal{M}(z,P,\Delta)$, can also be written as a function of Lorentz-invariant variables, that is, $\mathcal{M}(z^2,zP,\Delta^2)$, with the $zP=z_3P_3$ being the so-called Ioffe time. We will use notation $zP$ instead of $z_3P_3$ in what follows. If the momentum transfer $\Delta$ equals zero in ${\cal F}$, the $\mathcal{M}(z,P,\Delta=0)$ will reduce to the standard Ioffe-time pseudo-distribution~\cite{Radyushkin:2017cyf, Orginos:2017kos}.
In \fig{MisoVS}, we show the ratio scheme renormalized matrix elements for the iso-vector (upper panels) and iso-scalar (lower panels) quasi-GPDs. The matrix elements shown are from the Lorentz-invariant definition of quasi-GPDs (bare matrix elements presented in \fig{bmLI}, with the filled squared symbols for the real part and the circled open symbols for the imaginary part. It can be seen that the renormalized matrix elements have a good signal and show a clear dependence on the momentum transfer $-t$. The iso-scalar $\cal E$ quasi-GPDs are still mostly consistent with zero after renormalization, as also observed in the bare case.

\section{Short distance factorization and Mellin moments.}\label{sec:SDF}
\subsection{Short distance factorization}\label{sec:OPE}

The renormalized matrix elements can be expanded in $z^2$, namely the short distance factorization (SDF). In the case of zero skewness, $\xi=0$, the leading-twist SDF for the quasi-GPDs is the same as in the quasi-PDF case~\cite{Liu:2019urm}, which in the $\overline{\rm{MS}}$ scheme reads~\cite{Radyushkin:2017cyf, Orginos:2017kos, Izubuchi:2018srq},
\begin{align}\label{eq:OPE}
	\begin{split}
		\mathcal{F}^{\overline{\rm{MS}}}(z,P,\Delta)=\sum_{n=0}\frac{(-izP)^n}{n!}C^{\overline{\rm{MS}}}_n(\mu^2 z^2) \langle x^n\rangle+ {\cal O}(\Lambda_{\rm QCD}^2 z^2),
	\end{split}
\end{align}
where $C^{\overline{\rm{MS}}}_n(\mu^2 z^2)$ are Wilson coefficients calculated from perturbation theory, which are available up to NNLO~\cite{Chen:2020ody,Li:2020xml} for the iso-vector ($u-d$) case, but only up to NLO for the iso-scalar ($u+d$) one with complete calculation including the quark-gluon mixing~\cite{Ji:2022thb}. The $\langle x^n\rangle$ are the Mellin moments of GPDs, defined as
\begin{align}
\begin{split}
	\int_{-1}^1dxx^nH^q(x,\xi=0,t)&=\sum^n_{\stackrel{i=0}{\rm{even}}}A^q_{n+1,i}(t),\\
	\int_{-1}^1dxx^nE^q(x,\xi=0,t)&=\sum^n_{\stackrel{i=0}{\rm{even}}}B^q_{n+1,i}(t),\\
\end{split}
\end{align}
in which the second moments $\langle x\rangle$, namely $A_{20}$ and $B_{20}$, are of particular interest due to their connection to the QCD energy-momentum tensor as gravitational form factors and provide access to the angular momentum sum rule. 
The $\langle x^0\rangle$ are essentially the Dirac (for $H$) and Pauli (for $E$) form factors. 
This SDF formula suffers from higher-twist contamination ${\cal O}(\Lambda_{\rm QCD}^2 z^2)$ growing as a function of $z^2$. 
Thus, it is expected to be valid for small spatial separations $z^2$. As a consequence, the matrix elements will only be sensitive to the lower moments, as the higher ones are factorially suppressed by $(-izP)^n/n!$ for a finite hadron momentum $P_3$. 
Despite the abovementioned challenges, it is desirable to pursue the direction of extracting Mellin moments from an SDF, because traditional calculations of higher-order moments through local operators suffer from notorious mixing between higher-dimensional and lower-dimensional operators~\cite{Hagler:2003jd, QCDSF-UKQCD:2007gdl, Alexandrou:2011nr, Alexandrou:2013joa,Constantinou:2014tga,Green:2014xba,Alexandrou:2017ypw,Alexandrou:2017hac,Hasan:2017wwt,Gupta:2017dwj,Capitani:2017qpc,Alexandrou:2018sjm,Shintani:2018ozy,Jang:2018djx,Bali:2018qus,Bali:2018zgl,Alexandrou:2019ali,Constantinou:2020hdm,Alexandrou:2022dtc}. In contrast, quasi-GPDs involve only a non-local operator of dimension three, and therefore, the higher moments can be systematically accessed by increasing the hadron momentum. Inserting the SDF formula into the ratio scheme renormalized matrix elements in \sec{ratio}, one obtains~\cite{Fan:2020nzz, Gao:2020ito}
\begin{align}\label{eq:ratioOPE}
	\begin{split}
		\mathcal{M}(z,P,\Delta)=\sum_{n=0}\frac{(-izP)^n}{n!}\frac{C^{\overline{\rm{MS}}}_n(\mu^2 z^2)}{C^{\overline{\rm{MS}}}_0(\mu^2 z^2)} \langle x^n\rangle+ {\cal O}(\Lambda_{\rm QCD}^2 z^2)\,.
	\end{split}
\end{align}
The ratio scheme renormalization has the advantage of potentially reducing the higher-twist contamination, because of the latter's cancellation between the numerator and denominator. Nevertheless, it is still important to keep the value of $z$ moderate so that the SDF does not break down. For practical reasons, one needs to truncate \Eq{ratioOPE} up to $n_{\rm max}$ when performing the fit, and then minimizing the $\chi^2$ defined as,
\begin{align}\label{eq:chisqfit}
	\chi^2=\sum_{P_3}\sum_{z=z_{\rm min}}^{z_{\rm max}}\left(\frac{({\rm Re}[\mathcal{M}^{\rm SDF}(z,P,\Delta)]-{\rm Re}[\mathcal{M}^{\rm data}(z,P,\Delta)])^2}{\sigma_{\rm Re}^2}+\frac{(\rm{Im}[\mathcal{M}^{\rm SDF}(z,P,\Delta)]-{\rm Im}[\mathcal{M}^{data}(z,P,\Delta)])^2}{\sigma_{\rm Im}^2}\right)\,,
\end{align}
from which we can determine the moments up to $\langle x^{n_{\rm max}}\rangle$. It is worth mentioning that the real and imaginary parts of $\mathcal{M}(z,P,\Delta)$ will provide even and odd moments, respectively.

\subsection{Fixed-$z^2$ analysis and pQCD correction}\label{sec:momsKernel}

At leading-order approximation, that is, for $\mathcal{O}(\alpha_s^0)$ and with $C_n(\mu^2z^2)=1$, the factorization formula in \Eq{ratioOPE} reduces to a polynomial function of $zP$ with coefficients $\langle x^n\rangle$ for a fixed momentum transfer $-t$. Beyond the leading approximation, the renormalized matrix element or the moments non-trivially depend on the physical scale $z^2$. At short distance, the Wilson coefficients $C_n(\mu^2z^2)/C_0(\mu^2z^2)$ are expected to compensate for this scale dependence and evolve the moments to the factorization scale $\mu$. 
\begin{figure}[h!]
    \centering
    \includegraphics[width=0.4\textwidth]{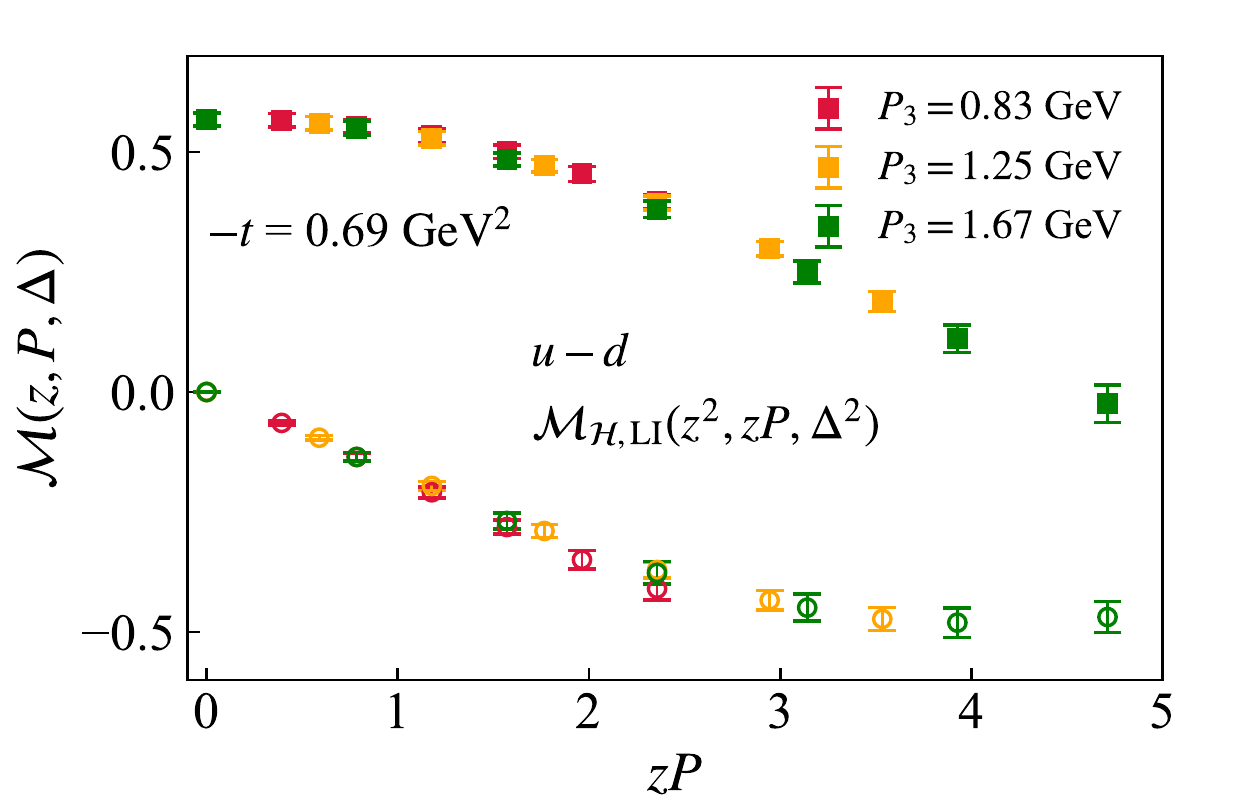}
    \includegraphics[width=0.4\textwidth]{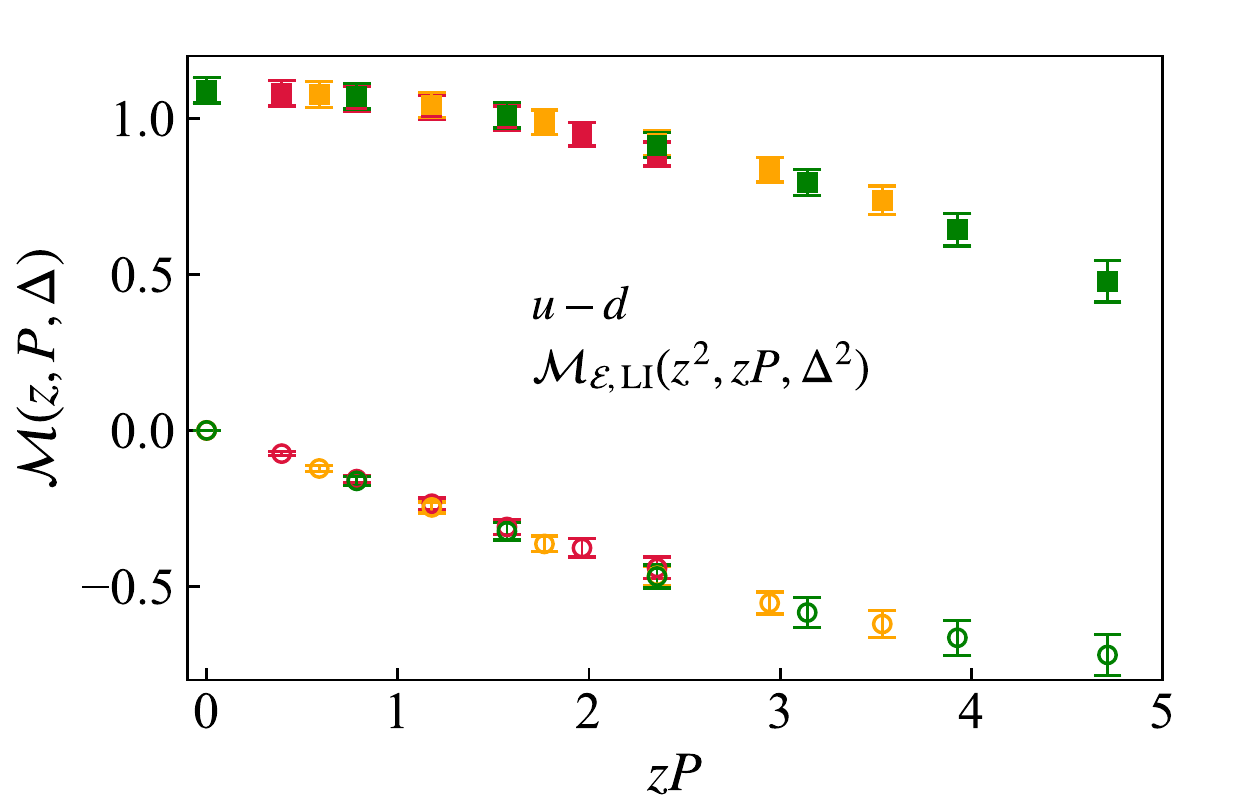}
    \includegraphics[width=0.4\textwidth]{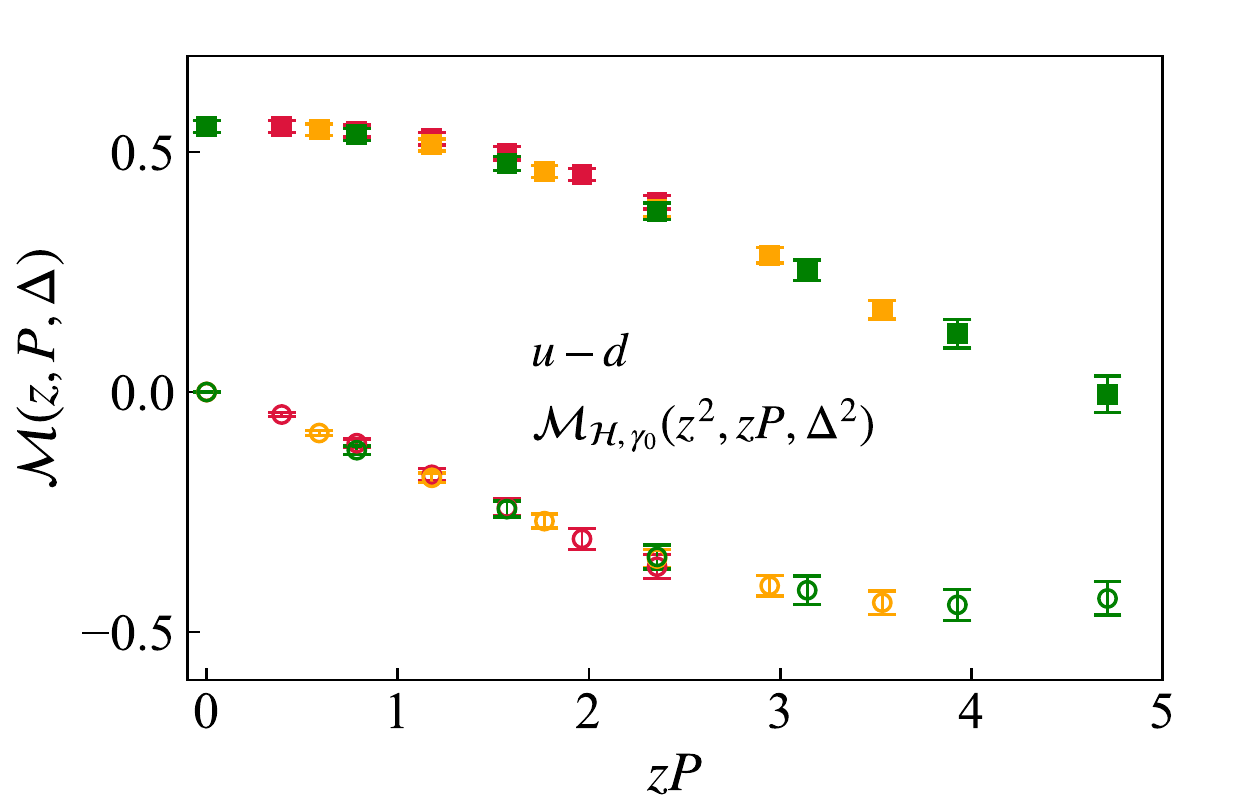}
    \includegraphics[width=0.4\textwidth]{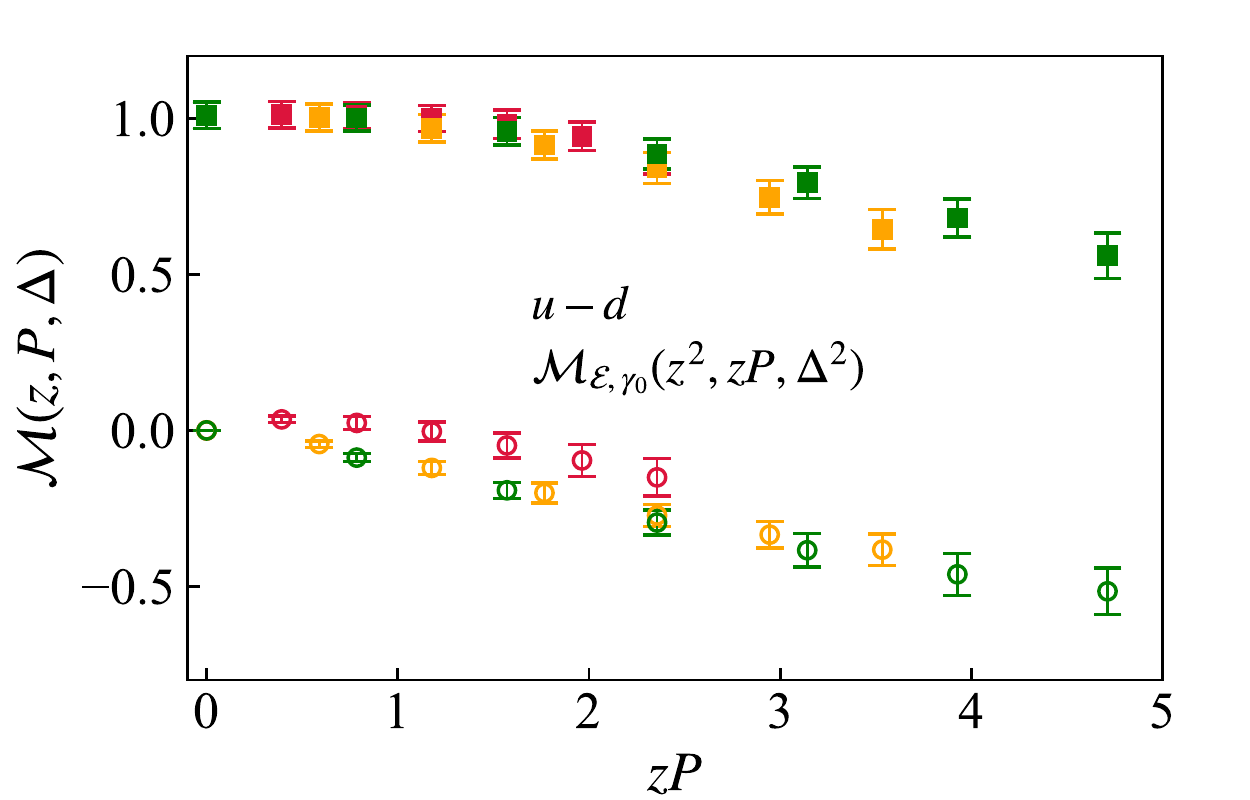}
	\caption{Iso-vector ratio scheme renormalized matrix elements from the Lorentz-invariant definition (upper panels) and the $\gamma_0$ definition (lower panels) at momentum transfer $-t=0.69~\rm{GeV}^2$ as a function of $zP$. The $z_3$ shown are in the range $[a,6a]$. We have three different values of the momentum $P_3=0.83,1.25$ and 1.67 GeV for this $-t$. The filled squared symbols are for the real part, while the circled open symbols are for the imaginary part.\label{fig:rGPDzpz}}
\end{figure}

\begin{figure}[h!]
    \centering
    \includegraphics[width=0.4\textwidth]{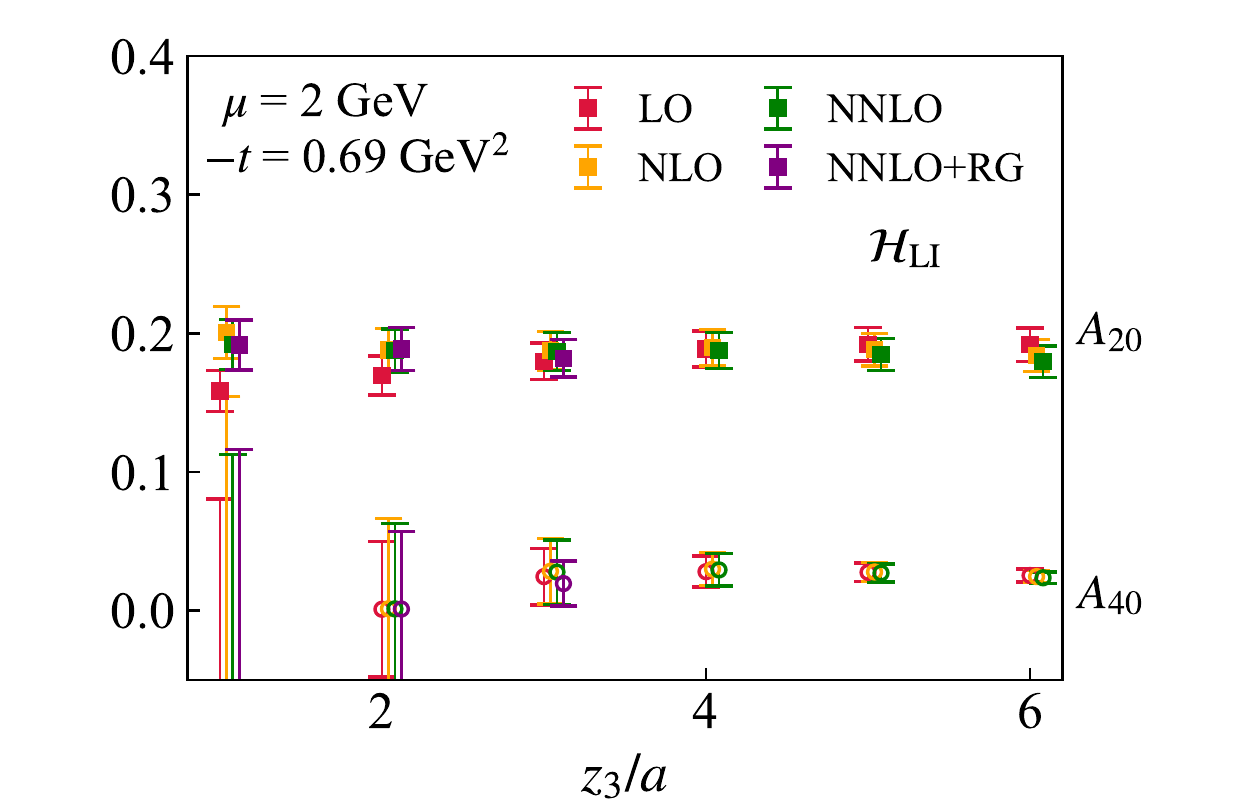}
    \includegraphics[width=0.4\textwidth]{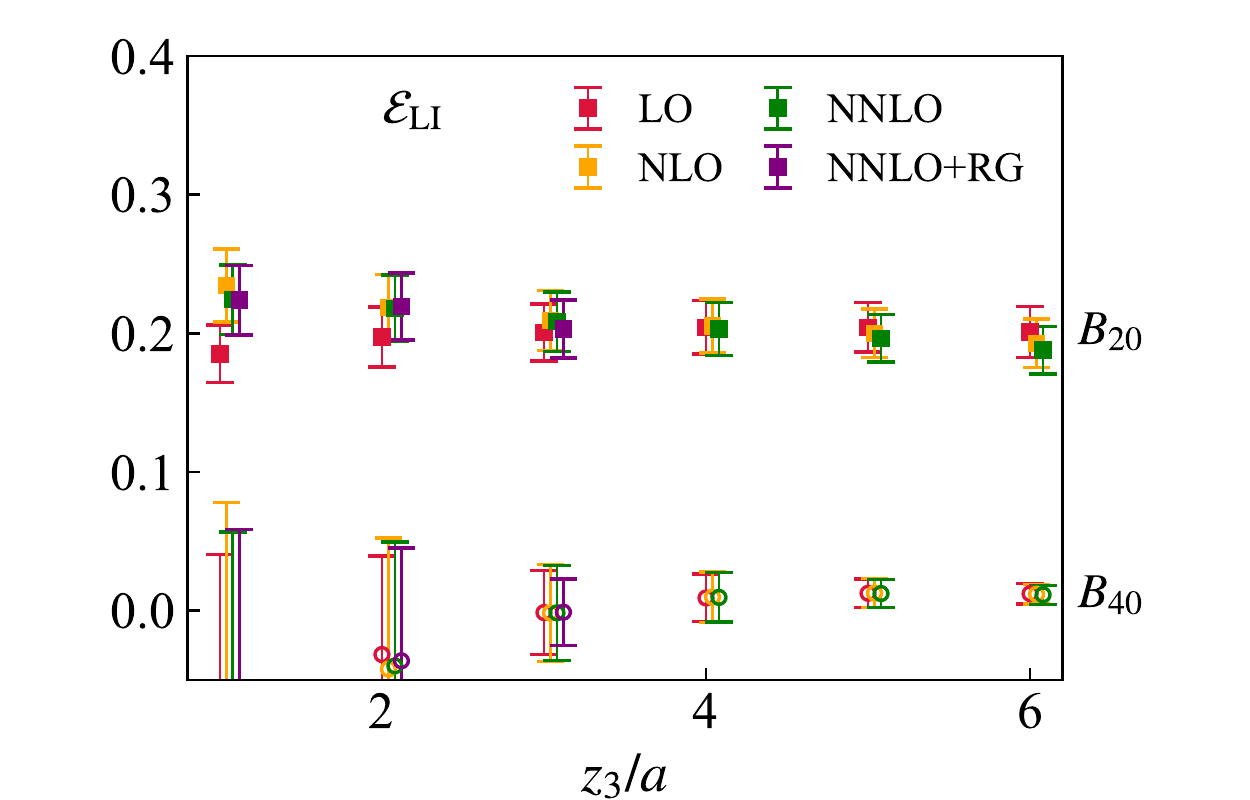}
    \includegraphics[width=0.4\textwidth]{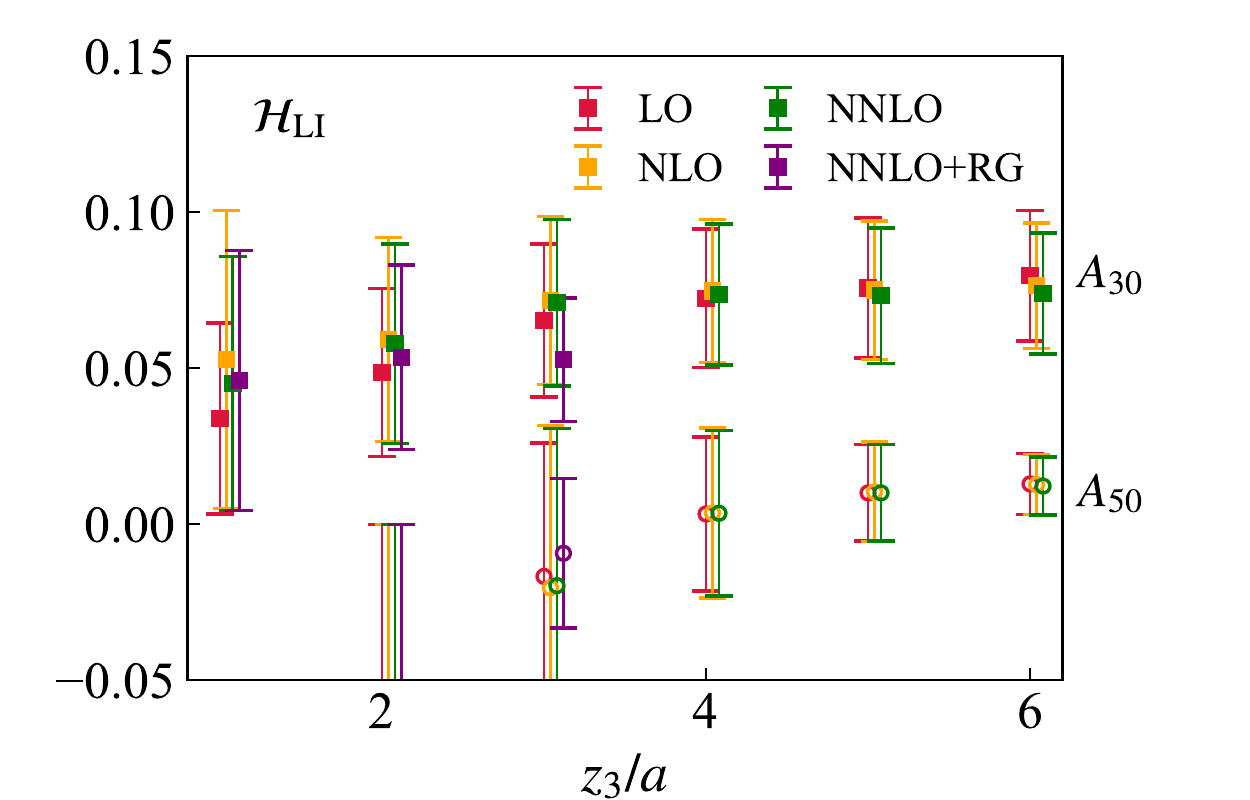}
    \includegraphics[width=0.4\textwidth]{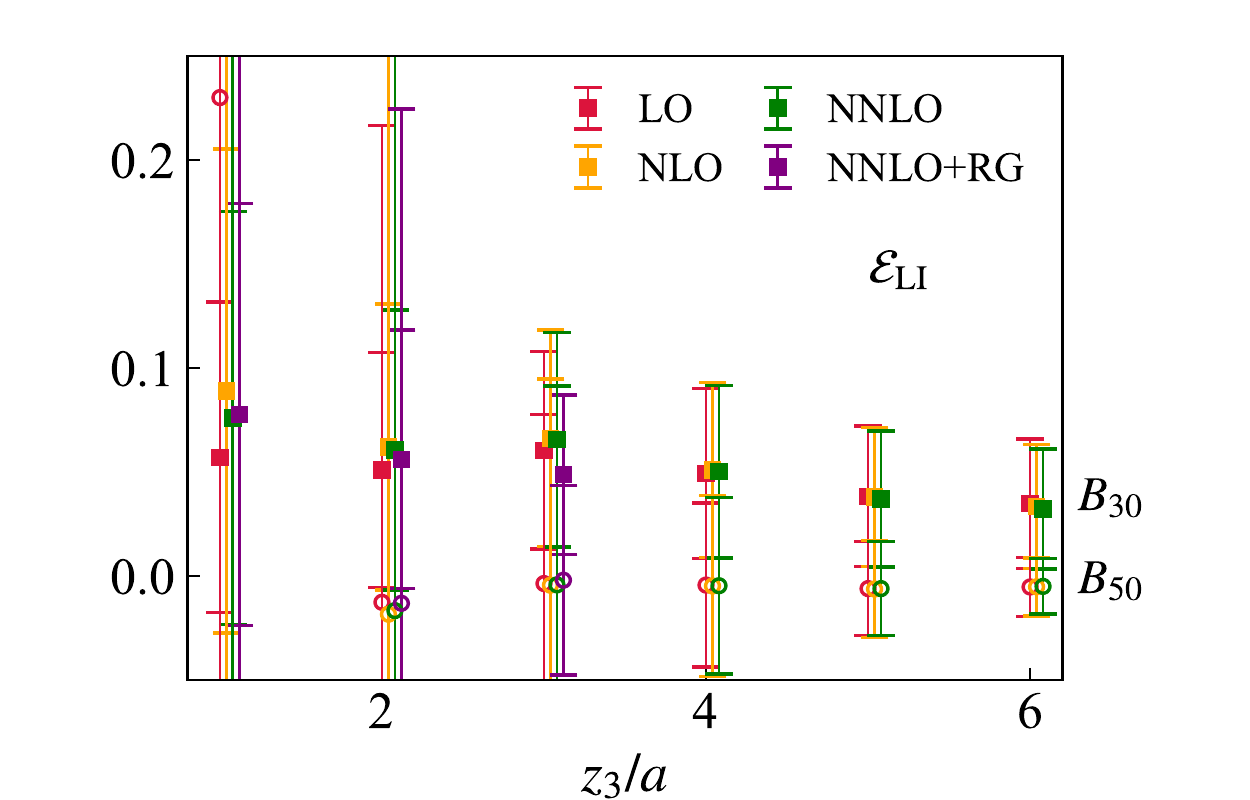}
	\caption{The left panels show the first few moments extracted at each $z$ from the iso-vector $\mathcal{M}_{\mathcal{H},\rm{LI}}$, while the right panels display the corresponding moments from $\mathcal{M}_{\mathcal{E},\rm{LI}}$, for the symmetric case of $P_3=0.83,1.25,1.67$ GeV and $-t=0.69~\rm{GeV}^2$, utilizing the data in \fig{rGPDzpz} and Wilson coefficients at LO, NLO, NNLO, and NNLO+RG order. The filled squared symbols are for the real part, while the circled open symbols are for the imaginary part.\label{fig:isoVmomsizLI}}
\end{figure}

\begin{figure}[h!]
    \centering
    \includegraphics[width=0.4\textwidth]{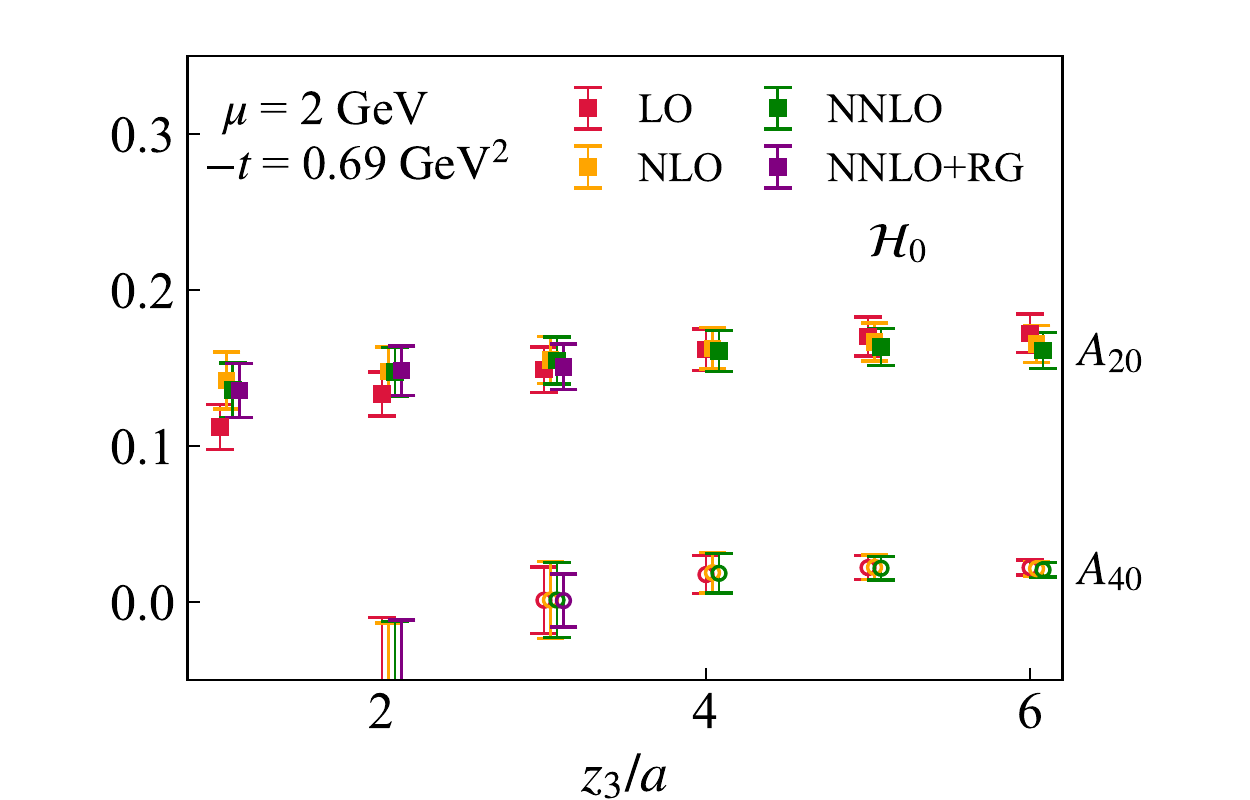}
    \includegraphics[width=0.4\textwidth]{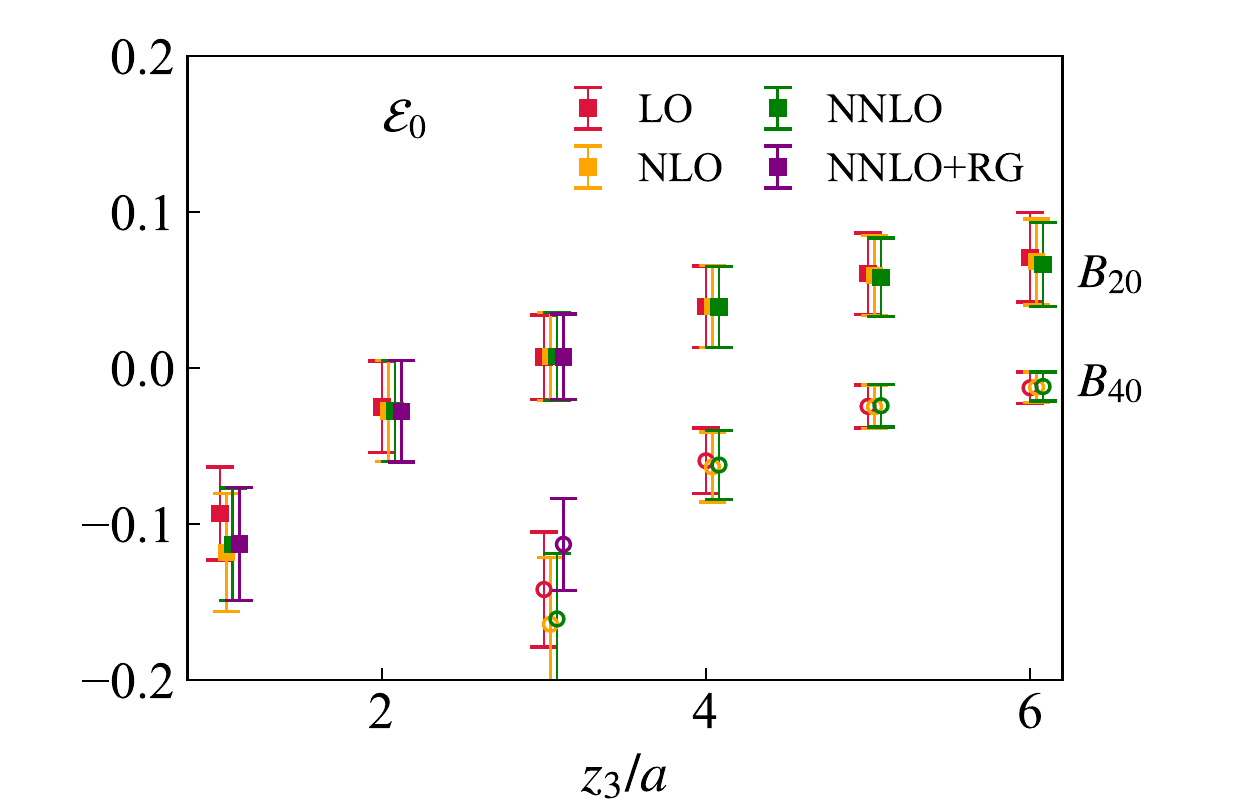}
    \includegraphics[width=0.4\textwidth]{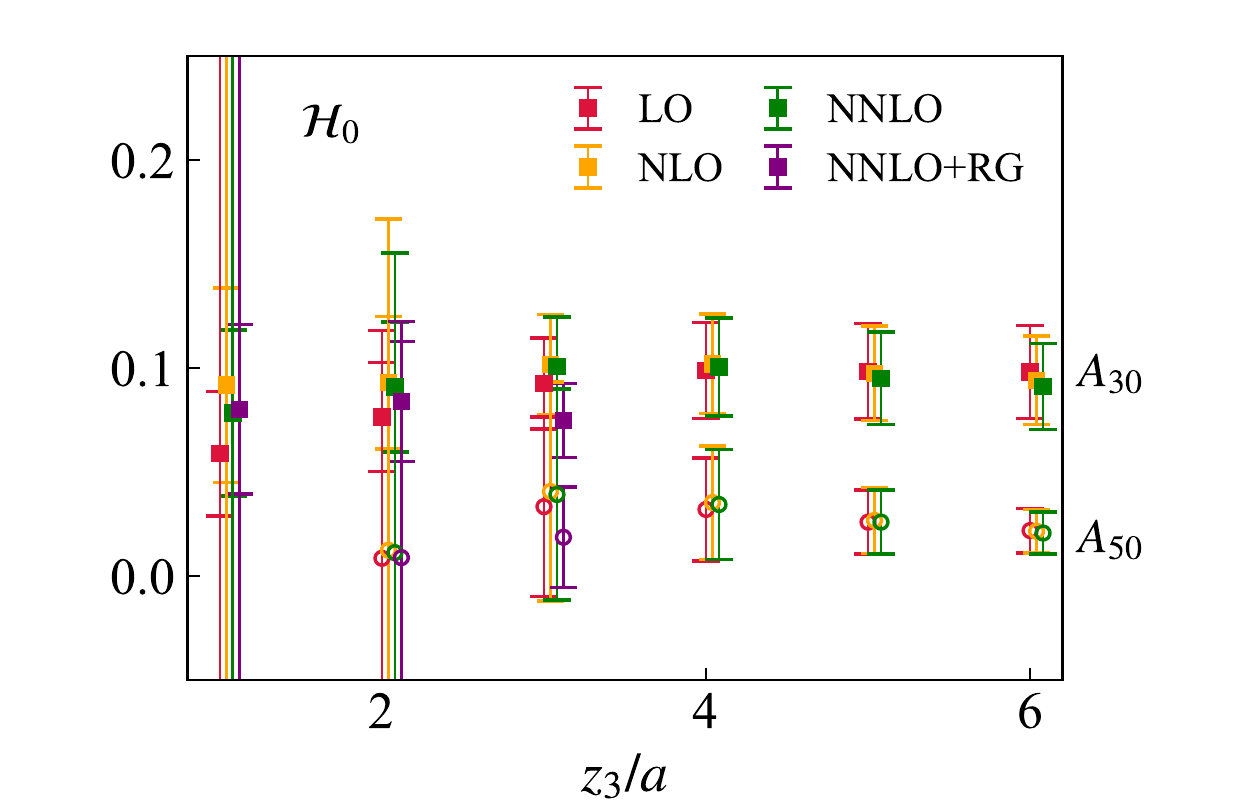}
    \includegraphics[width=0.4\textwidth]{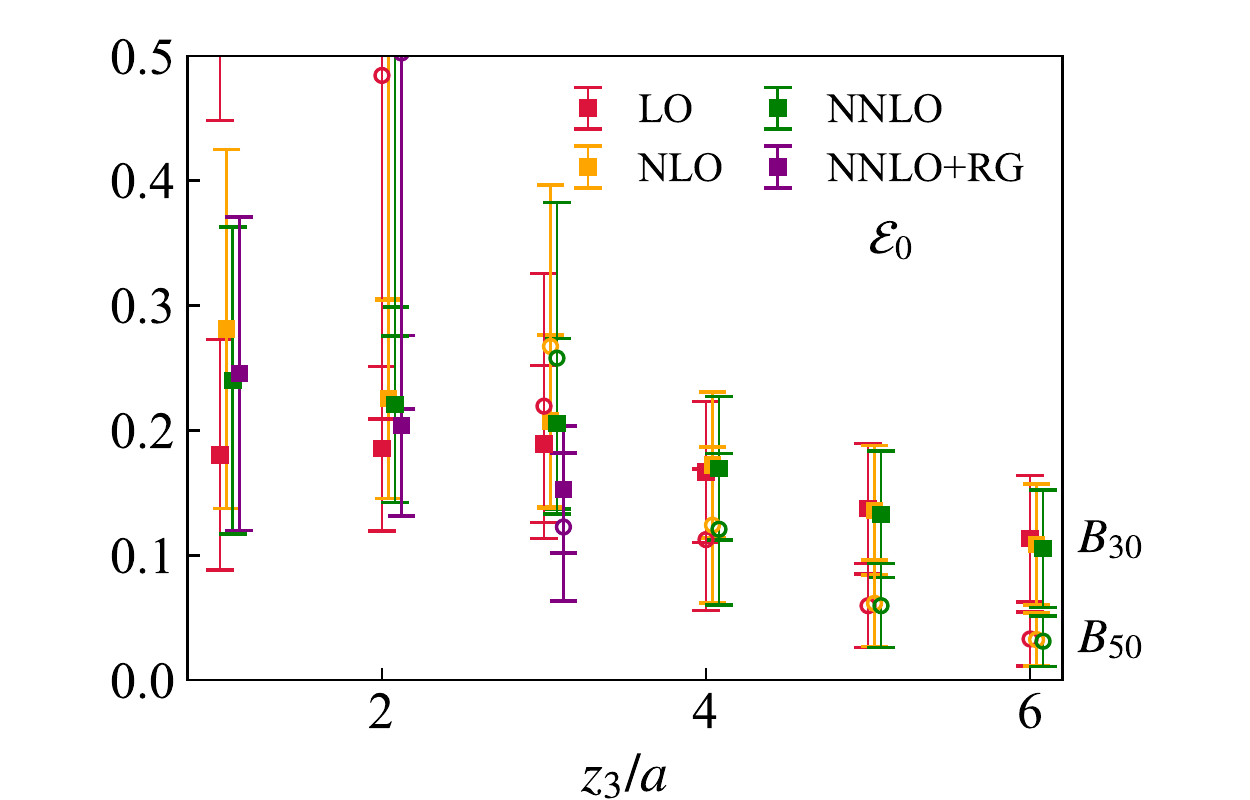}
	\caption{The left panels show the first few moments extracted at each $z$ from the iso-vector $\mathcal{M}_{\mathcal{H},\gamma_0}$, while the right panels display the corresponding moments from $\mathcal{M}_{\mathcal{E},\gamma_0}$, for a symmetric case of $P_3=0.83,1.25,1.67$ GeV and $-t=0.69~\rm{GeV}^2$, utilizing Wilson coefficients at LO, NLO, NNLO, and NNLO+RG order. The filled squared symbols are for the real part, while the circled open symbols are for the imaginary part.\label{fig:isoVmomsizstd}}
\end{figure}

At the momentum transfer of $-t=0.69~\rm{GeV}^2$, we have three different momenta: $P_3=0.83,1.25$ and 1.67 GeV, which can be used to assess the validity of the leading twist factorization.
Since the matrix elements of a local current should have no boost or frame dependence at a given $-t$, we averaged the $z_3=0$ matrix elements $\overline{\mathcal{M}_{\mathcal{H}}}(0,P,\Delta)$ of the three $P_3$ for each bootstrap sample, and normalized the $\mathcal{M}_{\mathcal{H}}$ by $\mathcal{M}_{\mathcal{H}}(0,P,\Delta)/\overline{\mathcal{M}_{\mathcal{H}}}(0,P,\Delta)$. We also do this for $\mathcal{M}_{\mathcal{E}}$. In the upper panels of \fig{rGPDzpz}, we show the iso-vector matrix elements $\mathcal{M}_{\mathcal{H},\rm{LI}}$ (left panels) and $\mathcal{M}_{\mathcal{E},\rm{LI}}$ (right panels) from the Lorentz-invariant definition with momentum transfer $-t=0.69~\rm{GeV}^2$ as a function of $zP$, while the ones of the $\gamma_0$ definition are shown in the lower panels for comparison. It is evident that, in the case of the Lorentz-invariant definition, the matrix elements for different $P_3$ values appear to overlap with each other as a function of $zP$. This observation is consistent with the leading-order approximation of \Eq{ratioOPE}, while the beyond-leading-order effects are only minor, as perturbative contributions fall within the current statistical errors. This finding is also applicable to $\mathcal{M}_{\mathcal{H},\gamma_0}$ under the $\gamma_0$ definition. However, when it comes to $\mathcal{M}_{\mathcal{E},\gamma_0}$, the matrix elements from our three momentum values display significant deviations, even at very short distances, especially in the imaginary part. These deviations are more pronounced for smaller momenta, indicating additional power corrections that can be suppressed by $P_3$, but remain significant for small momenta. To explore this further, we will extract the moments at each $z$ using the Wilson coefficients at different orders by fitting the $P_3$ dependence using \Eq{chisqfit}. We set $n_{\rm max}=4$ as the truncation order for the SDF formula, which is expected to be sufficient for the considered momentum range, given the higher-order terms are factorially suppressed, as we will see in the following discussion. 

In \fig{isoVmomsizLI}, we show the first few moments extracted from each fixed $z_3/a$ from iso-vector $\mathcal{M}_{\mathcal{H},\rm{LI}}$ and $\mathcal{M}_{\mathcal{E},\rm{LI}}$. We have used the Wilson coefficients from leading order (LO) to next-to-next-to-leading order (NNLO), including the one considering the renormalization group resummation (NNLO+RG) at short  distance~\cite{Gao:2022iex,Su:2022fiu}. The factorization scale was set to $\mu=$ 2 GeV. We leave out the first moments $A_{10}$ and $B_{10}$, which have no scale dependence as conserved charges. As for the second moments of $H$ and $E$, $A_{20}$ and $B_{20}$ have the best non-zero signal and exhibit small $z$-dependence at very short distances in the LO results. What is more, using the NLO, NNLO, and NNLO+RG Wilson coefficients, such $z$-dependence can be compensated, resulting in plateaus at the factorization scale $\mu=2$ GeV. The differences among NLO, NNLO, and NNLO+RG are compatible with zero within the statistical error. The results with RG are shown only for $z_3\lesssim0.3$ fm as $\alpha_s(1/z_3)$ will hit the Landau pole at large $z_3$, despite that the fixed-order Wilson coefficients are still finite. In the following analysis, we will use fixed-order Wilson coefficients and vary the range of $z_3$ to beyond $\sim0.3$ fm to have more data points for estimating the systemics, though a rigorous treatment would require the use of RG improved OPE with only small-$z_3$ matrix elements. This situation can be improved with finer lattices and larger momenta $P_3$ in the future.

As for the higher moments that are noisy at short $z_3$, no dependence on the perturbative order is observed. These findings are consistent with 
the fact that the ratio-scheme renormalized matrix elements mostly
depend on $z P$ and have very mild dependence on $P_3$
as observed in \fig{rGPDzpz}, for the Lorentz-invariant definition. For comparison, in \fig{isoVmomsizstd} we show the first few moments extracted from $\mathcal{M}_{\mathcal{H},\gamma_0}$ and $\mathcal{M}_{\mathcal{E},\gamma_0}$. For the moments from $\mathcal{M}_{\mathcal{H},\gamma_0}$ shown in the left panels, similar behavior can be observed as in the case of $\mathcal{M}_{\mathcal{H},LI}$, though the central values of moments are slightly shifted. However, strong $z_3$ dependence can be observed for moments from $\mathcal{M}_{\mathcal{E},\gamma_0}$ as the $zP$ dependence breaks down for the $\gamma_0$ definition seen in \fig{rGPDzpz}. It is clear that the perturbative kernels, even up to NNLO, cannot explain this $z_3$ dependence. We then conclude that the quasi-GPD matrix elements $\mathcal{H}$ and $\mathcal{E}$ under Lorentz-invariant definition, as a function of both $z^2$ and $zP$, can be well described by the perturbative kernels together with the ratio-scheme renormalization. However, the factorization formula is 
not applicable
for the quasi-GPD $\mathcal{E}$ with $\gamma_0$ definition
in the considered range of $z_3$ and $P_3$, because of additional power corrections in this case, as we discussed in \sec{HandE}. A similar observation holds for the iso-scalar cases, where we used Wilson coefficients only up to the NLO level and ignored the quark-gluon mixing. Therefore we will use the quasi-GPDs in Lorentz-invariant definition, which may converge to the light-cone GPDs faster, for the following analysis. Meanwhile, we will stick to the best-known Wilson coefficients, NNLO and NLO, for iso-vector and iso-scalar cases, respectively.

\subsection{Determination of Mellin moments}\label{sec:moms}

\begin{figure}[h!]
    \centering
    \includegraphics[width=0.4\textwidth]{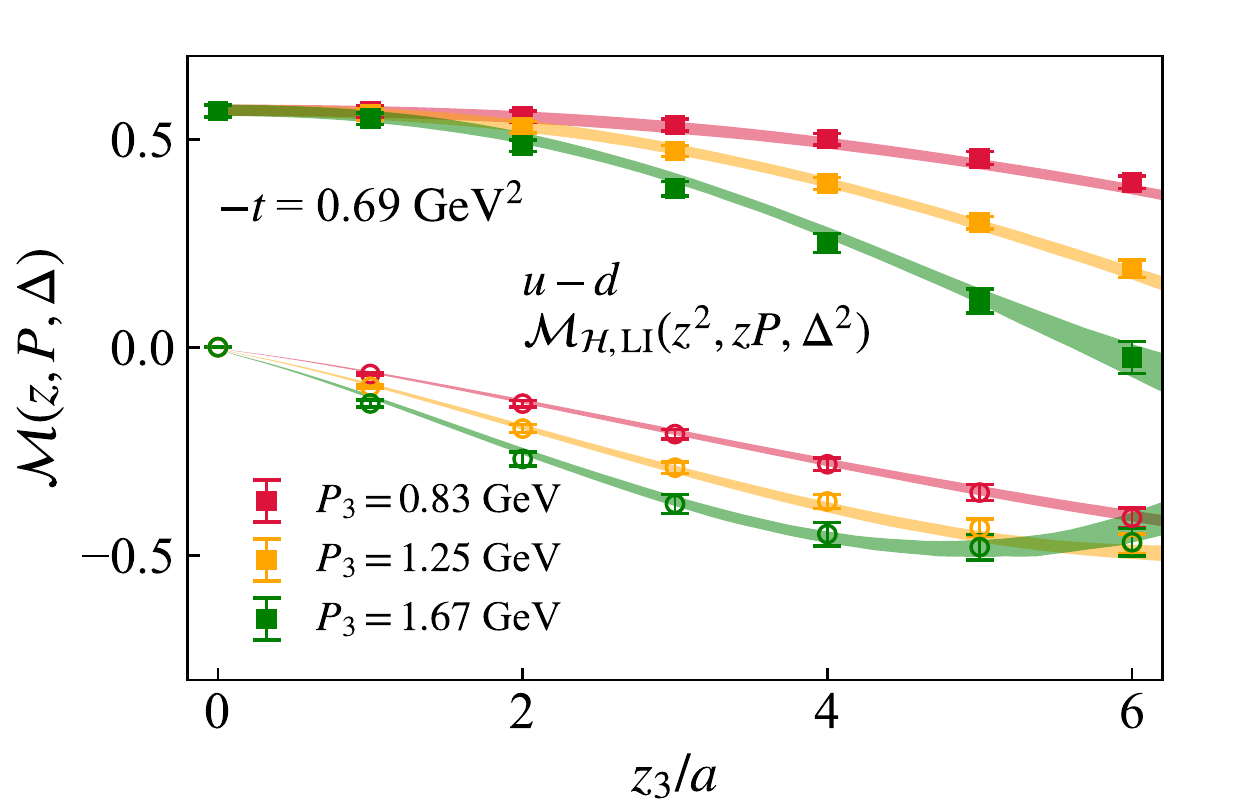}
    \includegraphics[width=0.4\textwidth]{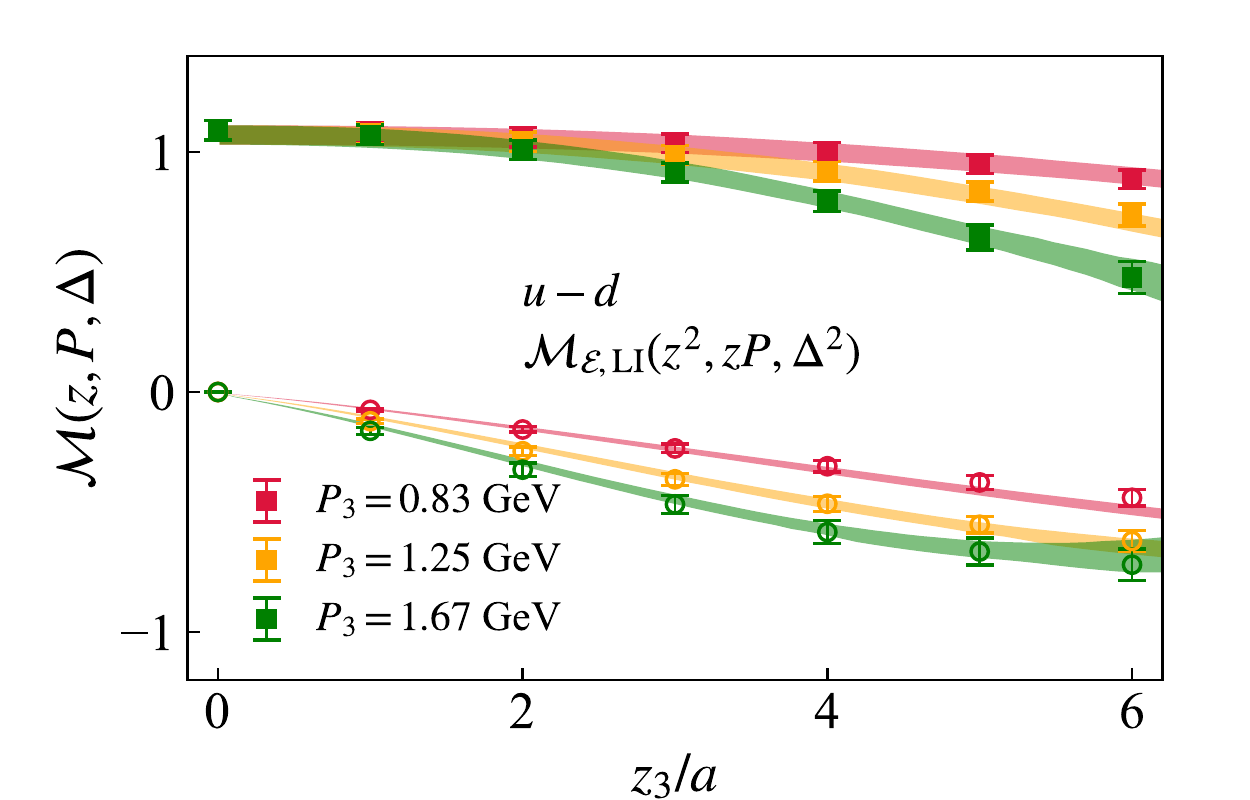}
	\caption{The combined fit results from $z_3\in[2a,6a]$ for $\mathcal{M}_{\mathcal{H},\rm{LI}}$ (left panel) and $\mathcal{M}_{\mathcal{E},\rm{LI}}$ (right panel) at the momentum transfer $-t=0.69~\rm{GeV}^2$, as a function of $z_3$ and using NNLO matching. We have three different momentum $P_3=0.83,1.25$ and 1.67 GeV for this $-t$. The filled squared symbols are for the real part, while the circled open symbols are for the imaginary part.\label{fig:isoVfitzmax}}
\end{figure}

\begin{figure}[h!]
    \centering
    \includegraphics[width=0.33\textwidth]{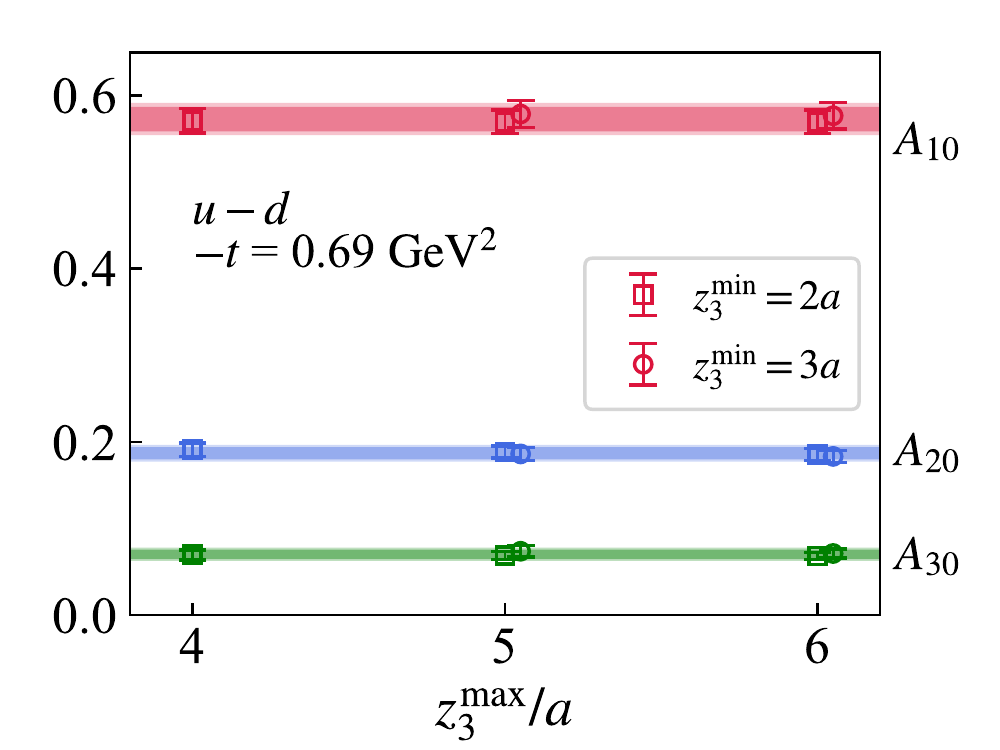}
    \includegraphics[width=0.33\textwidth]{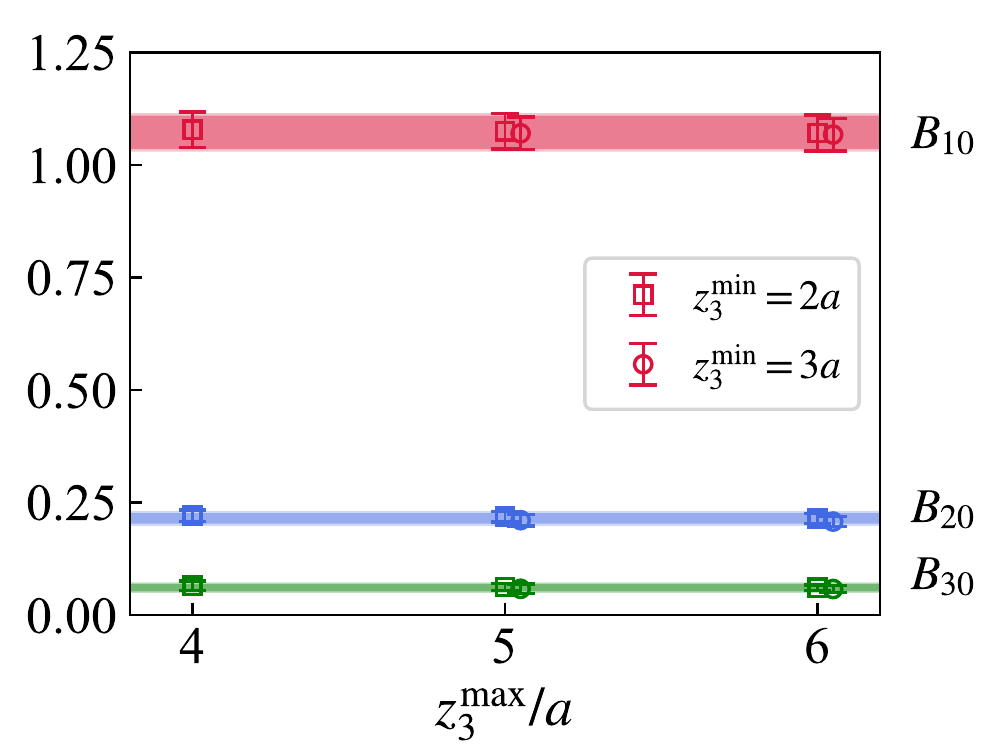}
    \includegraphics[width=0.33\textwidth]{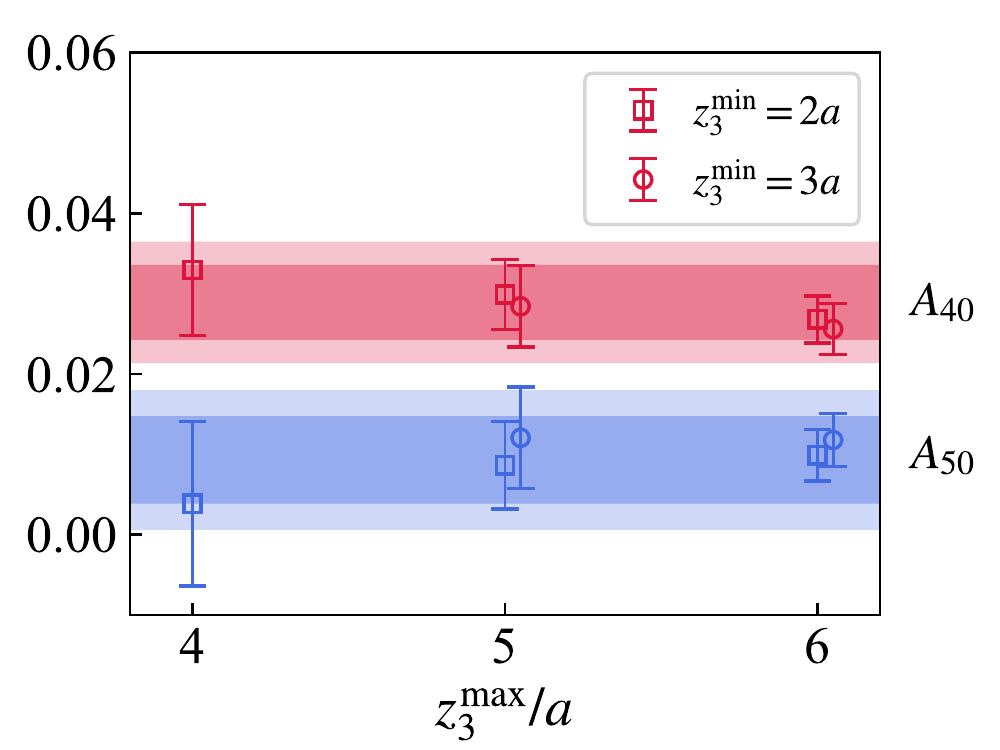}
    \includegraphics[width=0.33\textwidth]{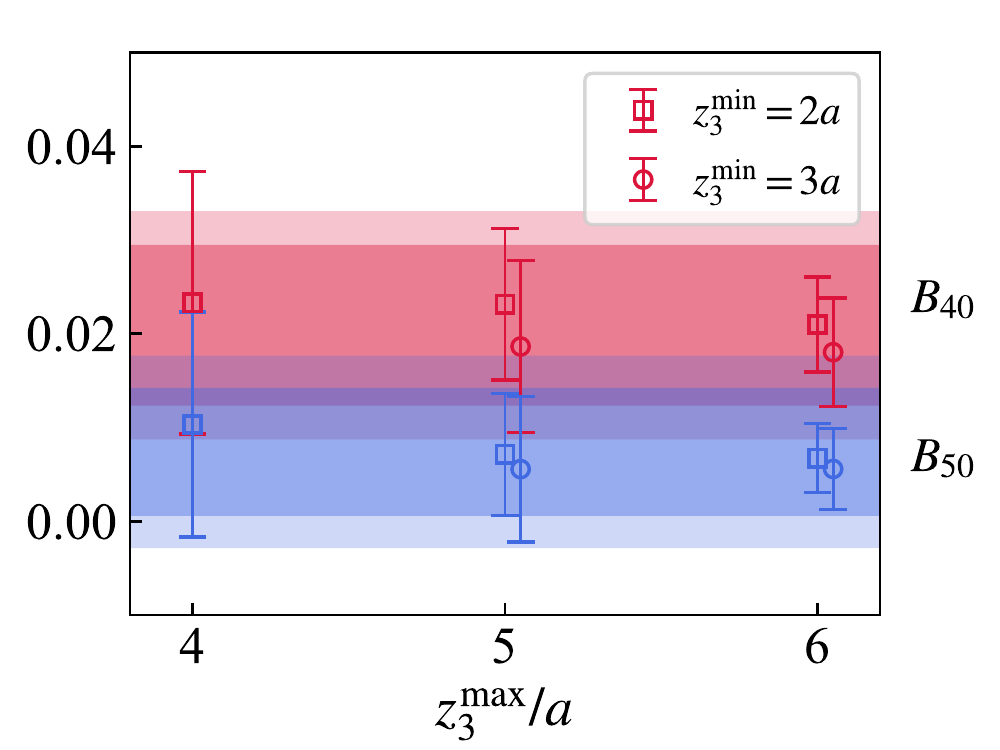}
	\caption{The first three moments of $A_{n+1,0}$ (left panels) and $B_{n+1,0}$ (right panels) extracted from $z_3^{\rm min}=2a,3a$ of $\mathcal{M}_{\mathcal{H},\rm{LI}}$ and $\mathcal{M}_{\mathcal{E},\rm{LI}}$ using NNLO matching, as a function of $z_3^{\rm max}$. We vary $z_3^{\rm min}=2a,3a$ and $z_3^{\rm max}=4a,5a,6a$ to estimate the statistical errors as the darker bands and the systematic errors as the lighter bands. The momentum transfer is $-t=0.69~\rm{GeV}^2$.\label{fig:isoVmomszmax}}
\end{figure}

It is found that the perturbative matching can well describe the ratio-scheme renormalized matrix elements under Lorentz-invariant definition. However, the higher moments extracted from fixed $z$ exhibit significant noise. To stabilize the fit, and also because we only have one momentum $P_3$ for many values of $-t$, we will perform combined fits of several renormalized matrix elements with $z_3 \in[z_3^{\rm min},z_3^{\rm max}]$. We note that matrix elements at small $z$ may suffer from discretization effects, while at large $z$ they may be affected by higher-twist effects. We omit $z_3=a$ to avoid the most severe discretization effects and vary $z_3^{\rm min} \in [2a, 3a$], $z_3^{\rm max} \in [4a, 6a]$ to estimate systematic errors related to discretization and higher-twist effects. To be specific, for an observable $\mathcal{X}$ and a given bootstrap sample, we average over the fit results with different $[z_3^{\rm min},z_3^{\rm max}]$ to obtain ${\rm Mean}(\mathcal{X})$ and estimate the systematic error as $\rm{err}(\mathcal{X})={\rm Mean}((\mathcal{X}-{\rm Mean}(\mathcal{X}))^2)$. Then, we consider all bootstrap samples to obtain the average value of the observable $\overline{\rm{Mean}(\mathcal{X})}$ and estimate  the statistical uncertainty. The final systematic error is obtained
as $\overline{\rm{err}(\mathcal{X})}$.

In \fig{isoVfitzmax}, we show the fit results from $z_3\in[2a,6a]$ at the momentum transfer $-t=0.69~\rm{GeV}^2$ for $\mathcal{M}_{\mathcal{H},\rm{LI}}$ (left panel) and $\mathcal{M}_{\mathcal{E},\rm{LI}}$ (right panel) as a function of $z_3$. The bands come from the combined fit with the NNLO kernel, which can describe well the matrix elements. In \fig{isoVmomszmax}, we summarize the first five moments of $A_{n+1,0}$ and $B_{n+1,0}$ extracted from $z_3^{\rm min}=2a,3a$ as a function of $z_3^{\rm max}$, where reasonable signal can be observed. Meanwhile, the extracted moments show little dependence on $z_3^{\rm min}$ or $z_3^{\rm max}$, suggesting the systematic errors (outer light bands) are small compared to the statistical errors (inner dark bands) at the current stage. It can also be found that the higher moments have smaller values than the lower ones. Together with the fact that they are factorially suppressed by $(-izP)^n/n!$ with finite hadron momentum $P_3$, the fourth and fifth moments are still noisy within the $zP \lesssim 5$ used in this work. To further constrain the higher moments, one need to increase the hadron momentum to achieve larger $zP$.

\section{$-t$ dependence of the Mellin moments}\label{sec:tdependence}

The Mellin moments of GPDs encapsulate a wealth of physics pertaining to the structure of the nucleon. The first moments $A_{10}$ and $B_{10}$ are Dirac and Pauli form factors. At zero momentum transfer, $-t=0$, the $A_{10}(0)$ is the total charge carried by the quarks. In addition, one can alternatively define the so-called Sachs electric and magnetic form factors $G_E(-t)=A_{10}(-t)+B_{10}(-t)$ and $G_M(-t)=A_{10}(-t)+-t/{(2m_N)^2}B_{10}(-t)$, then infer the electric and magnetic radius of the nucleon. It is the second moments $A_{20}$ and $B_{20}$ that have attracted a lot of interest in recent years due to their connection to the gravitational form factors (GFFs), which are the matrix elements of QCD energy-momentum tensor,
\begin{align}
	\langle p_f | \hat{T}_{\rm{QCD}}^{\mu\nu} | p_i\rangle=\bar{u}(p_f)\left[A_{20}(t)\gamma^{(\mu}\bar{P}^{\nu)}+B_{20}(t)\frac{\bar{P}^{(\mu}i\sigma^{\nu)\alpha}\Delta_\alpha}{2M} + C_{20}(t)\frac{\Delta^\mu\Delta^\nu-g^{\mu\nu}\Delta^2}{4M} \right]u(p_i),
\end{align}
where $C_{20}$ term can also be extracted from non-zero skewness GPDs. At zero momentum transfer, $-t=0$, $A_{20}$ provides information on the momentum fraction carried by the quarks inside the nucleon. Moreover, combining $A_{20}$ and $B_{20}$, one can infer the total angular momentum carried by quarks via the Ji sum rule~\cite{Ji:1996ek},
\begin{align}\label{eq:totalA}
	J=\frac{1}{2}(A_{20}(0)+B_{20}(0))\,.
\end{align}
To obtain these quantities, we need to extrapolate the moments to $-t\rightarrow0$, particularly for $B_{n+1,0}$ using a parametrization of choice. It is known that at low momentum transfer $-t$, the nucleon electromagnetic form factors can be well described by the dipole form, 
\begin{align}
	\langle x^n\rangle(-t)=\frac{\langle x^n\rangle(0)}{(1+\frac{-t}{M^2})^2} \,.
\end{align}
However, at large $-t$, there is no strong theoretical support for this simple form. A more flexible parameterization could be the $z$-expansion~\cite{Lee:2015jqa},
\begin{align}
	\langle x^n\rangle(-t)=\sum_{k=0}^{k_{\rm max}}a_kz(t)^k\,,
\end{align}
with
\begin{align}
	z(t)=\frac{\sqrt{t_{\rm cut}-t}-\sqrt{t_{\rm cut}-t_0}}{\sqrt{t_{\rm cut}-t}+\sqrt{t_{\rm cut}-t_0}}\,,
\end{align}
where the parameter $t_{\rm cut}$ is the timelike kinematic threshold for particle production: $t_{\rm cut}=9m_\pi^2$ for isoscalar form factors and $t_{\rm cut}=4m_\pi^2$ for isovector form factors. 
The $t_0$ is usually chosen to ensure $0 < -t < -t_{\rm max}$ corresponding to the smallest range of $z(-t)$, which is $t_0^{\rm opt}(t_{\rm cut},-t_{\rm max})=t_{\rm cut}(1-\sqrt{1-t_{\rm max}/t_{\rm cut}})$. 
The $k_{\rm max}$ needs to be truncated at a finite value according to the range of $-t$. 
We truncated $k_{\rm max}$ up to 2 in this work with reasonable $\chi^2/dof$.
To stabilize the fit, we imposed a Gaussian prior to the $|a_k/a_0|$ with central value of 0 and width $|a_k/a_0|_{\rm max}=5$. In our analysis, as discussed later, we found both dipole and $z$-expansion can reasonably describe all of our data with $\chi^2/dof$ ranging from 0.4 to 1.9. However, we note that these two functional forms are designed for the first moments, so they may not be optimal for the higher moments. Therefore, we vary the fit range of $-t$ as well as the model to estimate the systematic errors using the method described in \sec{moms}. In addition, we stick to the quasi-GPDs in the Lorentz-invariant definition in this section.

\subsection{The model fit and $-t\rightarrow0$ extrapolation}

\begin{figure}[h!]
    \centering
    \includegraphics[width=0.38\textwidth]{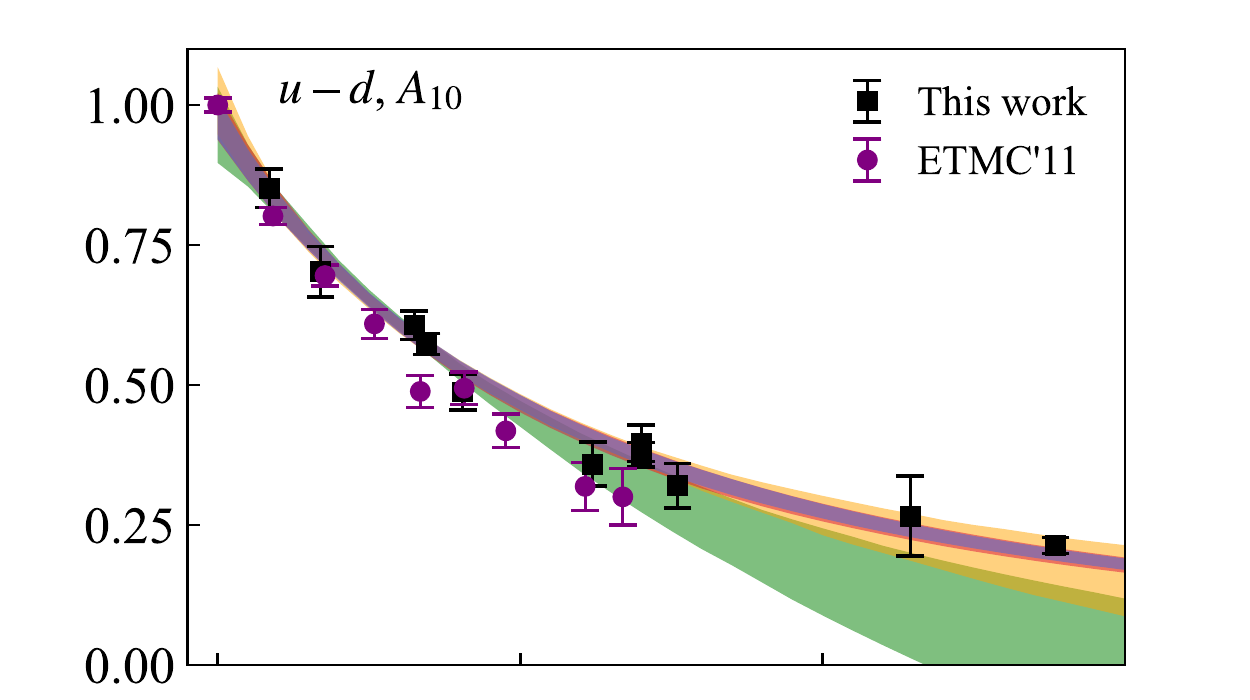}
    \includegraphics[width=0.38\textwidth]{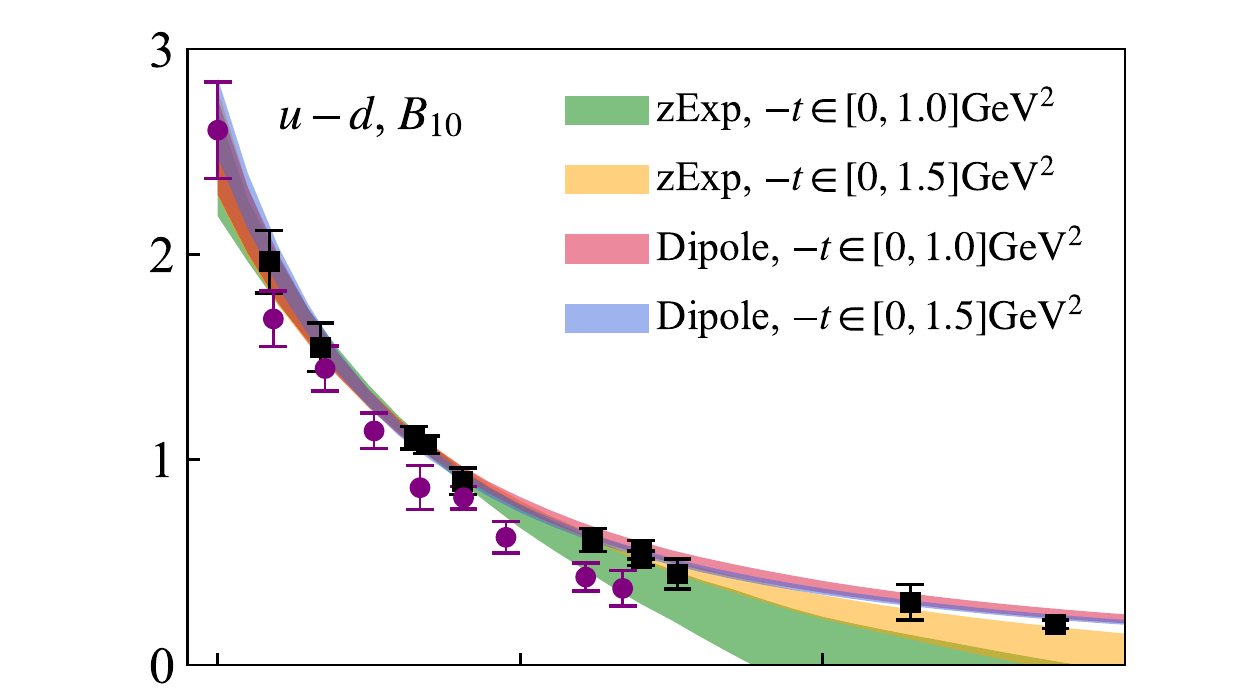}
    \includegraphics[width=0.38\textwidth]{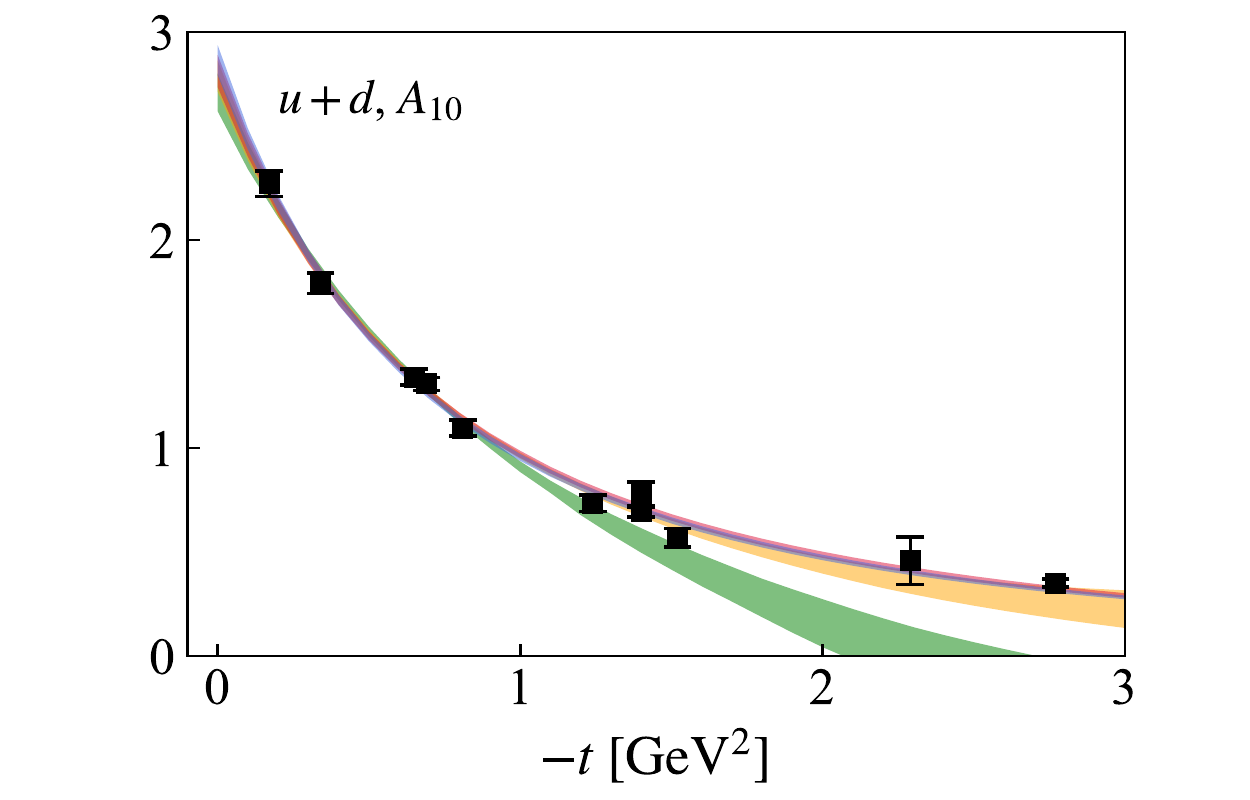}
    \includegraphics[width=0.38\textwidth]{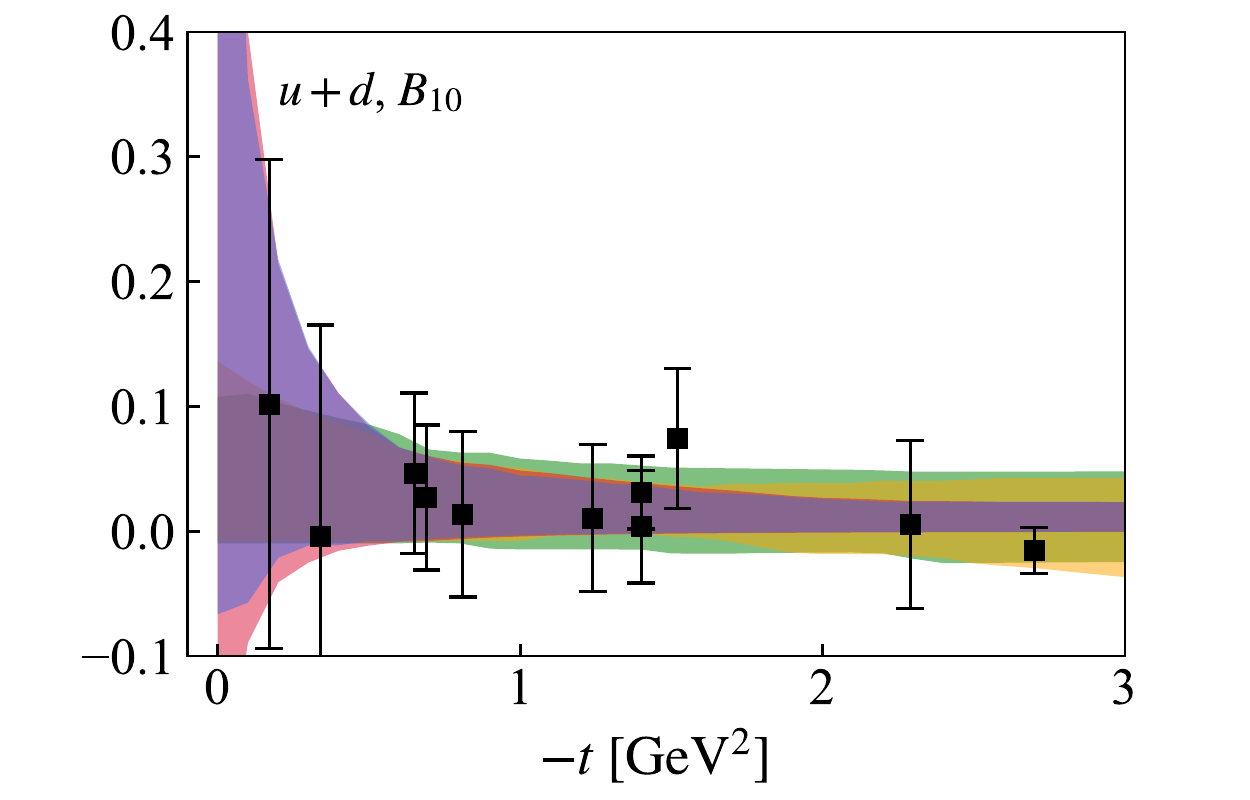}
	\caption{The first moments $A_{10}$ and $B_{10}$ for iso-vector (upper panels) and iso-scalar (lower panels) as a function of $-t$. The error bars include both statistical errors and systematic errors. The bands come from two different parametrizations using two ranges of $-t$. For comparison, we also show the ETMC determination of iso-vector moments with a similar lattice setup~\cite{Alexandrou:2011db}.\label{fig:fitQsqx0}}
\end{figure}

In \fig{fitQsqx0}, we show the first moments $A_{10}$ and $B_{10}$ extracted from Lorentz-invariant quasi-GPDs, for the iso-vector (upper panels) as well as the iso-scalar (lower panels) cases. The errors included are both statistical and systematic. We also show the results from a direct calculation~\cite{Alexandrou:2011db} by ETMC of matrix elements of the local current $\bar{q} (0) \gamma^\mu q (0)$, which have been obtained in the rest frame with similar lattice setup. We observe good agreement between the results. We fit the $-t$ dependence using the dipole form as well as the $z$-expansion form (zExp) shown as the bands. Two ranges of $-t$ are used with $-t_{\rm max}=1.0$ and $1.5~\rm{GeV}^2$, as our data are sparse in large $-t$ region, and we are more interested in the small $-t$ behavior. We will treat the difference from various $-t_{\rm max}$ as a source of systematic errors.  As can be seen, all the bands can describe well the data included in the fit and overlap with each other in the region of small $-t$. However, the dipole form, being a simpler expression, tends to yield smaller errors. It is interesting to note that the dipole form is also able to extrapolate to larger values of $-t$ effectively and remains consistent with data points beyond 2 $\rm{GeV}^2$. On the other hand, the $z$-expansion, which takes the form of polynomial functions of $z(-t)$, tends to become unstable rapidly during the extrapolation. This causes the $z$-expansion fit to show larger errors and potentially deviate from data points that require long-distance extrapolation. In \tb{fitQsqx01}, we summarize the $-t\rightarrow0$ extrapolation results, where we take the same strategy as described in \sec{moms} to estimate the statistical and systematic uncertainties from the variation of $-t_{\rm max}$ and two different models. As one can see, the $A_{10}^{u-d}(0)$, which measures the iso-vector charge of the nucleon, is close to the expected value of 1. However, our estimation of the iso-scalar charge $A_{10}^{u+d}$ using the large momentum matrix element has a 2-$\sigma$ deviation with a value of 3, whereas the result from the rest frame matrix elements as discussed in \app{pz0} is very close to the expectation. This discrepancy could be due to the discretization effect that occurs for highly boosted hadron states~\cite{Gao:2020ito}. In addition, one can observe good agreement between our $B_{10}^{u-d}(0)$ with the ETMC'11 results, while the $B_{10}^{u+d}(0)$ are consistent with zero within the errors, as has been observed at the level of bare matrix elements in \fig{bmLI}.

\begin{figure}[h!]
    \centering
    \includegraphics[width=0.38\textwidth]{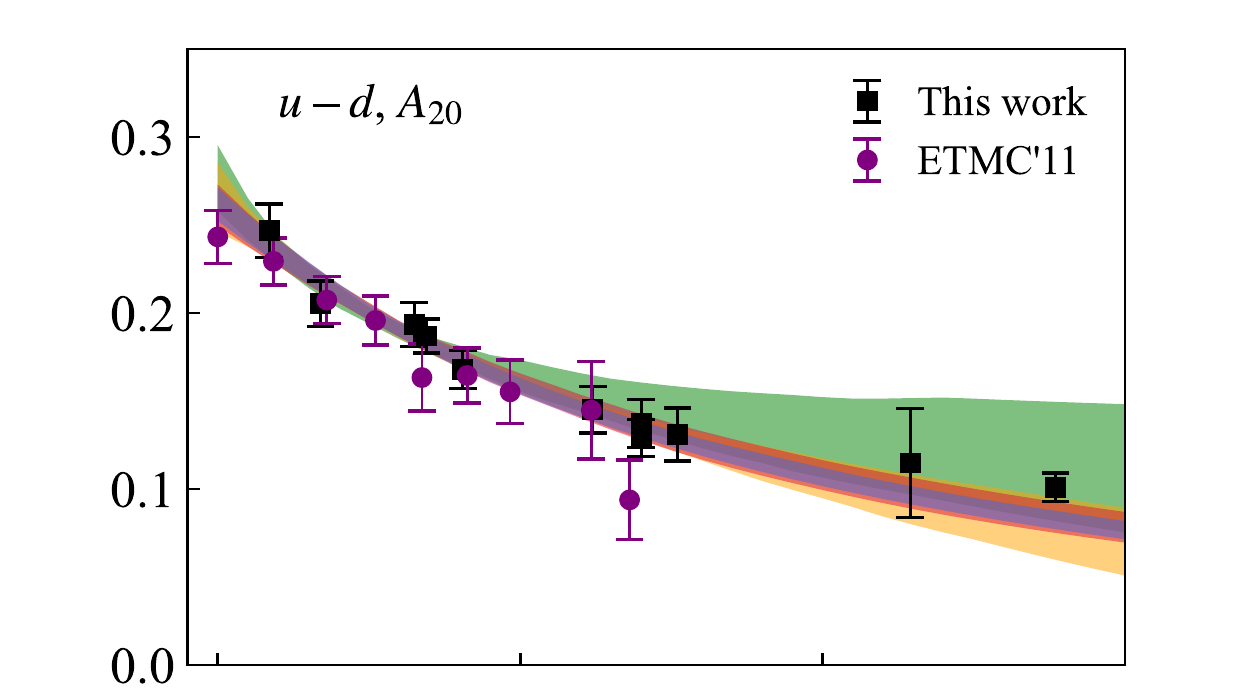}
    \includegraphics[width=0.38\textwidth]{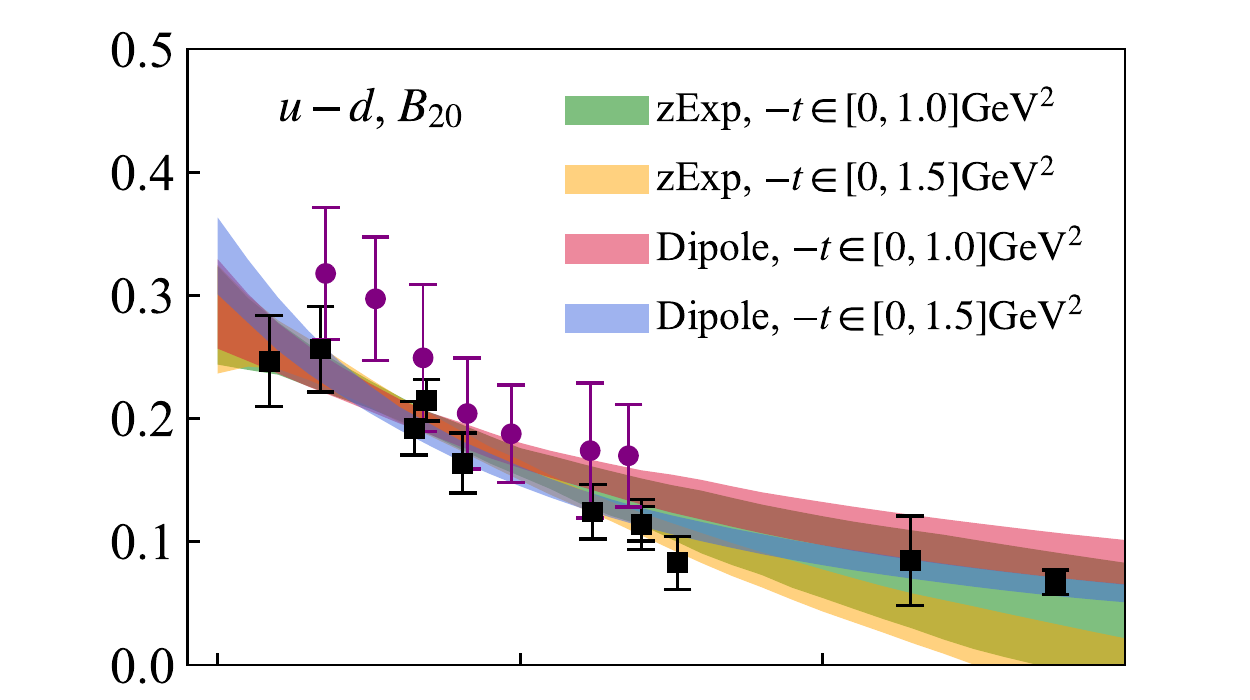}
    \includegraphics[width=0.38\textwidth]{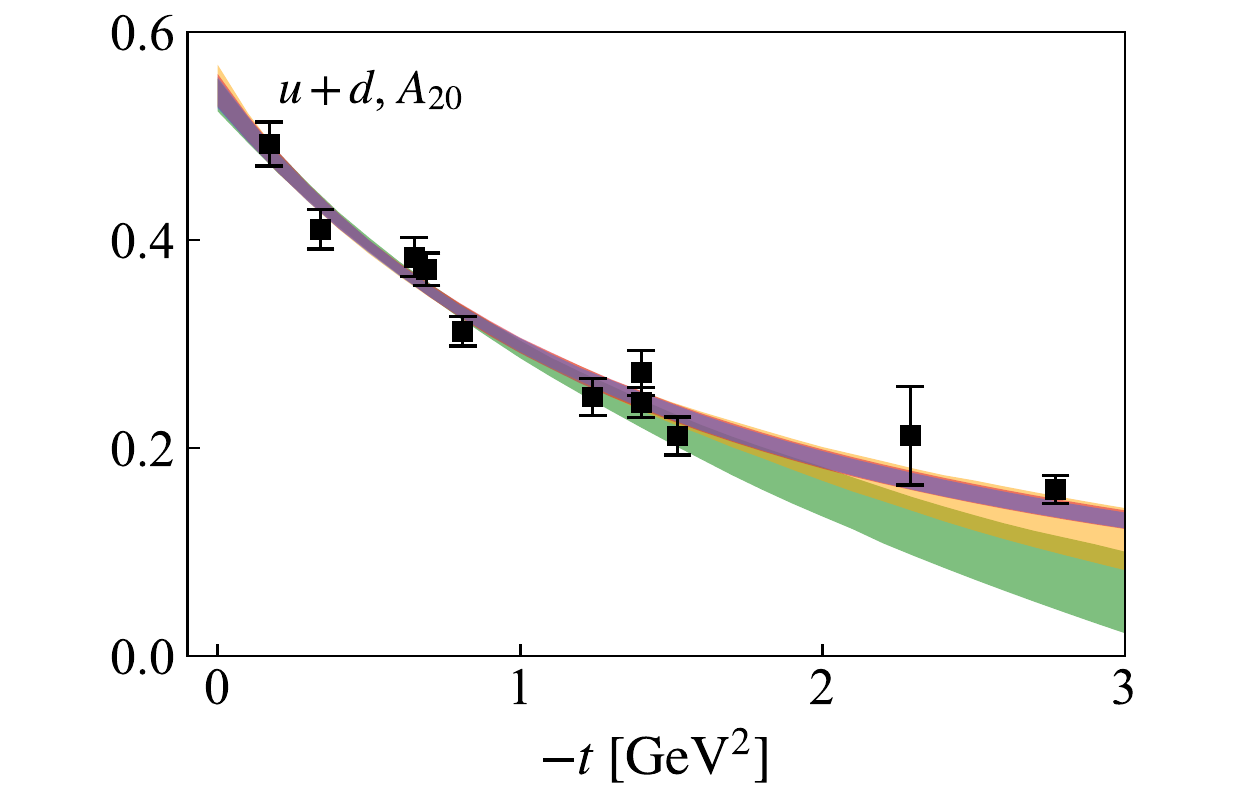}
    \includegraphics[width=0.38\textwidth]{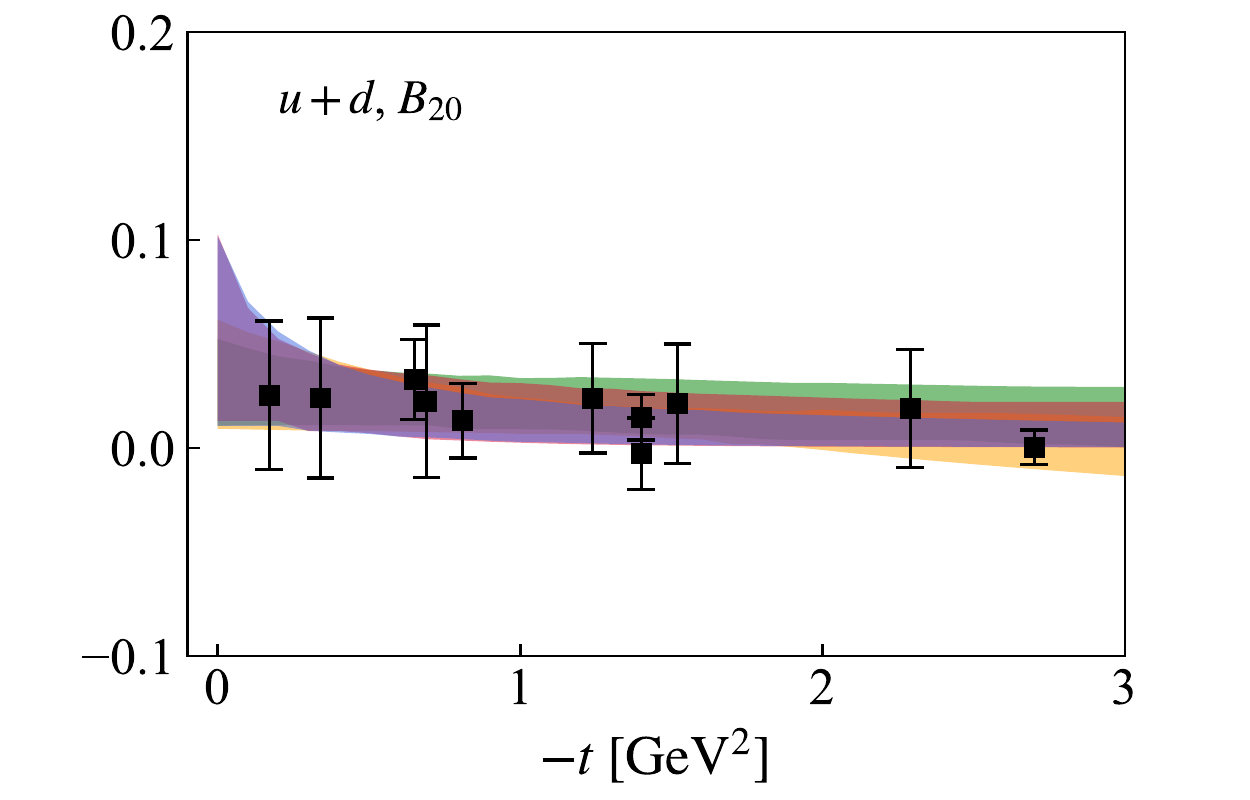}
	\caption{The second moments $A_{20}$ and $B_{20}$ for iso-vector (upper panels) and iso-scalar (lower panels) as a function of $-t$. The error bars include both statistical errors and systematic errors. The bands come from two different parametrizations using two ranges of $-t$. For comparison, we also show the ETMC determination of iso-vector moments with a similar lattice setup~\cite{Alexandrou:2011nr}.\label{fig:fitQsqx1}}
\end{figure}

In \fig{fitQsqx1}, we show the gravitational form factors $A_{20}$ and $B_{20}$ as a function of $-t$ for the iso-vector (upper panels) as well as the iso-scalar (lower panels) cases. Good agreement can be found by comparing our results to those obtained from traditional calculations of local operators with one covariant derivative and a similar lattice setup (ETMC'11)~\cite{Alexandrou:2011nr}. Building upon our findings in \sec{momsKernel}, this further strengthens our confidence for the quasi-GPDs, particularly for $\mathcal{E}$ quasi-GPDs, under our Lorentz-invariant definition~\cite{Bhattacharya:2022aob}, having smaller power corrections and good perturbative convergence to the light-cone GPDs. Moreover, the bands resulting from different fits can capture well all the data points within the fit, where the dipole form typically exhibits more stability and smaller errors with fewer parameters. The $-t\rightarrow0$ extrapolation results are summarized in \tb{fitQsqx01}, in which the $A_{20}(0)$ can be interpreted as the hadron momentum fraction carried by the iso-vector and iso-scalar quarks. Our results for iso-vector cases agree well with the determination from ETMC'11~\cite{Alexandrou:2011nr}, while a mild deviation is again observed for the $A_{20}^{u+d}$, possibly due to the discretization effect~\cite{Gao:2020ito}, and the mixing with gluons that is not accounted for in the SDF~\cite{Ji:2022thb}. Using the gravitational form factors extrapolated at $-t=0$, we evaluate the spin contribution of the nucleon from the iso-vector and iso-scalar quarks as indicated by \Eq{totalA}, which are,
\begin{align}
\begin{split}
	J^{u-d}&=0.281(21)(11),\\
	J^{u+d}&=0.296(22)(33),
\end{split}
\end{align}
The errors in the first and second parenthesis are the usual statistical errors and systematic errors described above. We note that the NNLO and NLO matching kernels are used for iso-vector and iso-scalar cases, respectively. Our determination is consistent with the existing results of ETMC~\cite{Alexandrou:2011nr}; however, we note that the quark masses of this exploratory calculation are not at the physical point, and we also need more studies to address the lattice discretization errors and the excited state contamination in the future.

\begin{table}[h!]
\centering
\begin{tabular}{c c c | c c c}
\hline
\hline
$-t\rightarrow0$&This work &ETMC'11~\cite{Alexandrou:2011db}&$-t\rightarrow0$&This work &ETMC'11~\cite{Alexandrou:2011nr}\cr
\hline
$A_{10}^{u-d}$ &0.982(47)(28)&1&$A_{20}^{u-d}$ &0.267(13)(10)&0.264(13)\cr
$A_{10}^{u+d}$ &2.786(78)(64)&$-$&$A_{20}^{u+d}$ &0.544(15)(06)&0.613(14)\cr
$B_{10}^{u-d}$ &2.540(221)(108)&2.61(23)&$B_{20}^{u-d}$ &0.295(38)(23)&0.301(47)\cr
$B_{10}^{u+d}$ &0.170(299)(290)&$-$&$B_{20}^{u+d}$ &0.047(33)(65)&-0.046(43)\cr
\hline
\hline
\end{tabular}
\caption{The first two moments extrapolated to $-t\rightarrow0$ are shown. For comparison, we also show the first and second moments from the ETMC determination with similar lattice setup~\cite{Alexandrou:2011db,Alexandrou:2011nr}.
}
\label{tb:fitQsqx01}
\end{table}

\begin{figure}[h!]
    \centering
    \includegraphics[width=0.38\textwidth]{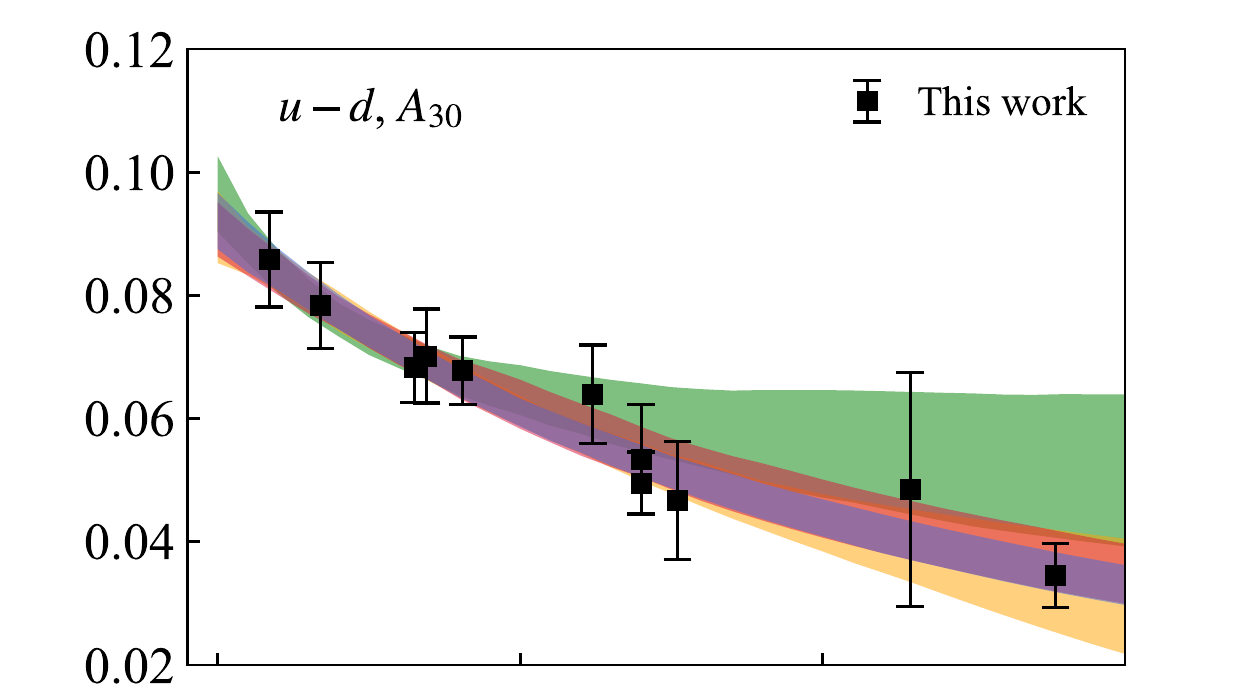}
    \includegraphics[width=0.38\textwidth]{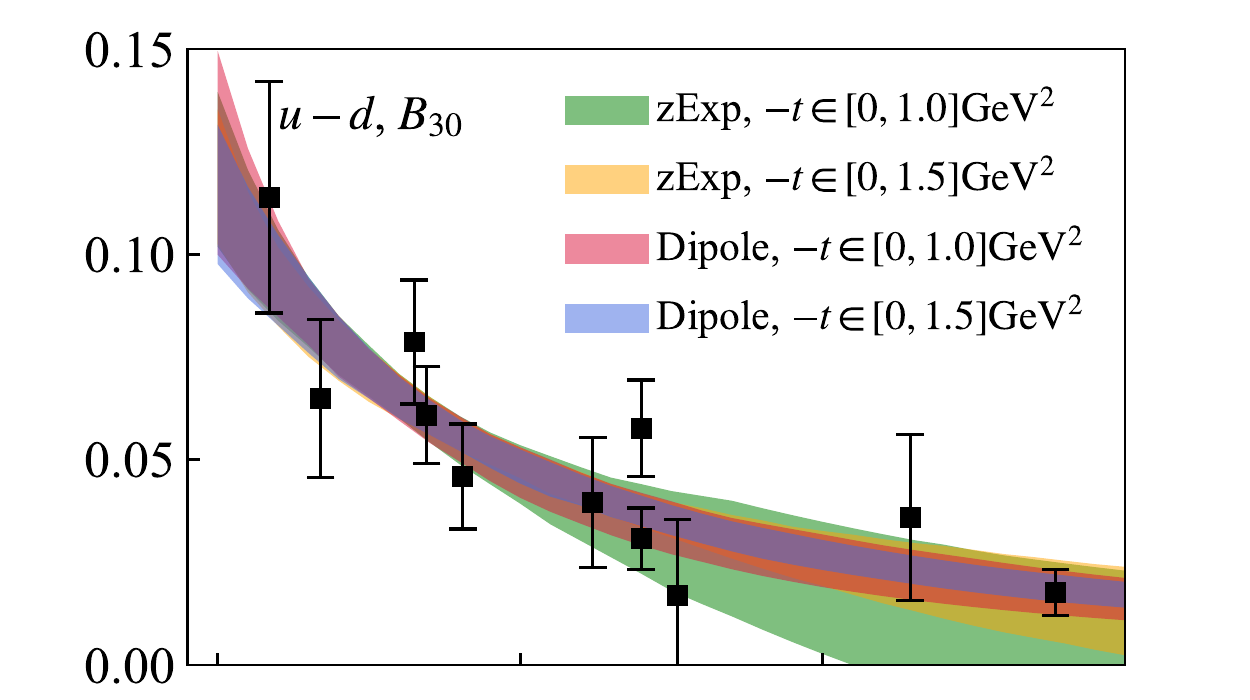}
    \includegraphics[width=0.38\textwidth]{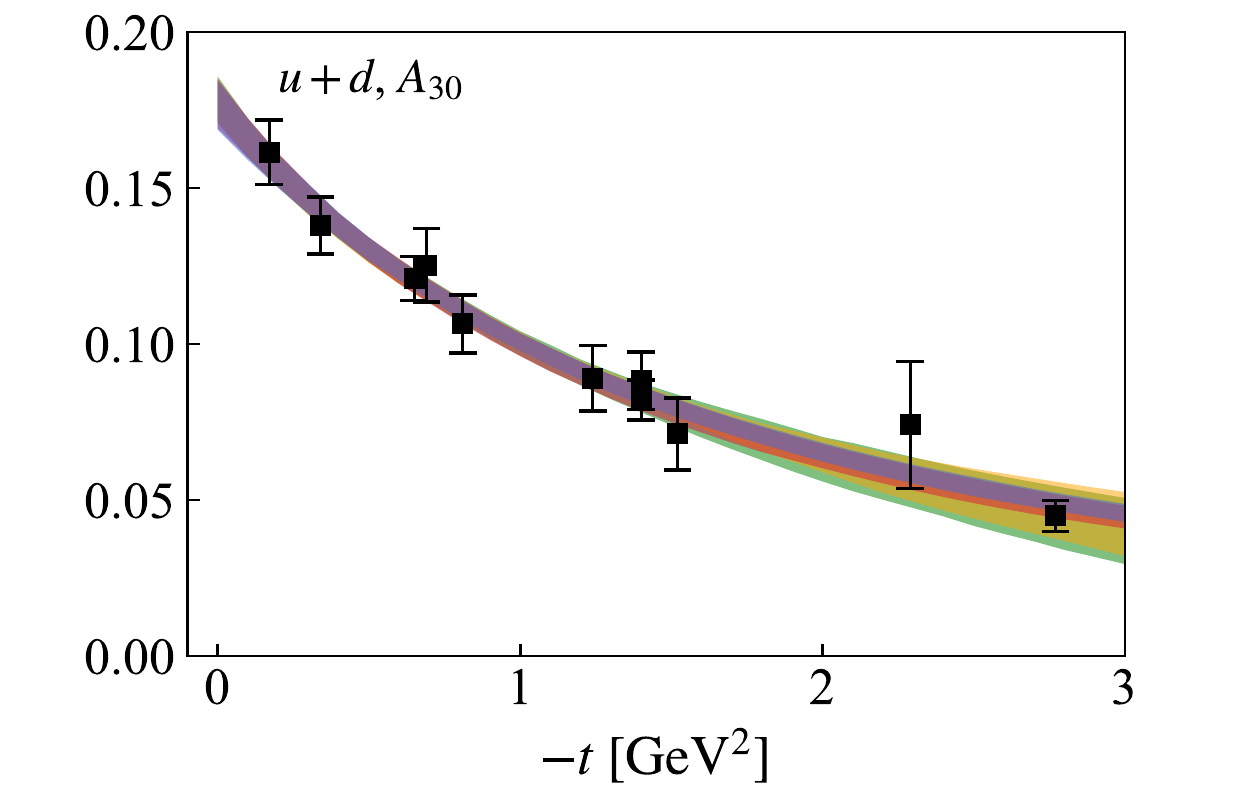}
    \includegraphics[width=0.38\textwidth]{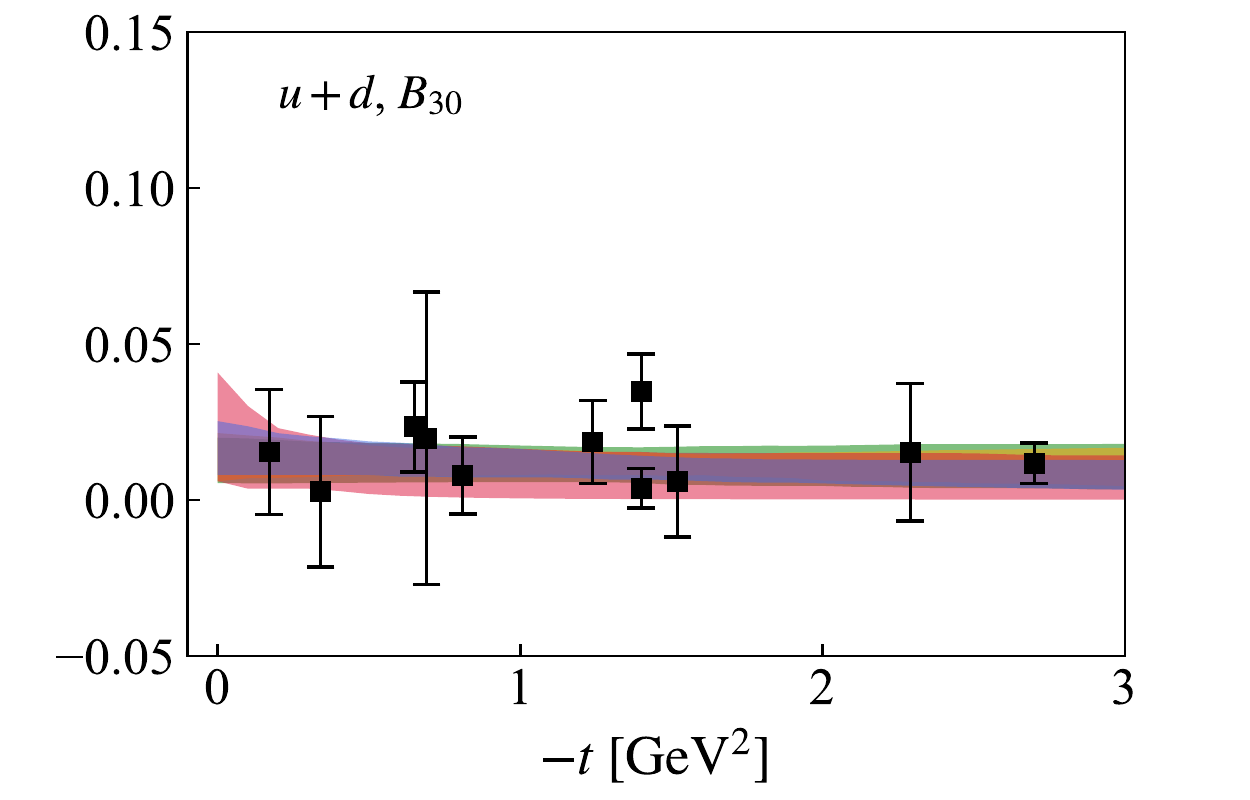}
	\caption{The third moments $A_{30}$ and $B_{30}$ for iso-vector (upper panels) and iso-scalar (lower panels) as a function of $-t$. The error bars include both statistical errors and systematic errors. The bands come from two different parametrizations using two ranges of $-t$. \label{fig:fitQsqx2}}
\end{figure}

\begin{figure}[h!]
    \centering
    \includegraphics[width=0.38\textwidth]{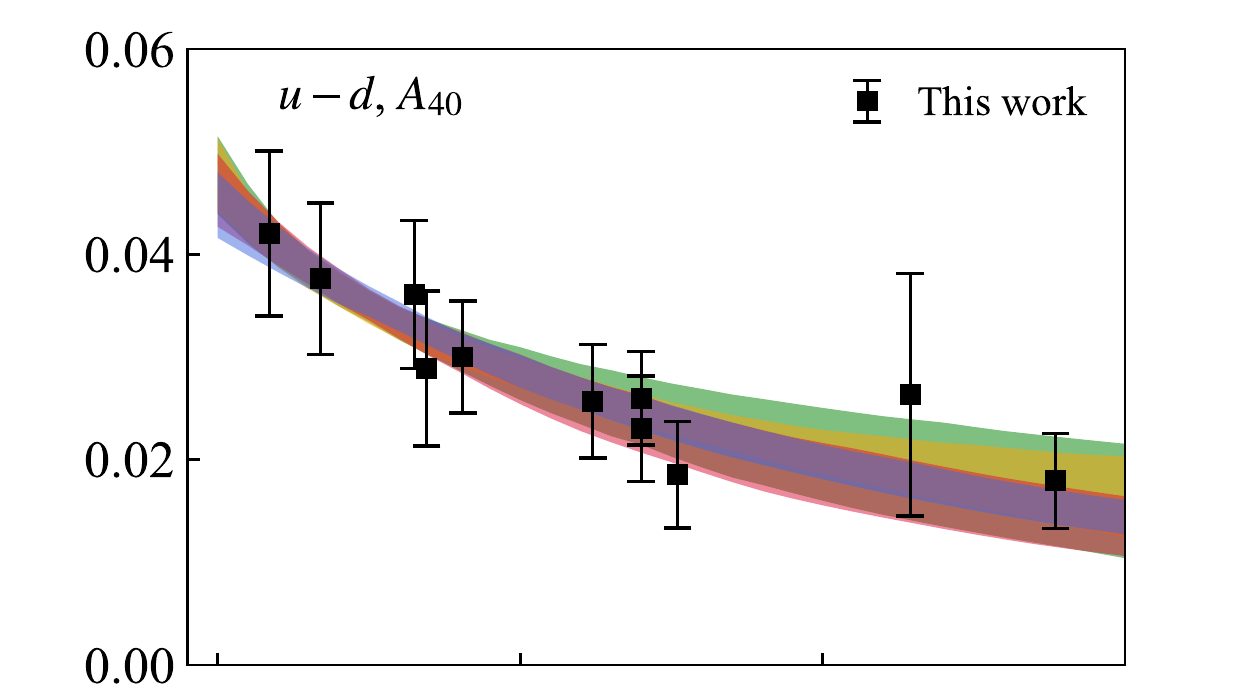}
    \includegraphics[width=0.38\textwidth]{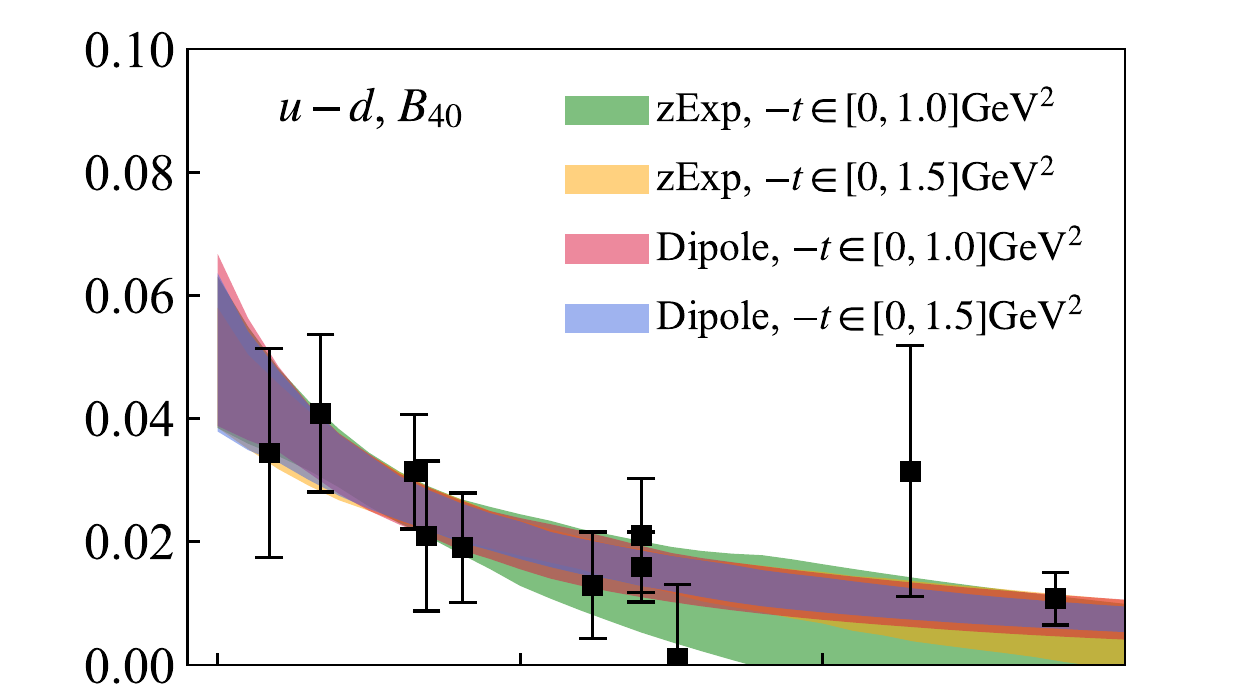}
    \includegraphics[width=0.38\textwidth]{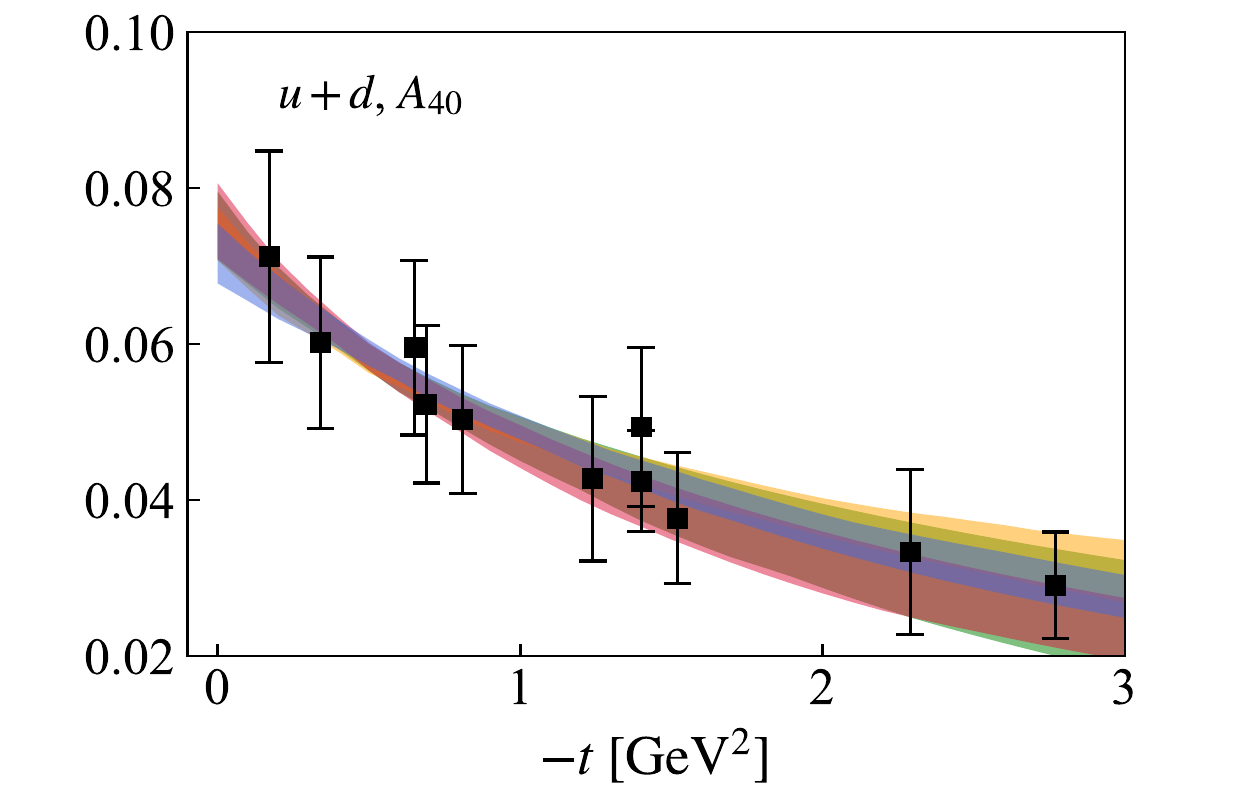}
    \includegraphics[width=0.38\textwidth]{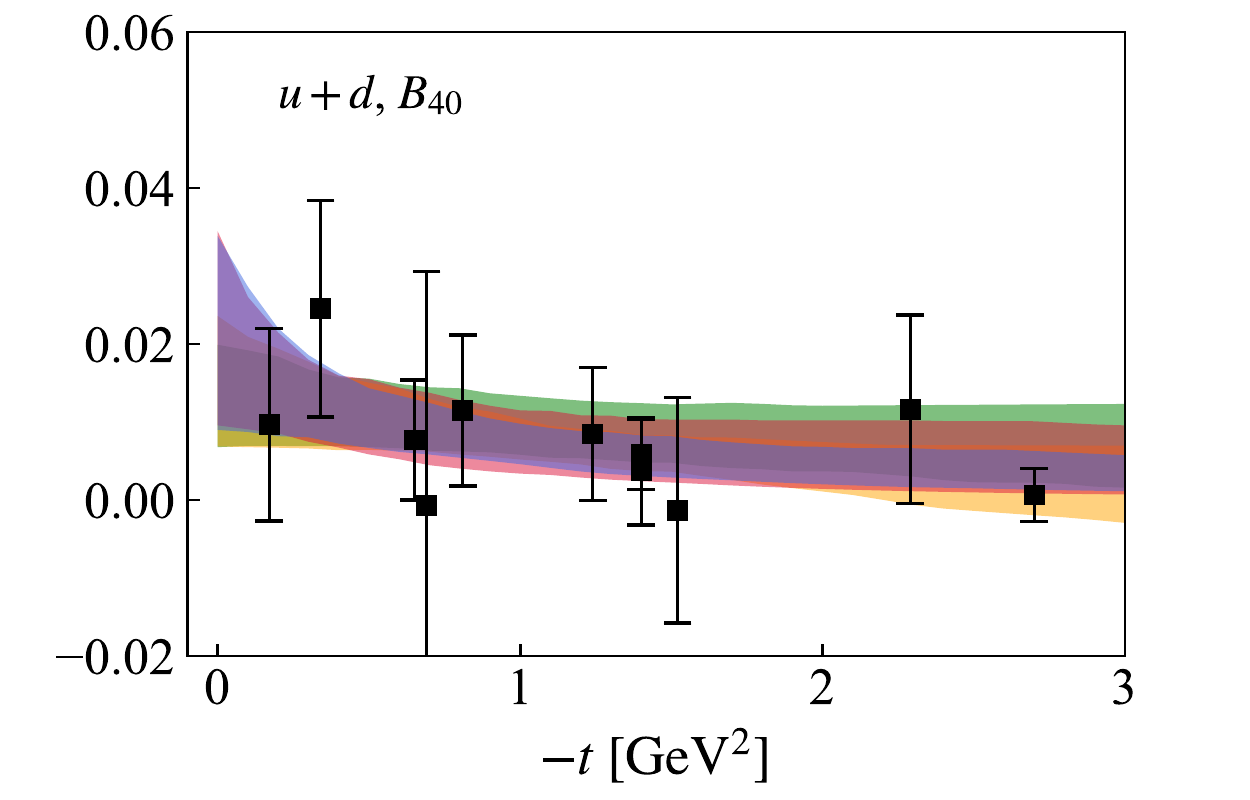}
	\caption{The fourth moments $A_{40}$ and $B_{40}$ for iso-vector (upper panels) and iso-scalar (lower panels) as a function of $-t$. The error bars include both statistical errors and systematic errors. The bands come from two different parametrizations using two ranges of $-t$. \label{fig:fitQsqx3}}
\end{figure}

\begin{figure}[h!]
    \centering
    \includegraphics[width=0.38\textwidth]{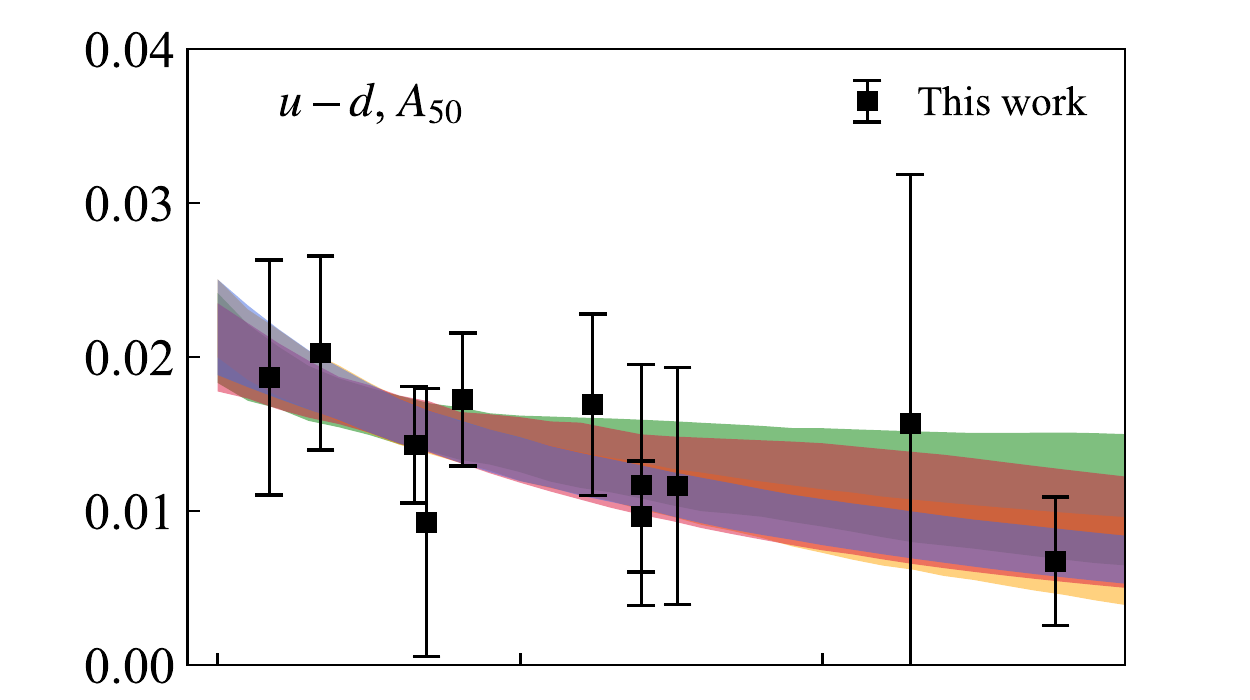}
    \includegraphics[width=0.38\textwidth]{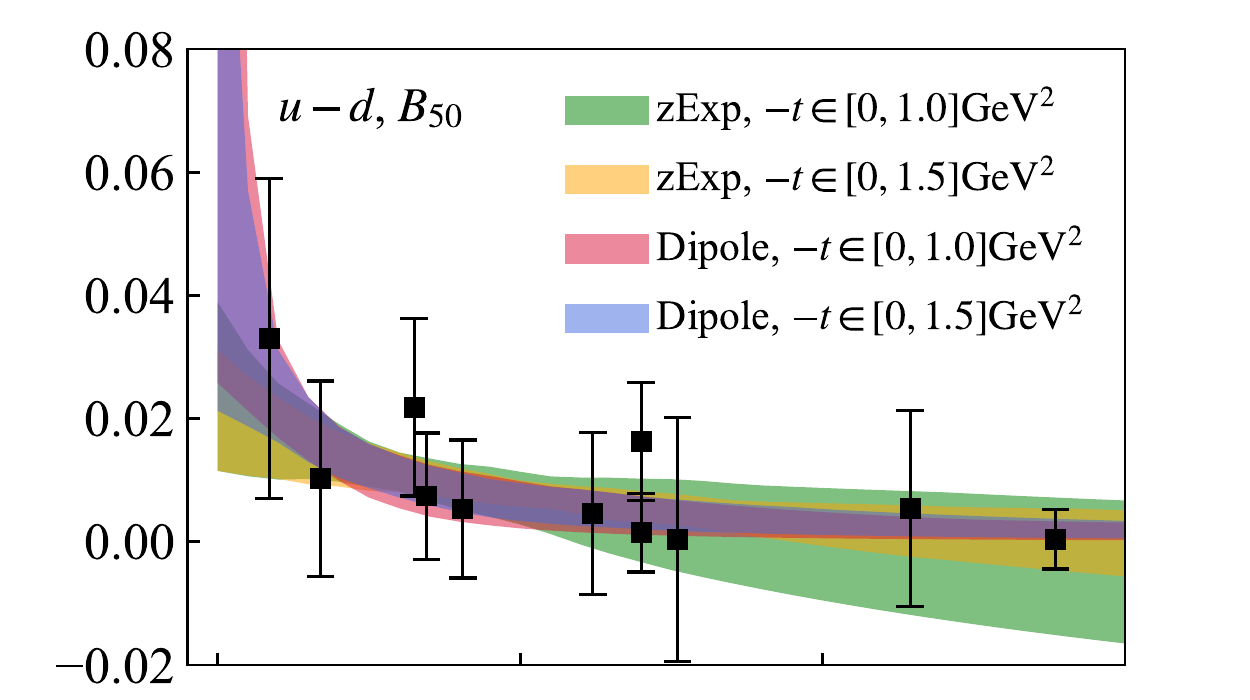}
    \includegraphics[width=0.38\textwidth]{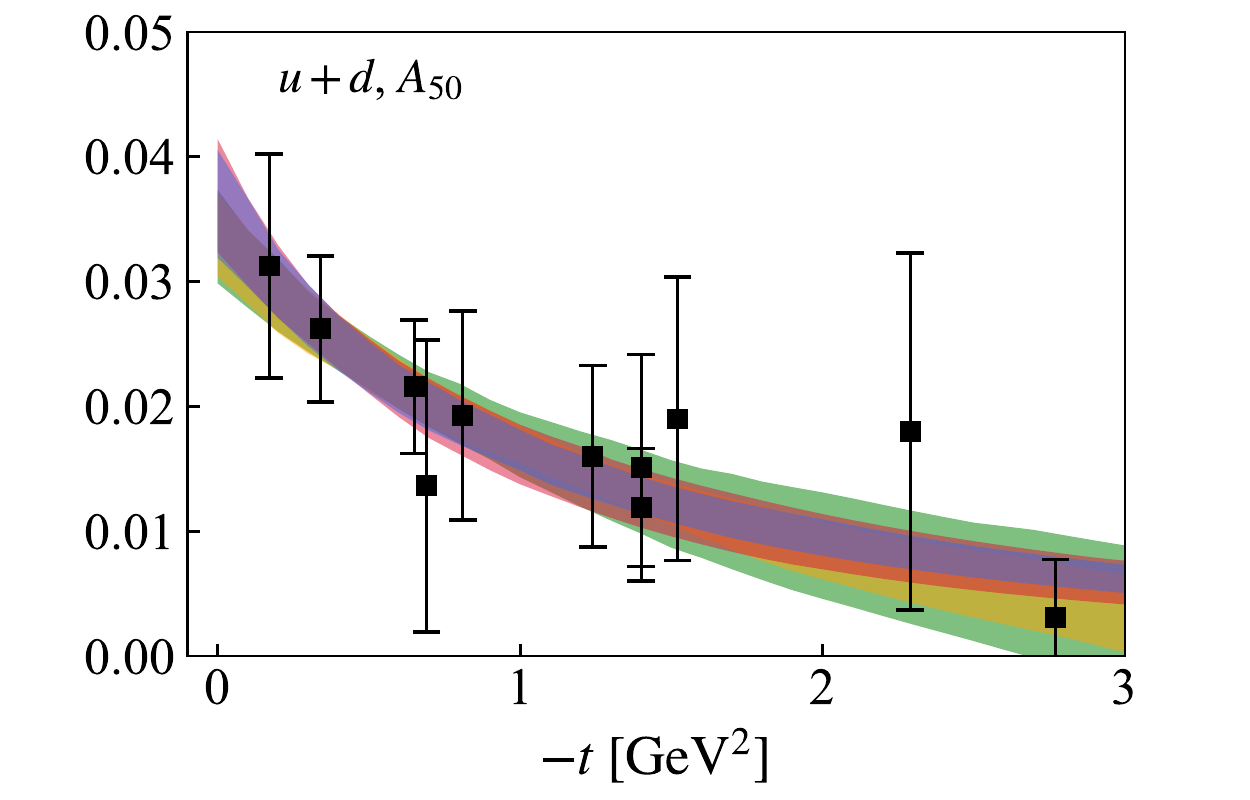}
    \includegraphics[width=0.38\textwidth]{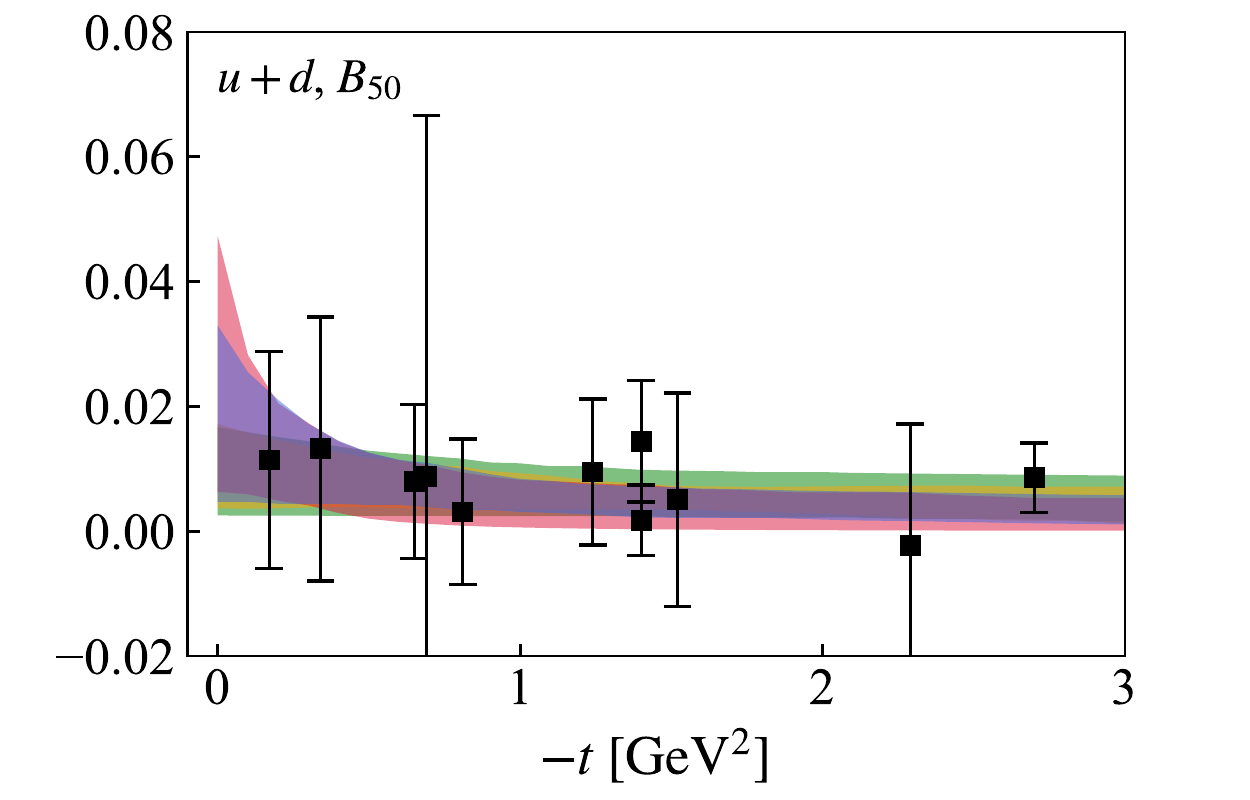}
	\caption{The fifth moments $A_{50}$ and $B_{50}$ for iso-vector (upper panels) and iso-scalar (lower panels) as a function of $-t$. The error bars include both statistical errors and systematic errors. The bands come from two different parametrizations using two ranges of $-t$. \label{fig:fitQsqx4}}
\end{figure}

We show the third, fourth, and fifth moments, for the first time, as a function of $-t$ in Figs.~\ref{fig:fitQsqx2},~\ref{fig:fitQsqx3} and \ref{fig:fitQsqx4}. Reasonable signal and smooth $-t$ dependence can be observed for the $A_{n+1,0}$ moments, while the iso-vector $B_{50}$ and all the iso-scalar $B_{n+1,0}$ moments are mostly consistent with zero. All the results are also summarized in \app{table}. There are no existing results for comparison, which, however, is essentially the advantage of the method used in this work, namely the short distance factorization of quasi-GPD matrix element, that one can systematically get access to the higher moments with larger hadron momentum. On the contrary, the traditional moments calculations are limited to the lowest few moments, as discussed previously. Moreover, it is notable that the moments of each type of $A_{n+1,0}$ and $B_{n+1,0}$ show a qualitative hierarchy that is consistent with the large $N_c$ counting rules~\cite{Goeke:2001tz},
\begin{align}
	|A_{n+1,0}^{u+d}|\sim N_c^2 \gg |A_{n+1,0}^{u-d}|\sim N_c,\qquad |B_{n+1,0}^{u-d}|\sim N_c^3 \gg |B_{n+1,0}^{u+d}|\sim N_c^2.
\end{align}
However, we also note a breaking of the prediction between different types, consistent with previous findings~\cite{LHPC:2007blg},
\begin{align}
	|B_{n+1,0}^{u-d}|\sim N_c^3 \gg |A_{n+1,0}^{u+d}|\sim N_c^2.
\end{align}
Specifically, our results show that $A_{n+1,0}^{u-d}\gtrsim B_{n+1,0}^{u-d}$.

\subsection{Impact parameter space interpretation}

The GPDs can be interpreted in the impact parameter space, with the zero-skewness GPDs being particularly relevant. Specifically, when considering unpolarized quarks inside an unpolarized proton, the Fourier transform of the $H$ GPDs provides information on how partons carrying a momentum fraction $x$ are distributed in the transverse plane,
\begin{align}
	q(x,\vec{b}_\perp)=\int \frac{d^2\vec{\Delta}_\perp}{(2\pi)^2}H(x,-\vec{\Delta}^2_\perp)e^{-i\vec{b}_\perp\cdot \vec{\Delta}_\perp}.
\end{align}
Taking into account the $E$ GPDs, one can explore the unpolarized quarks distribution inside a transversely polarized proton~\cite{Burkardt:2002hr}, defined as,
\begin{align}
\begin{split}
	q^T(x,\vec{b}_\perp)&=\int \frac{d^2\vec{\Delta}_\perp}{(2\pi)^2}[H(x,-\vec{\Delta}_\perp^2)+i\frac{\Delta_y}{2M}E(x,-\vec{\Delta}_\perp^2)]e^{-i\vec{b}_\perp\cdot \vec{\Delta}_\perp}\\
	&=q(x,\vec{b}_\perp)-\frac{1}{2M}\frac{\partial}{\partial b_y}q_E(x,\vec{b}_\perp),
\end{split}
\end{align}
where we denoted the Fourier transform of $E(x,-\vec{\Delta}^2)$ by $q_E(x,\vec{b}_\perp)$, and the proton is transversely polarized in the $x$ direction. These impact parameter-dependent parton distributions (IPDs) allow us to visualize the three-dimensional structure of the parton distribution inside the proton, taking into account both longitudinal momentum and transverse position. The $q^T(x,\vec{b}_\perp)$ also have a relation to the Sivers distributions~\cite{Burkardt:2003uw,Burkardt:2005hp,Meissner:2007rx}. In this work, we derived the moments of the $H$ and $E$ GPDs, which enabled us to infer the moments of $q(x,\vec{b}_\perp)$ and $q^T(x,\vec{b}_\perp)$,
\begin{align}
\begin{split}
	\rho_{n+1} (\vec{b}_\perp) &= \int \frac{d^2\vec{\Delta}_\perp}{(2\pi)^2} A_{n+1,0}(-\vec{\Delta}_\perp^2)e^{-i\vec{b}_\perp\cdot \vec{\Delta}_\perp},\\
	\rho_{n+1}^T (\vec{b}_\perp) &= \int \frac{d^2\vec{\Delta}_\perp}{(2\pi)^2} [A_{n+1,0}(-\vec{\Delta}_\perp^2)+i\frac{\Delta_y}{2M}B_{n+1,0}(-\vec{\Delta}_\perp^2)]e^{-i\vec{b}_\perp\cdot \vec{\Delta}_\perp}.
\end{split}
\end{align}
Since Fourier transforms require full information of $-t\in[0, \infty]$, we utilized our dipole fit result from $-t\in[0, 1.5]~\rm{GeV}^2$, which was found to describe the data up to 2.77 $\rm{GeV}^2$ and will model the $-t\rightarrow\infty$ behavior. We excluded the contribution from $B_{n+1,0}^{u+d}$, as these data were mostly consistent with zero in our calculations. We then will compute the impact space distribution for up and down quarks as a function of $\vec{b}_\perp$ for both $\rho_{n+1} (\vec{b}_\perp)$ and $\rho_{n+1}^T (\vec{b}_\perp)$, with $n$ ranging from 0 to 3. We excluded the $n=4$ case since $B_{50}^{u+d}$ and $B_{50}^{u-d}$ are both noisy and consistent with 0, although a reasonable signal was observed for $A_{50}^{u+d}$ and $A_{50}^{u-d}$. The flavor separation is derived by a linear combination of our iso-vector and iso-scalar results. We note again that our analysis did not account for the disconnected diagram contributions or mixing with gluons in the iso-scalar case, which are expected to be small and not significantly impact our qualitative findings~\cite{Alexandrou:2021oih}. Additionally, it is worth noting that the moments of the IPDs receive contributions from both quarks and anti-quarks, with an integral taken from $x=-1$ to 1. The difference or sum of the quark and anti-quark density distributions can be denoted as $\rho_{n+1}=\rho_{n+1}^q+(-1)^{n+1}\rho_{n+1}^{\bar q}$ for even or odd $n$, respectively~\cite{QCDSF:2006tkx}. 

\begin{figure}[h!]
    \centering
    \includegraphics[width=0.45\textwidth]{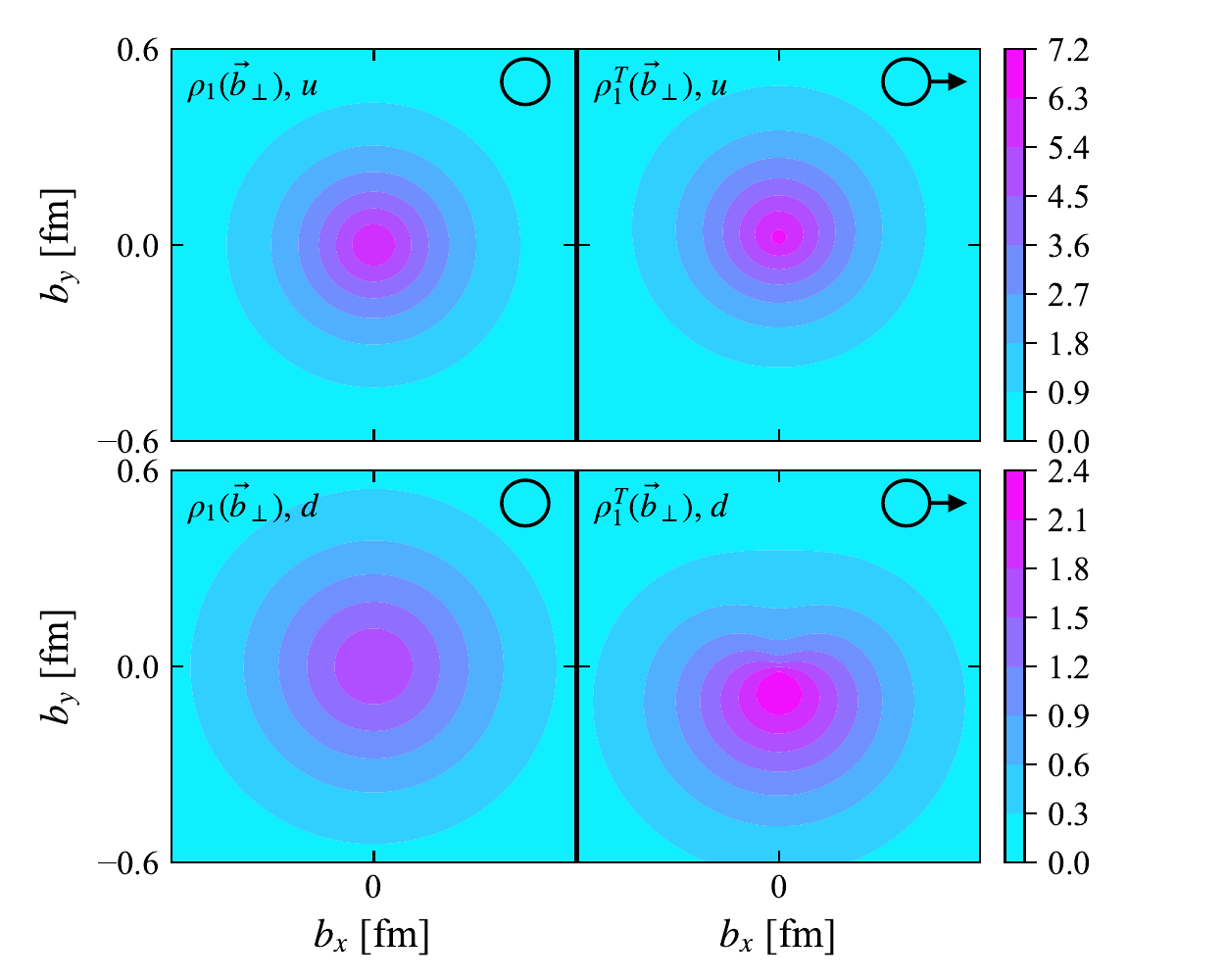}
    \includegraphics[width=0.45\textwidth]{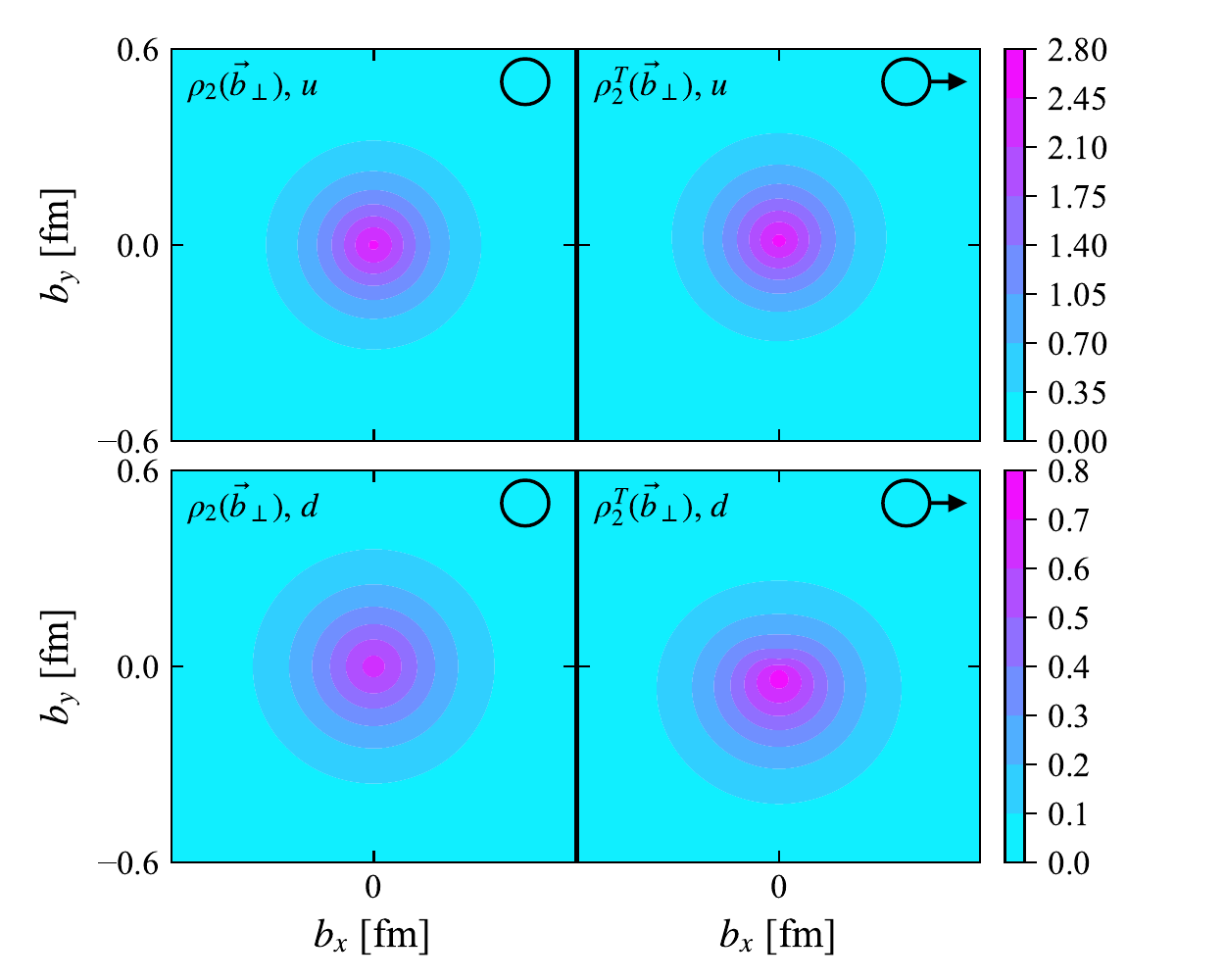}
    \includegraphics[width=0.45\textwidth]{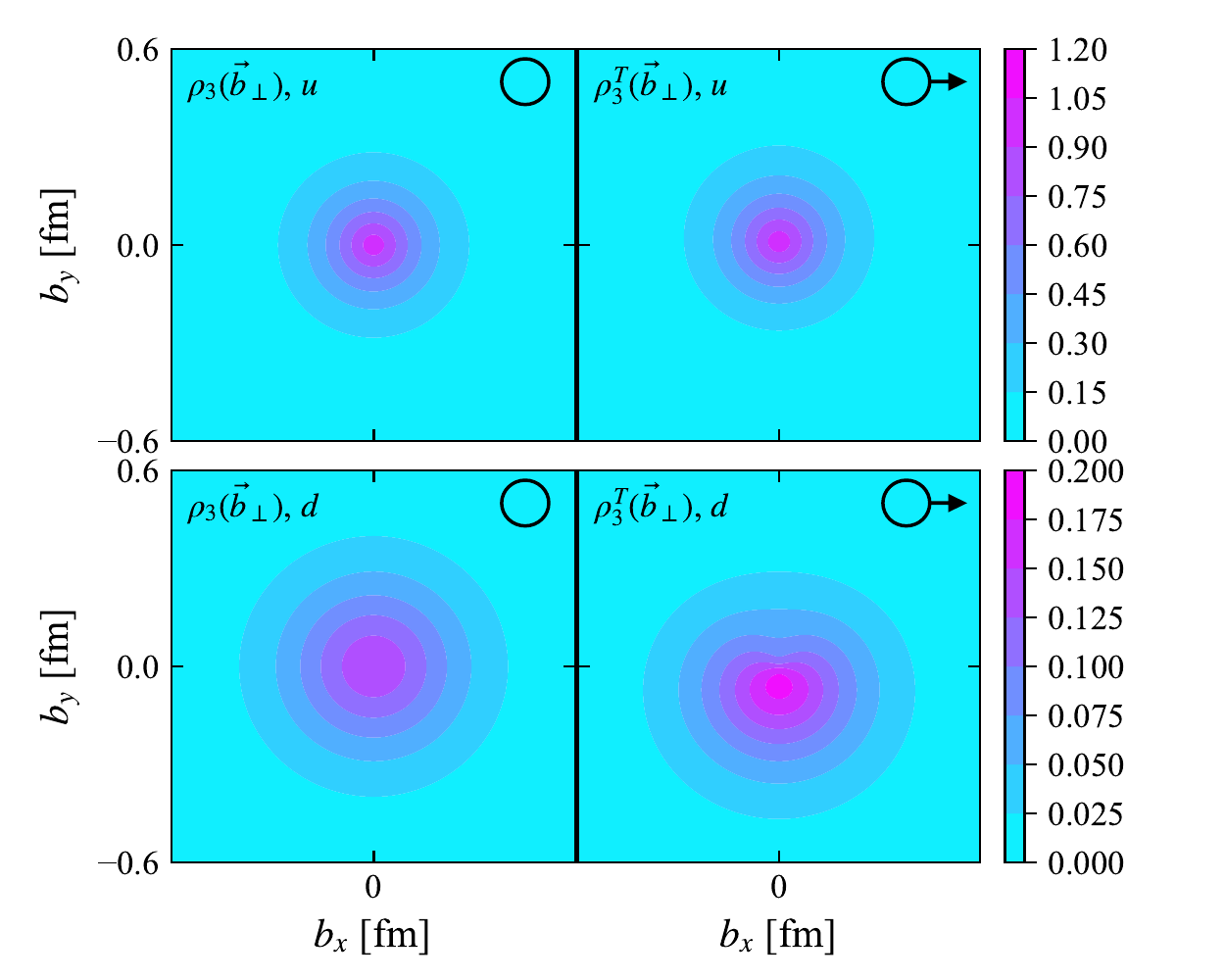}
    \includegraphics[width=0.45\textwidth]{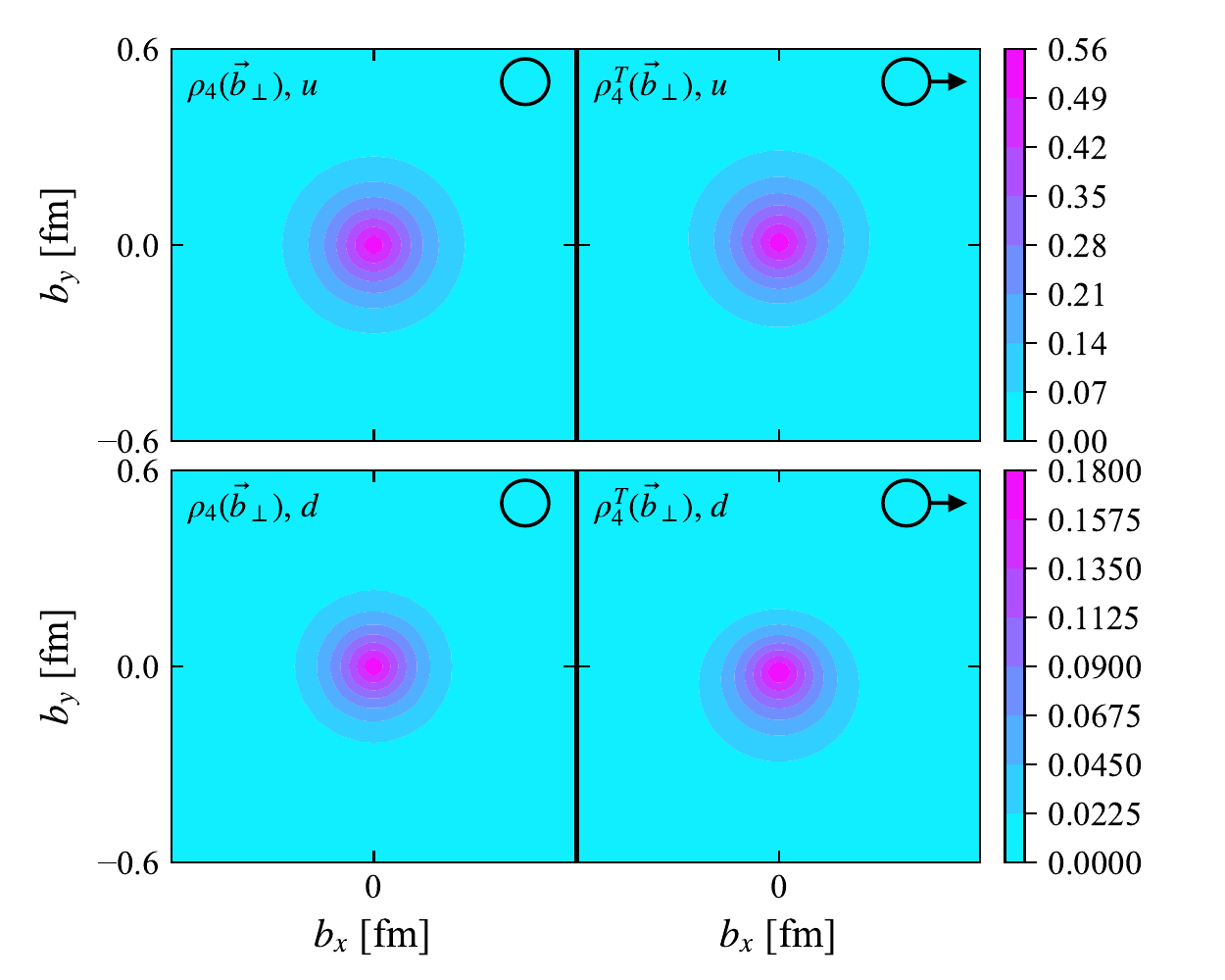}
	\caption{The first four moments of impact parameter parton distributions in the transverse plane. The unpolarized and transversely polarized (in $x$ direction) proton are both shown in the left and right parts of each panels, denoted by $\rho_{n+1}~[\rm fm^{-2}]$ and $\rho_{n+1}^T~[\rm fm^{-2}]$ for both up ($u$) and down ($d$) quarks.\label{fig:density}}
\end{figure}

In the upper left panel of \fig{density}, we present the first moments $\rho_{1}$ (left) and $\rho_{1}^T$ (right), which describe the density distribution of up and down quarks in the transverse plane. As one can see, the down quark exhibits a broader distribution and smaller amplitudes compared to the up quark in the unpolarized proton. When the proton is transversely polarized, the up and down quarks shift in different directions, with the down quarks showing larger distortion. This observation holds true for cases when $n>0$, where the density distribution is weighted by $x^{n}$. The second moment, in particular, is of interest as it describes how momentum is distributed in the transverse plane. The larger distortion of down quarks can be attributed to the fact that they constitute a smaller fraction of the proton, as indicated by the $|A_{n+1,0}^{u}|>|A_{n+1,0}^{d}|$, but are subject to similar distortion forces, as evidenced by $B_{n+1,0}^{u}\approx -B_{n+1,0}^{d}$. This also ensures that the total contribution of $u$ and $d$ quarks to the transverse center of momentum, $\sum_{u,d}\int d^2\vec{b}_\perp b_y\rho_{2}(\vec{b}_\perp)=1/(2M)B_{20}^{u+d}(-t=0)$, is approximately 0. Therefore, it is reasonable to expect that the shift of gluon distribution in a transversely polarized proton would be either close 0 or counterbalanced by heavier sea quarks and the contributions from the disconnected diagrams of light quarks. What is more, one can find that $\rho_{n+1}$ (and $\rho_{n+1}^T$) with higher $n$ exhibit a sharper drop in the transverse distance for even and odd moments, indicating that active quarks with larger $x$ may be more concentrated around the center than small-$x$ quarks inside the proton.

\section{Conclusion}\label{sec:conclusion}

In this work, we present the first lattice calculation of the Mellin moments of nucleon unpolarized quark generalized parton distributions (GPDs) up to the fifth order. We utilize the short distance factorization of quasi-GPD matrix elements, calculating both iso-vector and iso-scalar combinations, while ignoring the contribution of disconnected diagrams and quark-gluon mixing to the iso-scalar case. Our calculation was carried out using one ensemble of $N_f=2+1+1$ twisted mass fermions with clover improvement, where the pion mass was set to 260 MeV and the lattice spacing was $a=0.093$ fm. We calculate the bare quasi-GPDs in both asymmetric and symmetric frames using the $\gamma_0$ definition and construct the ones of a recently proposed Lorentz-invariant definition~\cite{Bhattacharya:2022aob}. We renormalize the bare matrix elements using the zero-momentum quasi-PDF matrix elements through the ratio scheme. By inserting the SDF formula into the RG-invariant ratio-scheme matrix elements, we can extract the Mellin moments with perturbatively calculated Wilson coefficients up to NNLO for the iso-vector case and up to NLO for the iso-scalar case. 
To test the validity of the approach we first performed analysis at fixed 
$z_3$ varying only $P_3$, and studied how the results change when we choose different values of $z_3$.
In the case of the Lorentz-invariant definition, within the statistical error our results on the moments
show no $z_3$-dependence with NLO/NNLO Wilson coefficients as compared to the slight $z_3$-dependence at LO, implying that matching can account for the small-$z_3$ evolution effects and that the higher-twist contributions are negligible.
In the iso-vector case there is no apparent difference between NLO and NNLO during the numerical implementation, given the current statistics, implying that
the higher order corrections are small compared to the statistical errors.
However, we find that for the  quasi GPDs $\mathcal{E}$ with the $\gamma_0$
definition the short distance factorization does not seem to work even
if NNLO matching coefficients are used. 
We then conducted a combined analysis by fitting a range of $z$ values and incorporating the evolution kernels, which enabled us to obtain reliable results for the moments up to the fifth order using the Lorentz-invariant quasi-GPDs, as detailed in \sec{moms}. Notably, the first two moments under the Lorentz-invariant definition consistently agree with traditional moment calculations. This may be an indication that the quasi-GPDs, at least for $\cal E$, from Lorentz-invariant definition have potentially smaller power corrections and converge to the light-cone GPDs faster. In section \sec{tdependence}, we presented, for the first time up to fifth order, moments from both the iso-vector and iso-scalar quasi-GPDs, with various values of momentum transfer $-t$. We benefited greatly from the short-distance factorization of the quasi-GPD matrix elements, which allowed us to systematically access higher moments by increasing the hadron momentum. Then we applied two methods of fit using the dipole model and $z$-expansion formula and found them both describe well the $-t$ dependence of the moments we got. From the fit results, we were able to infer the moments for $-t\rightarrow0$, as well as the quark charges, momentum fraction, and total angular momentum contributions to the proton spin. These are all crucial quantities for understanding the structure of the nucleon. Finally, we discuss the impact parameter space interpretation of the GPDs and their moments by Fourier transform the dipole fit results, which provides insights into the spatial distribution of quarks and their correlations in the transverse plane of an unpolarized or transversely polarized proton. In future work, we will refine calculations at the physical point and investigate systematic uncertainties such as discretization effects and excited state contamination, and possibly further constrain the higher moments with better data quality.

\begin{acknowledgements}

This material is based upon work supported by the U.S. Department of Energy, Office of Science, Office of Nuclear Physics through Contract No.~DE-SC0012704, No.~DE-AC02-06CH11357 and within the framework of Scientific Discovery through Advance Computing (SciDAC) award Fundamental Nuclear Physics at the Exascale and Beyond. S.~B.\ is supported by  Laboratory Directed Research and Development (LDRD) funds from Brookhaven Science Associates. 
K.~C.\ is supported by the National Science Centre (Poland) grants SONATA BIS no.\ 2016/22/E/ST2/00013 and OPUS no.\ 2021/43/B/ST2/00497. M.~C., J. D. and A.~S. acknowledge financial support by the U.S. Department of Energy, Office of Nuclear Physics, Early Career Award under Grant No.\ DE-SC0020405.
The work of A.~M. is supported by the National Science Foundation under grant number PHY-2110472, and also by the U.S. Department of Energy, Office of Science, Office of Nuclear Physics, within the framework of the TMD Topical Collaboration. 
F.~S.\ was funded by the NSFC and the Deutsche Forschungsgemeinschaft (DFG, German Research Foundation) through the funds provided to the Sino-German Collaborative Research Center TRR110 “Symmetries and the Emergence of Structure in QCD” (NSFC Grant No. 12070131001, DFG Project-ID 196253076 - TRR 110). 
The authors also acknowledge partial support by the U.S. Department of Energy, Office of Science, Office of Nuclear Physics under the umbrella of the Quark-Gluon Tomography (QGT) Topical Collaboration with Award DE-SC0023646.
This publication is partially funded by the Gordon and Betty Moore Foundation through the Grant GBMF6210 to APS to support the participation and presentation of X.G. at the 10$^{\rm th}$ APS GHP meeting (April 12-14, 2023).
Computations for this work were carried out in part on facilities of the USQCD Collaboration, which are funded by the Office of Science of the U.S. Department of Energy. 
This research used resources of the Oak Ridge Leadership Computing Facility, which is a DOE Office of Science User Facility supported under Contract DE-AC05-00OR22725.
This research was supported in part by PLGrid Infrastructure (Prometheus supercomputer at AGH Cyfronet in Cracow).
Computations were also partially performed at the Poznan Supercomputing and Networking Center (Eagle supercomputer), the Interdisciplinary Centre for Mathematical and Computational Modelling of the Warsaw University (Okeanos supercomputer), and at the Academic Computer Centre in Gda\'nsk (Tryton supercomputer).
The gauge configurations have been generated by the Extended Twisted Mass Collaboration on the KNL (A2) Partition of Marconi at CINECA, through the Prace project Pra13\_3304 ``SIMPHYS".
Inversions were performed using the DD-$\alpha$AMG solver~\cite{Frommer:2013fsa} with twisted mass support~\cite{Alexandrou:2016izb}. 

\end{acknowledgements}

\appendix

\section{Tables of the moments}\label{app:table}
We list all our determination of GPD moments in Tables~\ref{tb:allisovA}-\ref{tb:allisosB}. The factorization scale is set to be $\mu$ = 2 GeV.

\begin{table}[h!]
\centering
\begin{tabular}{c c c c c c}
\hline
$-t~\rm{GeV}^2$&$A_{10}^{u-d}$&$A_{20}^{u-d}$&$A_{30}^{u-d}$&$A_{40}^{u-d}$&$A_{50}^{u-d}$\cr
\hline
0.17 & 0.851(31)(03) & 0.247(11)(04) & 0.086(04)(03) & 0.042(03)(05) & 0.019(03)(05)\cr
0.34 & 0.702(43)(02) & 0.205(01)(03) & 0.078(05)(02) & 0.038(03)(04) & 0.020(03)(03)\cr
0.65 & 0.607(24)(01) & 0.193(09)(04) & 0.068(05)(01) & 0.036(02)(05) & 0.014(02)(02)\cr
0.69 & 0.573(14)(04) & 0.187(07)(03) & 0.070(05)(02) & 0.029(05)(03) & 0.009(05)(03)\cr
0.81 & 0.487(31)(01) & 0.168(09)(02) & 0.068(04)(01) & 0.030(03)(03) & 0.017(03)(02)\cr
1.24 & 0.359(37)(02) & 0.145(12)(01) & 0.064(06)(02) & 0.026(04)(02) & 0.017(03)(03)\cr
1.38 & 0.396(03)(03) & 0.137(12)(02) & 0.053(06)(03) & 0.023(03)(02) & 0.012(04)(04)\cr
1.38 & 0.376(21)(01) & 0.129(09)(02) & 0.050(04)(01) & 0.026(02)(02) & 0.010(02)(01)\cr
1.52 & 0.320(37)(03) & 0.131(14)(01) & 0.047(07)(03) & 0.019(04)(01) & 0.012(04)(04)\cr
2.29 & 0.266(66)(05) & 0.115(28)(03) & 0.048(14)(05) & 0.026(08)(04) & 0.016(08)(08)\cr
2.77 & 0.214(13)(01) & 0.101(07)(01) & 0.035(04)(01) & 0.018(03)(02) & 0.007(03)(02)\cr
\hline
\end{tabular}
\caption{The table of iso-vector GPD moments $A_{n+1,0}^{u-d}$.
}
\label{tb:allisovA}
\end{table}

\begin{table}[h!]
\centering
\begin{tabular}{c c c c c c}
\hline
$-t~\rm{GeV}^2$&$B_{10}^{u-d}$&$B_{20}^{u-d}$&$B_{30}^{u-d}$&$B_{40}^{u-d}$&$B_{50}^{u-d}$\cr
\hline
0.17 & 1.964(142)(011) & 0.247(33)(04) & 0.114(17)(11) & 0.034(11)(05) & 0.033(09)(17) \cr
0.34 & 1.547(113)(006) & 0.256(32)(03) & 0.065(13)(06) & 0.041(08)(04) & 0.010(08)(08) \cr
0.65 & 1.107(49)(06) & 0.192(18)(03) & 0.079(09)(06) & 0.031(05)(04) & 0.022(06)(09) \cr
0.69 & 1.073(37)(06) & 0.215(12)(05) & 0.061(08)(03) & 0.021(09)(04) & 0.007(07)(03) \cr
0.81 & 0.895(61)(03) & 0.164(23)(02) & 0.046(09)(04) & 0.019(07)(02) & 0.005(06)(05) \cr
1.24 & 0.609(52)(04) & 0.124(20)(02) & 0.040(11)(05) & 0.013(06)(03) & 0.005(06)(07) \cr
1.38 & 0.561(42)(03) & 0.114(18)(02) & 0.058(09)(03) & 0.021(06)(03) & 0.016(05)(04) \cr
1.38 & 0.520(33)(02) & 0.114(13)(01) & 0.031(05)(02) & 0.016(04)(02) & 0.001(03)(03) \cr
1.52 & 0.443(65)(07) & 0.083(19)(02) & 0.017(11)(07) & 0.001(09)(03) & 0.000(09)(11) \cr
2.29 & 0.306(81)(05) & 0.085(32)(04) & 0.036(15)(05) & 0.031(15)(05) & 0.005(09)(07) \cr
2.77 & 0.199(19)(02) & 0.067(09)(01) & 0.018(04)(02) & 0.011(03)(01) & 0.000(03)(02) \cr
\hline
\end{tabular}
\caption{The table of iso-vector GPD moments $B_{n+1,0}^{u-d}$.
}
\label{tb:allisovB}
\end{table}

\begin{table}[h!]
\centering
\begin{tabular}{c c c c c c}
\hline
$-t~\rm{GeV}^2$&$A_{10}^{u+d}$&$A_{20}^{u+d}$&$A_{30}^{u+d}$&$A_{40}^{u+d}$&$A_{50}^{u+d}$\cr
\hline
0.17 & 2.271(58)(04) & 0.492(14)(07) & 0.161(06)(04) & 0.071(03)(01) & 0.031(04)(05) \cr
0.34 & 1.792(47)(01) & 0.410(13)(06) & 0.138(08)(01) & 0.060(03)(08) & 0.026(04)(02) \cr
0.65 & 1.341(36)(02) & 0.383(12)(06) & 0.121(06)(02) & 0.060(03)(09) & 0.022(03)(02) \cr
0.69 & 1.308(20)(12) & 0.372(11)(05) & 0.125(06)(05) & 0.052(06)(05) & 0.014(06)(05) \cr
0.81 & 1.095(36)(03) & 0.312(01)(04) & 0.106(06)(03) & 0.050(04)(06) & 0.019(04)(05) \cr
1.24 & 0.735(37)(02) & 0.249(13)(04) & 0.089(08)(02) & 0.043(04)(06) & 0.016(04)(03) \cr
1.38 & 0.780(53)(03) & 0.272(17)(04) & 0.088(06)(03) & 0.049(05)(06) & 0.015(04)(05) \cr
1.38 & 0.692(23)(02) & 0.244(12)(03) & 0.082(05)(02) & 0.042(03)(04) & 0.012(02)(02) \cr
1.52 & 0.568(40)(04) & 0.212(16)(02) & 0.071(08)(04) & 0.038(05)(03) & 0.019(05)(06) \cr
2.29 & 0.457(110)(05) & 0.212(46)(02) & 0.074(16)(04) & 0.033(08)(02) & 0.018(08)(06) \cr
2.77 & 0.350(18)(01) & 0.160(11)(03) & 0.045(04)(01) & 0.029(03)(04) & 0.003(03)(02) \cr
\hline
\end{tabular}
\caption{The table of iso-scalar GPD moments $A_{n+1,0}^{u+d}$.
}
\label{tb:allisosA}
\end{table}

\begin{table}[h!]
\centering
\begin{tabular}{c c c c c c}
\hline
$-t~\rm{GeV}^2$&$B_{10}^{u+d}$&$B_{20}^{u+d}$&$B_{30}^{u+d}$&$B_{40}^{u+d}$&$B_{50}^{u+d}$\cr
\hline
0.17 & 0.102(191)(05) & 0.025(34)(02) & 0.015(15)(05) & 0.010(10)(03) & 0.011(10)(07) \cr
0.34 & -0.004(162)(08) & 0.024(35)(04) & 0.003(17)(08) & 0.025(09)(05) & 0.013(10)(11) \cr
0.65 & 0.046(61)(03) & 0.033(18)(02) & 0.023(11)(03) & 0.008(05)(02) & 0.008(07)(05) \cr
0.69 & 0.027(41)(17) & 0.022(30)(07) & 0.02(31)(16) & -0.001(19)(11) & 0.009(29)(28) \cr
0.81 & 0.014(63)(04) & 0.013(16)(02) & 0.008(08)(04) & 0.011(07)(02) & 0.003(06)(06) \cr
1.24 & 0.011(56)(03) & 0.024(24)(02) & 0.019(10)(03) & 0.008(06)(03) & 0.009(07)(05) \cr
1.38 & 0.004(42)(03) & -0.003(16)(01) & 0.035(09)(03) & 0.004(05)(02) & 0.014(06)(04) \cr
1.38 & 0.031(27)(02) & 0.015(10)(01) & 0.004(04)(02) & 0.006(04)(01) & 0.002(03)(03) \cr
1.52 & 0.074(50)(06) & 0.021(25)(04) & 0.006(12)(06) & -0.001(09)(06) & 0.005(08)(09) \cr
2.29 & 0.005(61)(06) & 0.019(26)(03) & 0.015(16)(06) & 0.012(09)(03) & -0.002(10)(10) \cr
2.77 & -0.016(16)(02) & 0.000(08)(01) & 0.012(04)(02) & 0.001(02)(01) & 0.009(02)(03) \cr
\hline
\end{tabular}
\caption{The table of iso-scalar GPD moments $B_{n+1,0}^{u+d}$.
}
\label{tb:allisosB}
\end{table}

\section{Quasi-PDF matrix elements at $P=0$}\label{app:pz0}

\begin{figure}[h!]
    \centering
    \includegraphics[width=0.4\textwidth]{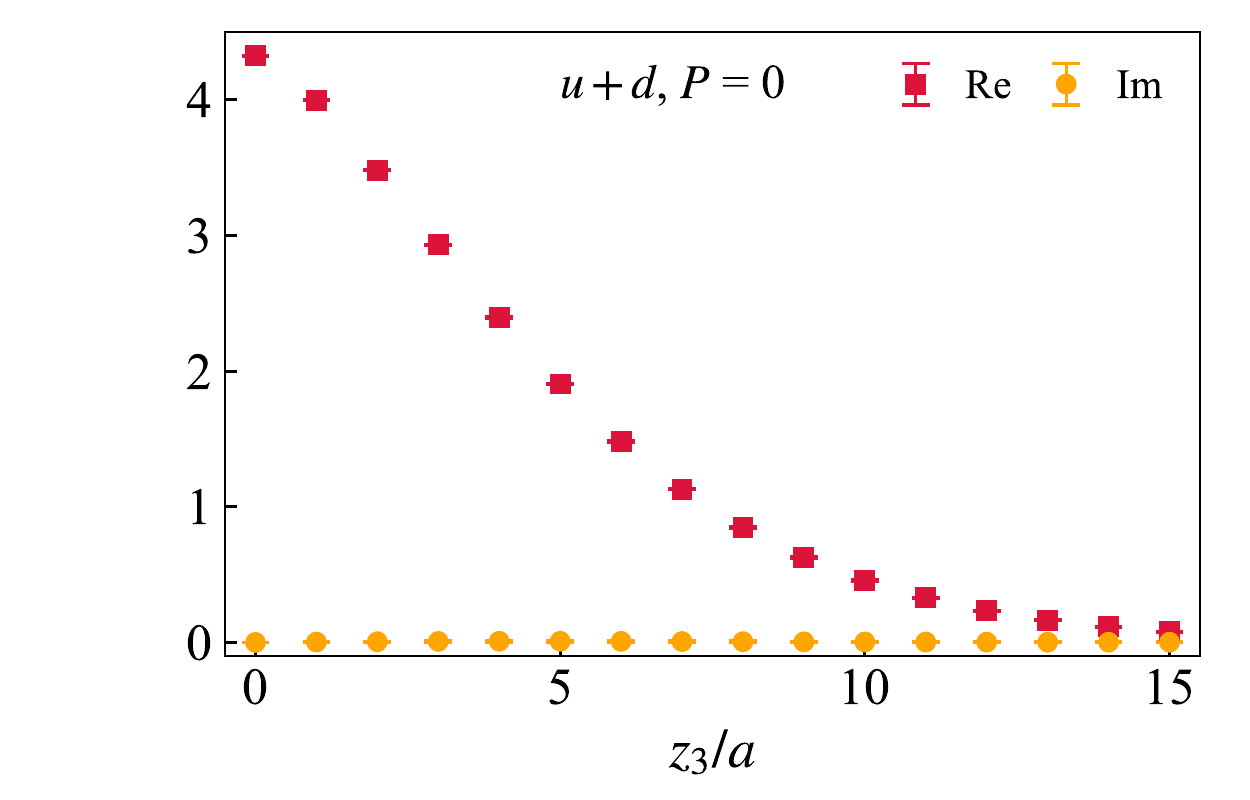}
	\caption{Bare matrix elements of iso-scalar zero-momentum quasi-PDF matrix elements $\mathcal{F}(z,P=0,\Delta=0)$.\label{fig:bmqPDFP0_isosc}}
\end{figure}

\begin{figure}[h!]
    \centering
    \includegraphics[width=0.4\textwidth]{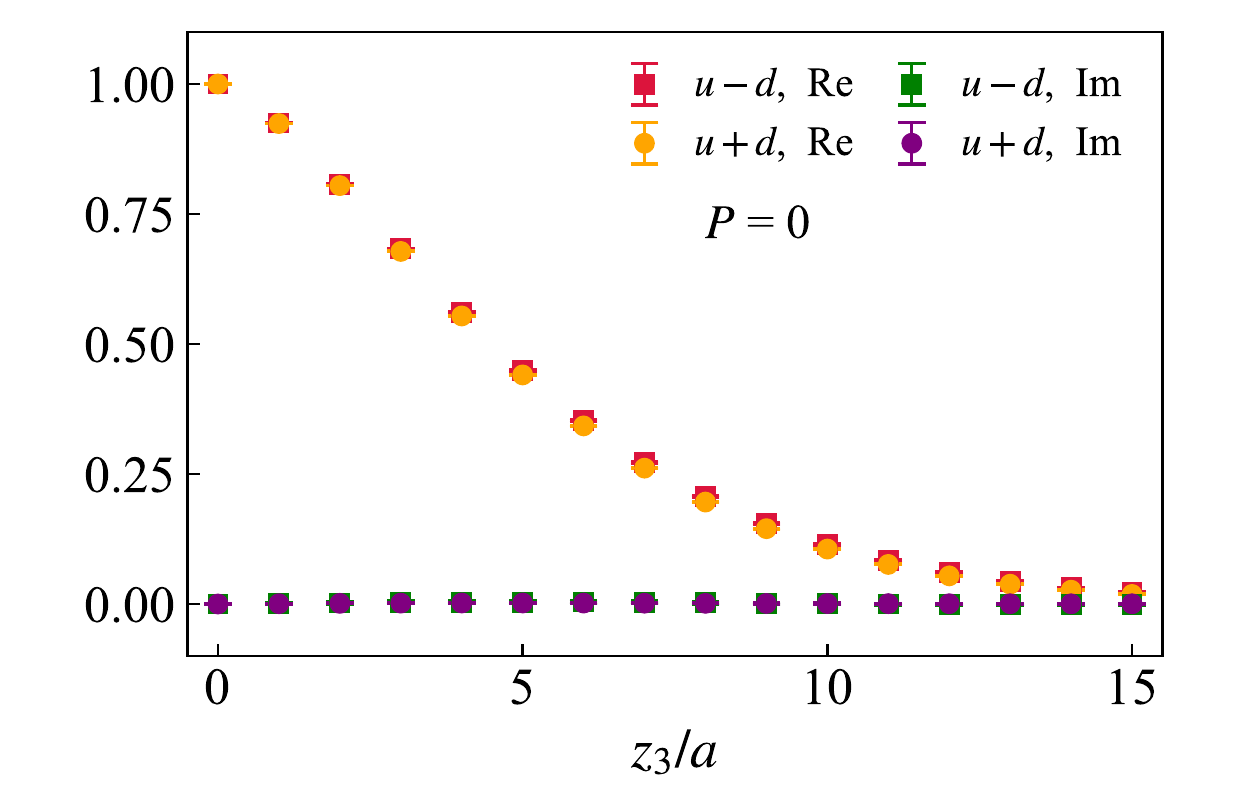}
	\caption{Normalized bare matrix elements of zero-momentum quasi-PDF matrix elements $\mathcal{F}_{u-d}(\vec{z},\vec{P}=0,\vec{\Delta}=0)/\mathcal{F}_{u-d}(\vec{z}=0,\vec{P}=0,\vec{\Delta}=0)$ and $\mathcal{F}_{u+d}(\vec{z},\vec{P}=0,\vec{\Delta})/\mathcal{F}_{u+d}(\vec{z}=0,\vec{P}=0,\vec{\Delta}=0)$ for both iso-vector and iso-scalar cases.\label{fig:rbmqPDFP0}}
\end{figure}

In \sec{ratio}, we take the iso-vector quasi-PDF matrix elements $\mathcal{F}(\vec{z},\vec{P}=0,\vec{\Delta})$ for the ratio scheme renormalization. For comparison, we show the iso-scalar bare quasi-PDF matrix elements in \fig{bmqPDFP0_isosc}. It's known that the iso-vector local matrix element $\mathcal{F}_{u-d}(\vec{z}=0,\vec{P}=0,\vec{\Delta}=0)$ by definition is the vector current renormalization factor, which is $Z_V^{-1}$ = 1.440(1) as shown in \fig{bmqPDFP0}. By utilizing the value $\mathcal{F}_{u+d}(\vec{z}=0,\vec{P}=0,\vec{\Delta}=0)=4.325(1)$, we obtain $Z_V\mathcal{F}_{u+d}(\vec{z}=0,\vec{P}=0,\vec{\Delta}=0)=3.004(2)$, a value which closely approximates the nucleon's iso-scalar charge of 3, even though the disconnected diagrams are not accounted for. 

Putting aside the difference of over all charges, one can normalize the bare matrix elements for both iso-vector and iso-scalar cases as $\mathcal{F}_{u-d}(\vec{z},\vec{P}=0,\vec{\Delta})/\mathcal{F}_{u-d}(\vec{z}=0,\vec{P}=0,\vec{\Delta}=0)$ and $\mathcal{F}_{u+d}(\vec{z},\vec{P}=0,\vec{\Delta})/\mathcal{F}_{u+d}(z=0,P=0,\Delta=0)$ shown in \fig{rbmqPDFP0}. As expected, the normalized matrix elements from iso-vector and iso-scalar cases are close to each other. This can be attributed to two reasons: firstly, the renormalization factor is identical for both cases; and secondly, after renormalization, the leading order perturbation calculation indicates that they are also approximately the same, disregarding the mixing between iso-scalar and gluon matrix elements and considering the Wilson coefficient $C_0$ at short distances. However, this argument may not hold beyond leading order or when $z$ increases significantly, at which point the iso-vector and iso-scalar cases may begin to diverge.

\section{Quasi-GPD matrix elements in $\gamma_0$ definition}\label{app:g0bm}

\begin{figure}[h!]
    \centering
    \includegraphics[width=0.4\textwidth]{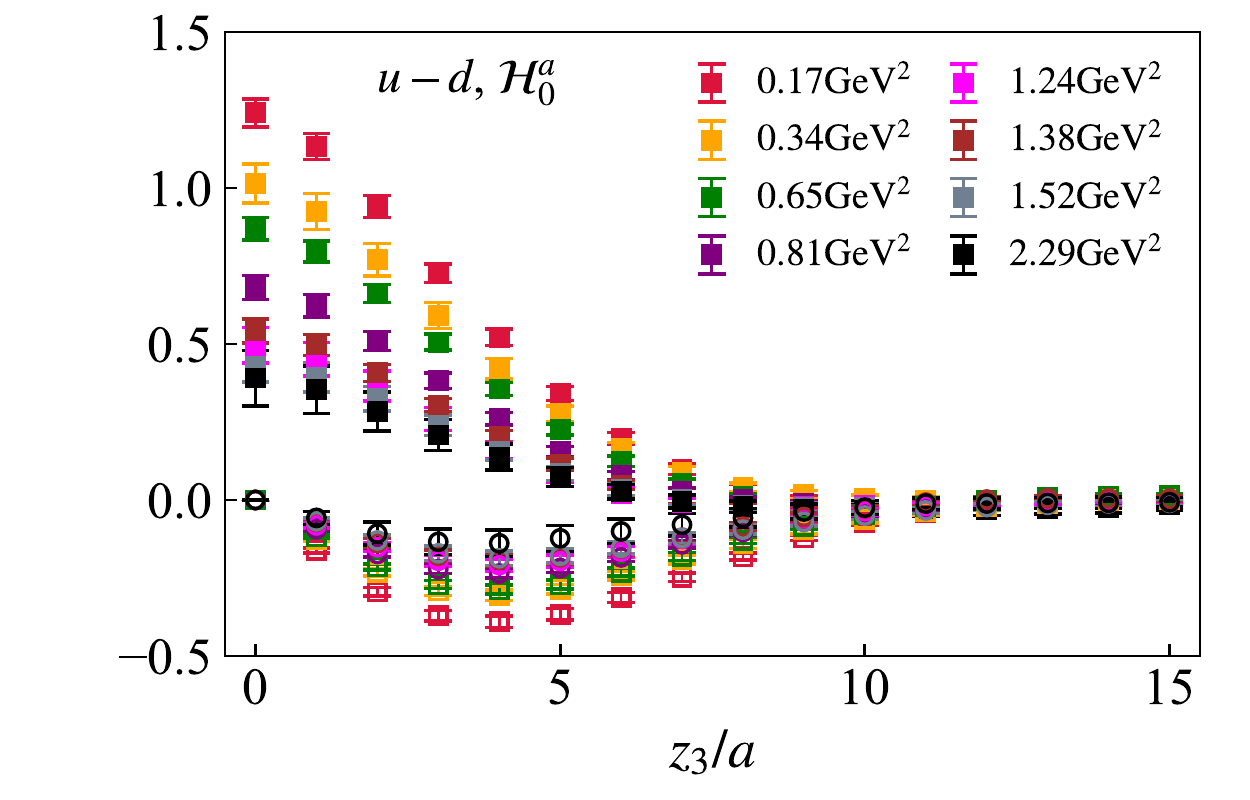}
    \includegraphics[width=0.4\textwidth]{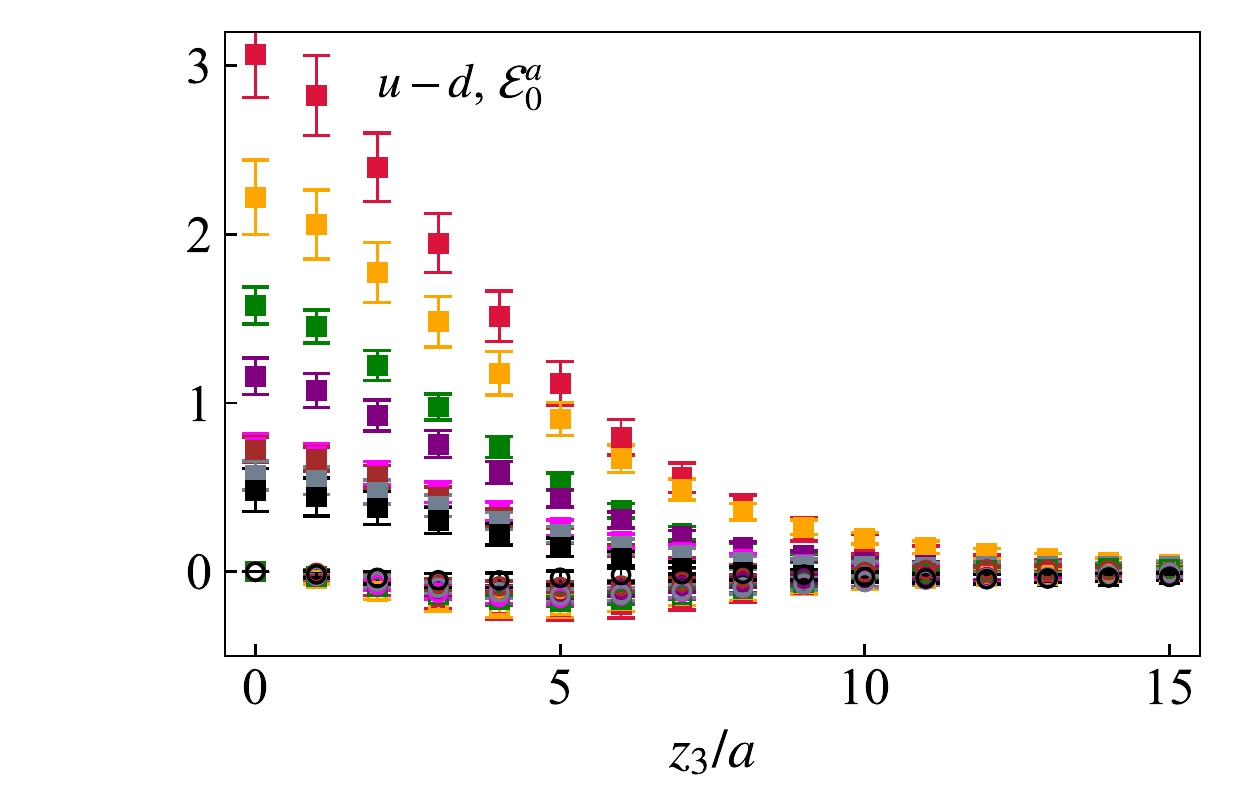}
    \includegraphics[width=0.4\textwidth]{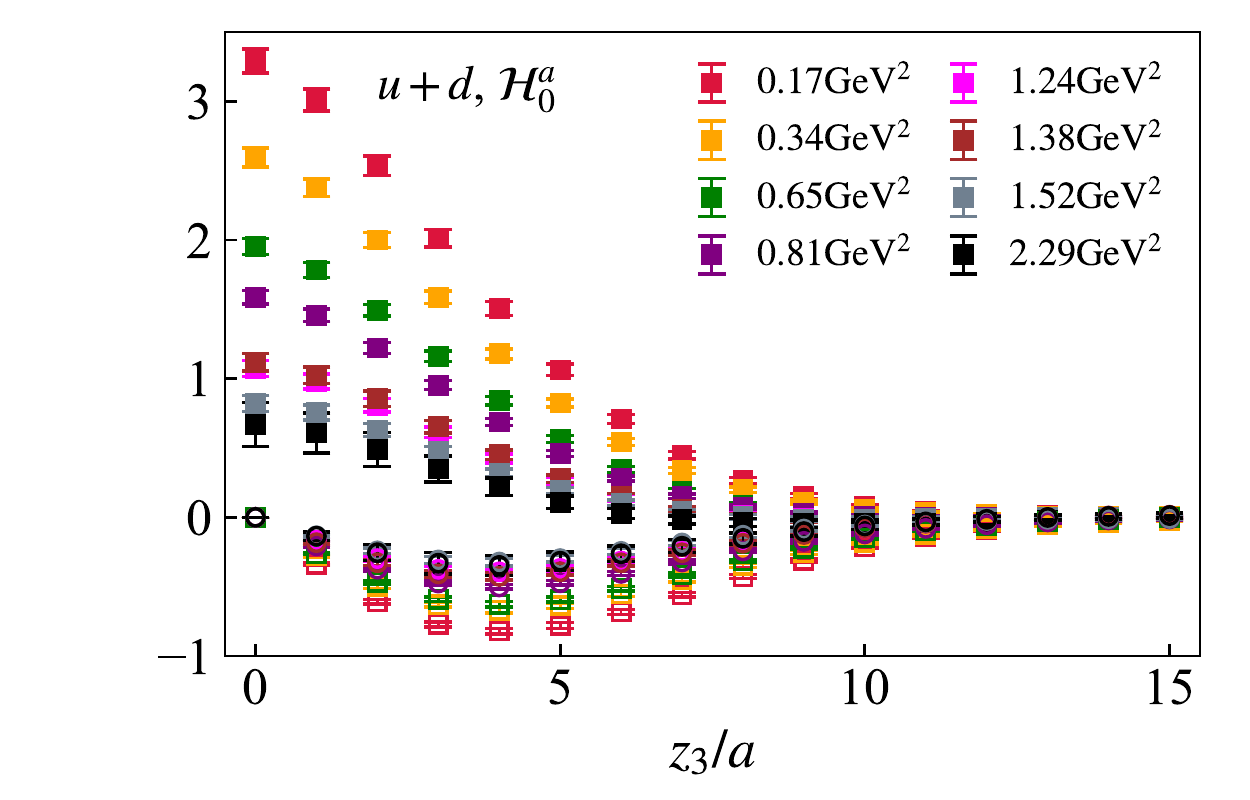}
    \includegraphics[width=0.4\textwidth]{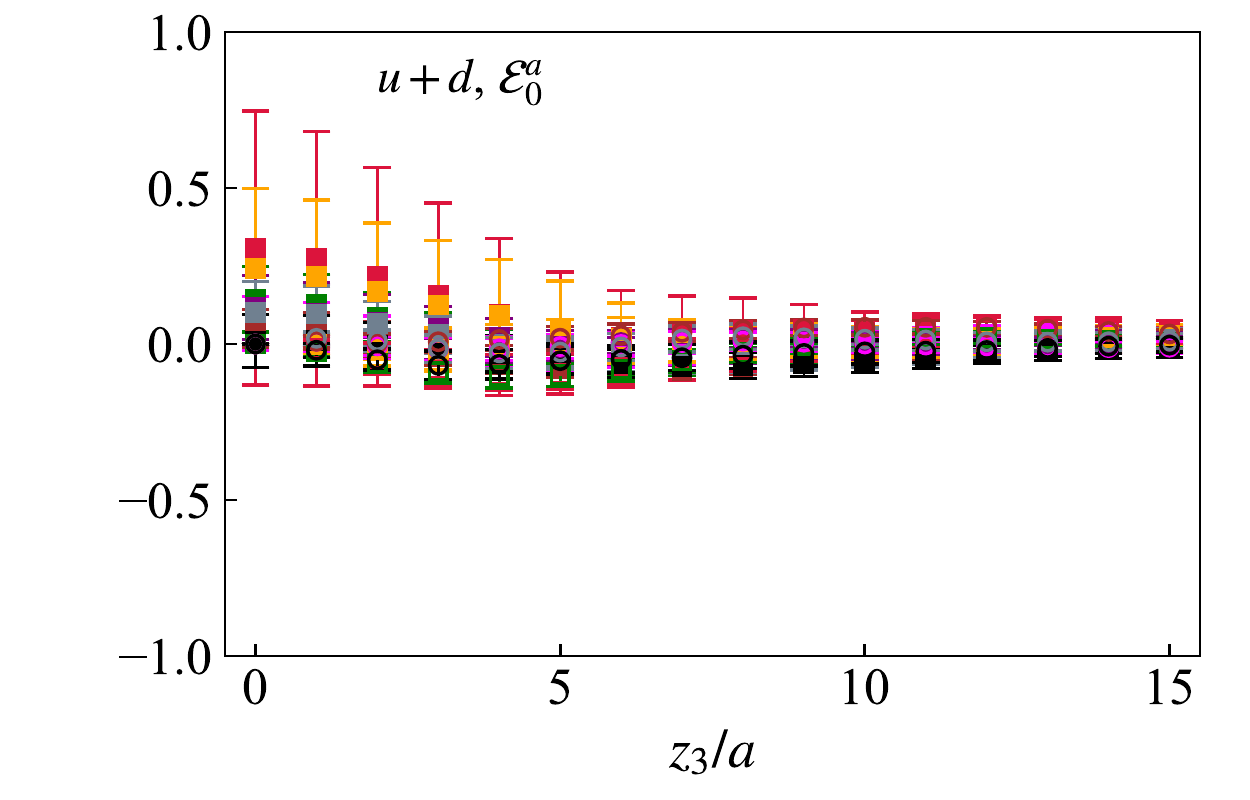}
	\caption{Bare matrix elements under $\gamma_0$ definition from asymmetric frame. The upper and lower panels are for the iso-vector and iso-scalar cases, with squared points for the real part and circled points for the imaginary part. The data shown are from hadron momentum $P_3$ = 1.25 GeV with all the momentum transfer $-t$ we have.\label{fig:bmstdasymm}}
\end{figure}

\begin{figure}[h!]
    \centering
    \includegraphics[width=0.4\textwidth]{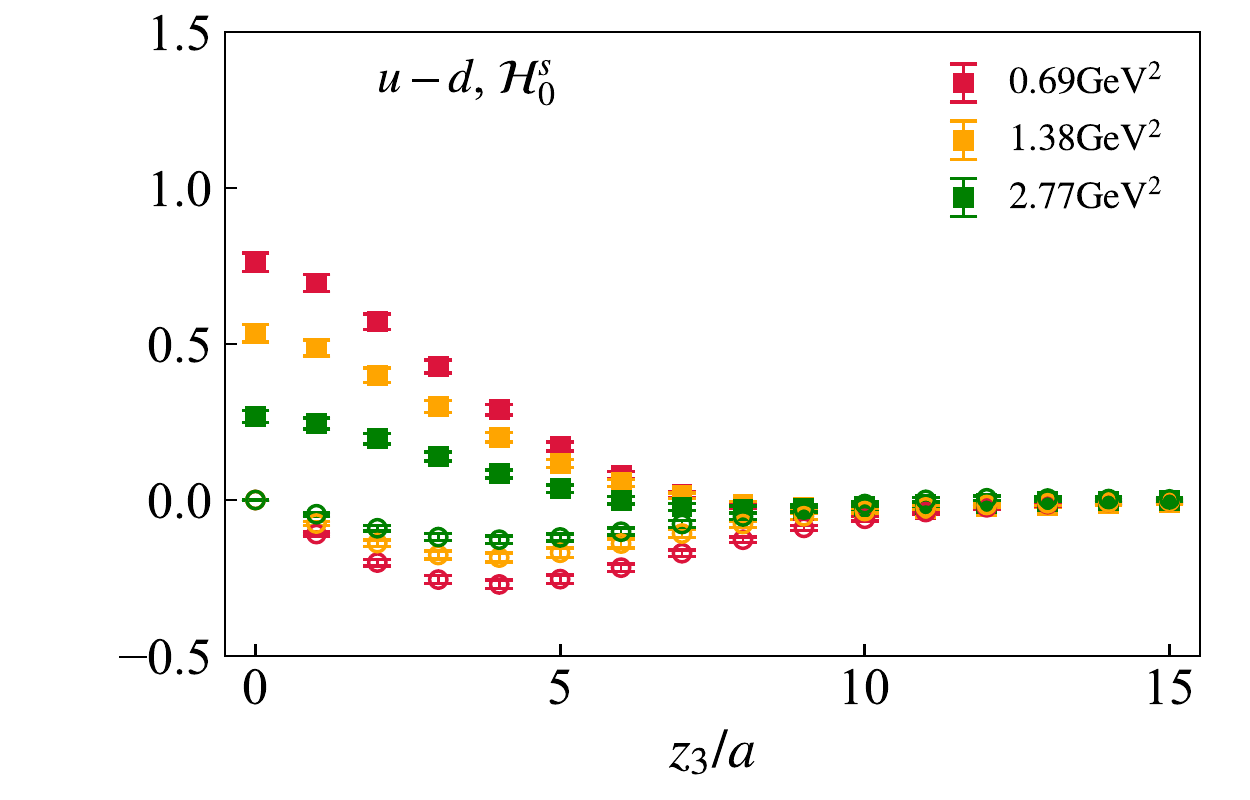}
    \includegraphics[width=0.4\textwidth]{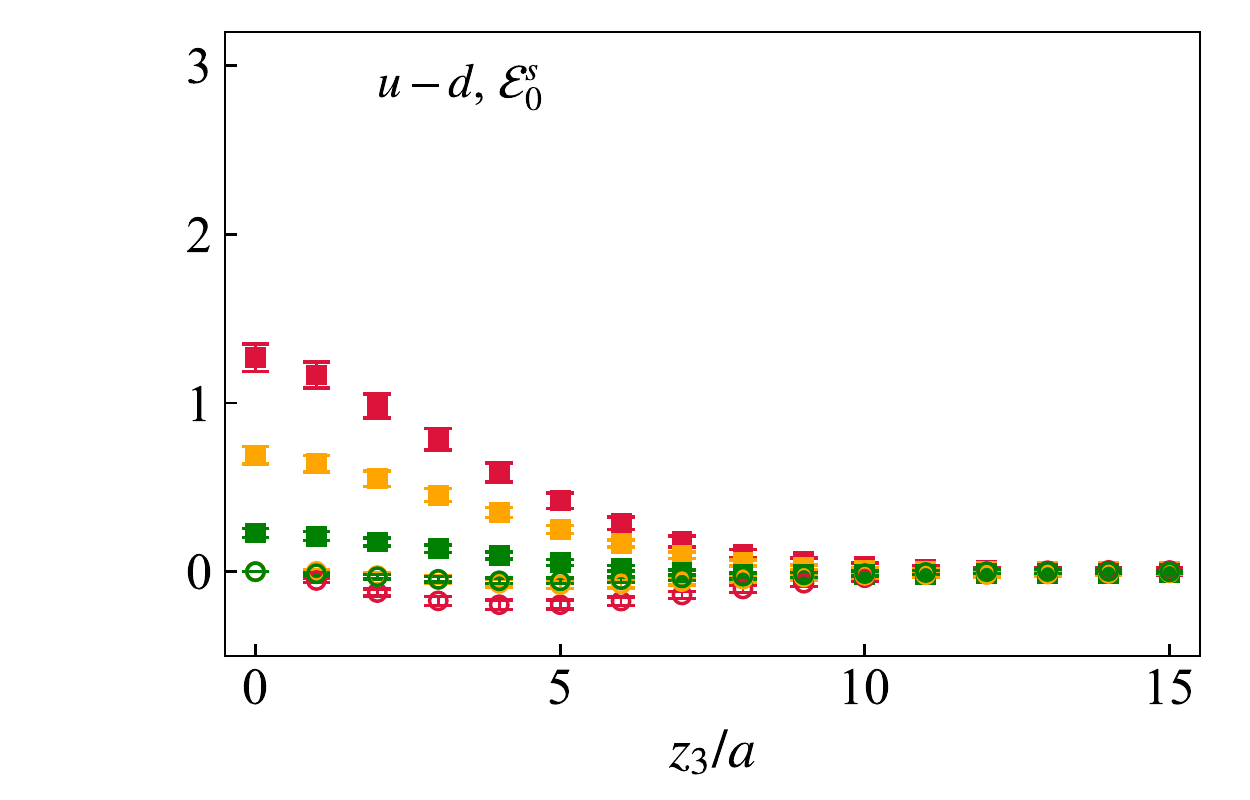}
    \includegraphics[width=0.4\textwidth]{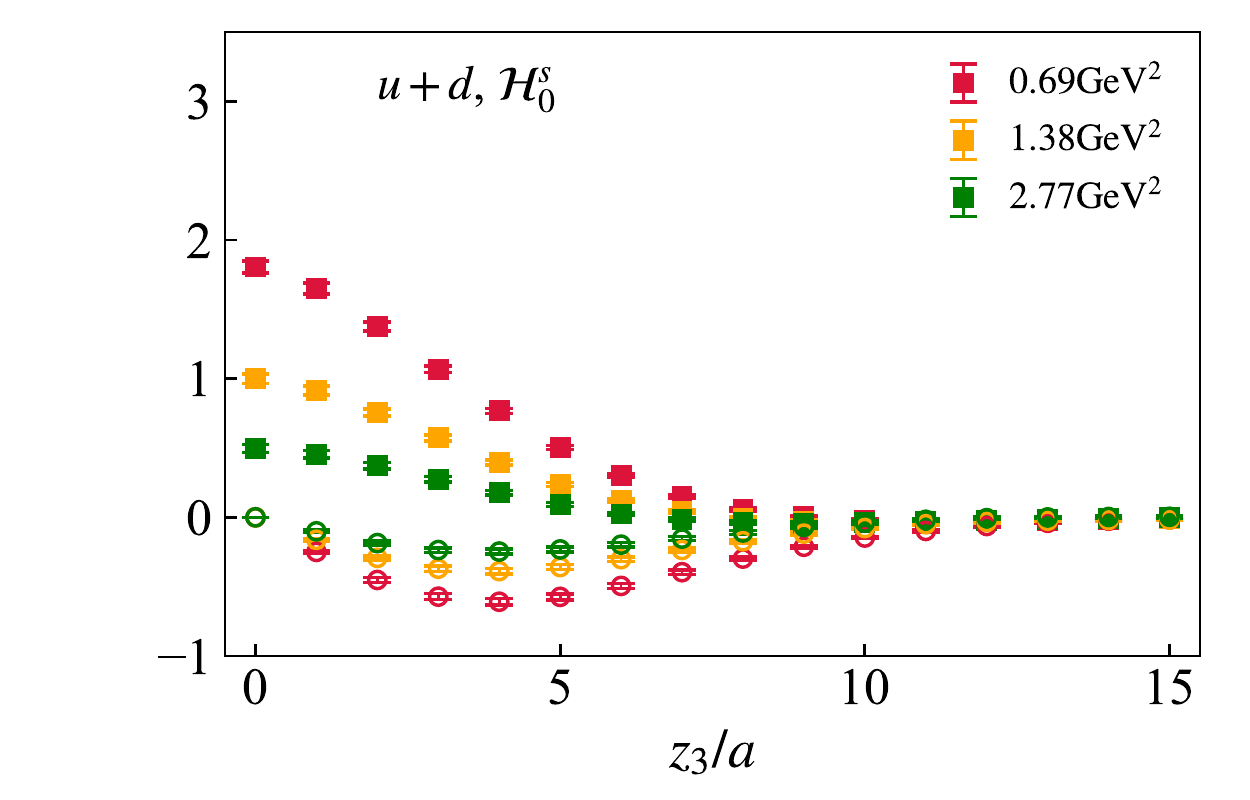}
    \includegraphics[width=0.4\textwidth]{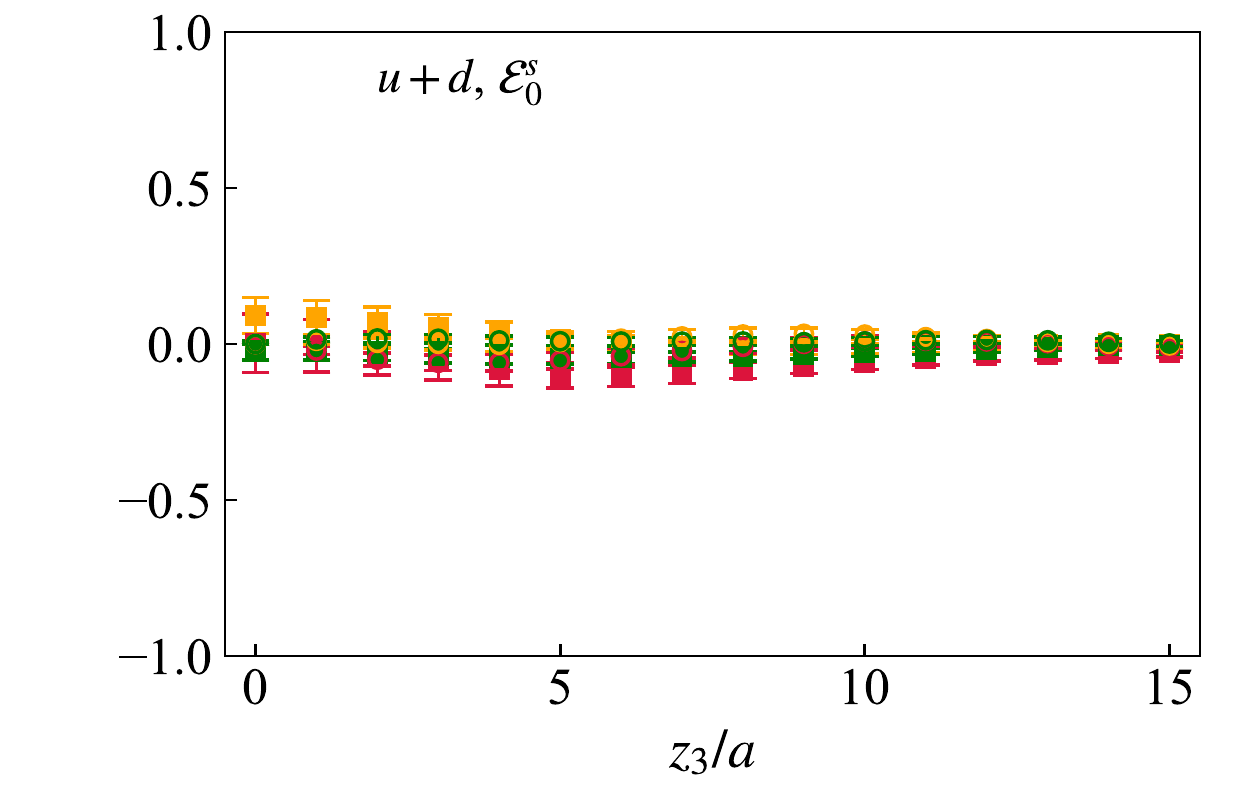}
	\caption{Bare matrix elements under $\gamma_0$ definition from symmetric frame. The upper and lower panels are for the iso-vector and iso-scalar cases, with squared points for real part and circled points for imaginary part. The data shown are from hadron momentum $P_3$ = 1.25 GeV with all the momentum transfer $-t$ we have.\label{fig:bmstdsymm}}
\end{figure}

In \sec{bm+ratio}, we presented the quasi-GPD matrix elements in Lorentz-invariant definition, while here we show all the matrix elements from the $\gamma_0$ definition, which have been used in the main text. 
The bare matrix elements in the asymmetric frame with $\vec{p}_f=\vec{P}^a$ and $\vec{p}_i=\vec{P}^a-\vec{\Delta}^a$ are shown in \fig{bmstdasymm}, and the ones in the symmetric frames with $\vec{p}_f=\vec{P}^s+1/2\vec{\Delta}^s$ and $\vec{p}_i=\vec{P}^s-1/2\vec{\Delta}^s$ are shown in \fig{bmstdsymm}. 
Compared to the Lorentz-invariant definition in \fig{bmLI}, one can find the quasi-GPD matrix elements in $\gamma_0$ definitions show some difference, which will result in a deviation of renormalized matrix and subsequently the extracted moments as discussed in the main text.

\section{Moments extraction including a leading-order power correction}

\begin{figure}[h!]
    \centering
    \includegraphics[width=0.3\textwidth]{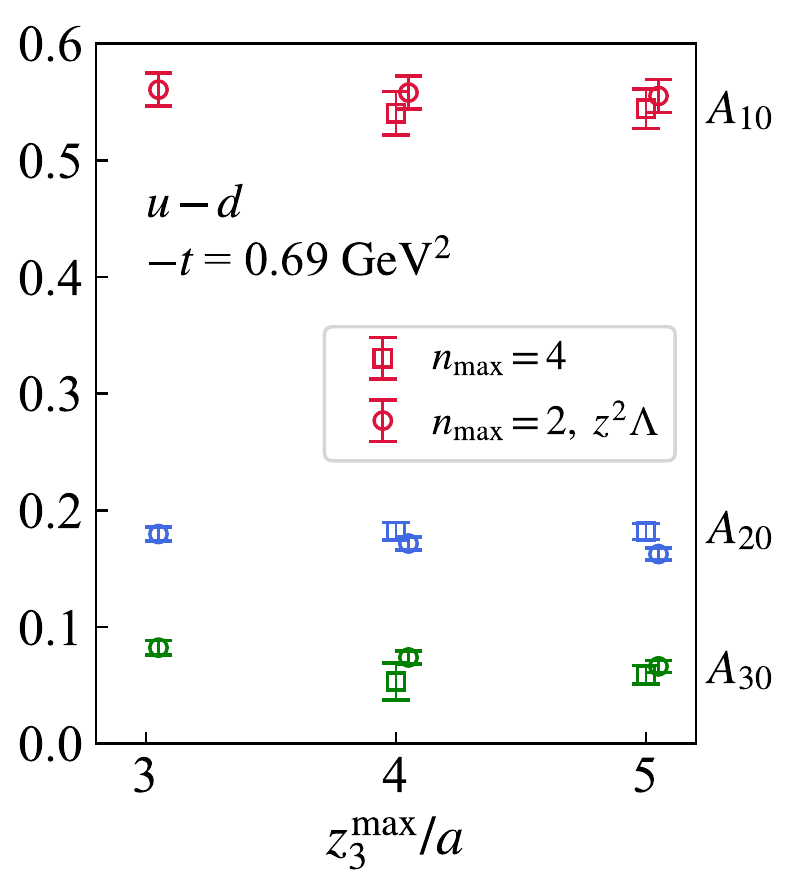}
    \includegraphics[width=0.3\textwidth]{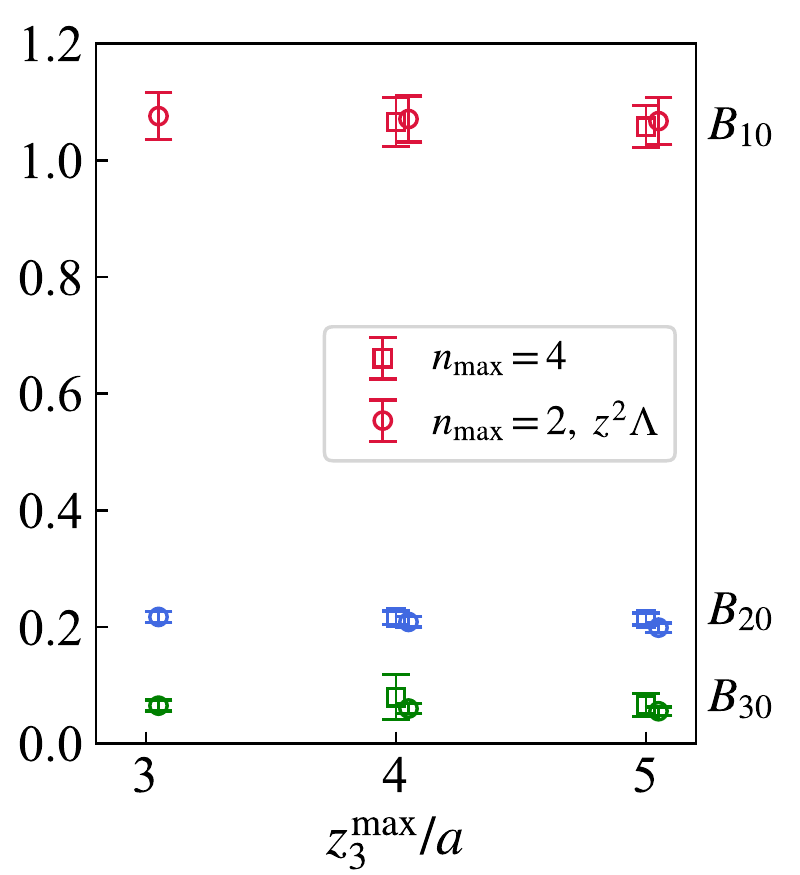}
	\caption{The first three moments extracted from leading-twist $n_{\rm max}=4$ SDF and $n_{\rm max}=2$ with $\Lambda^2 z^2$ term are shown as a function of $z_{\rm max}$ for the case of $-t=0.69~\rm{GeV}^2$.\label{fig:momshtw}}
\end{figure}

\begin{figure}[h!]
    \centering
    \includegraphics[width=0.3\textwidth]{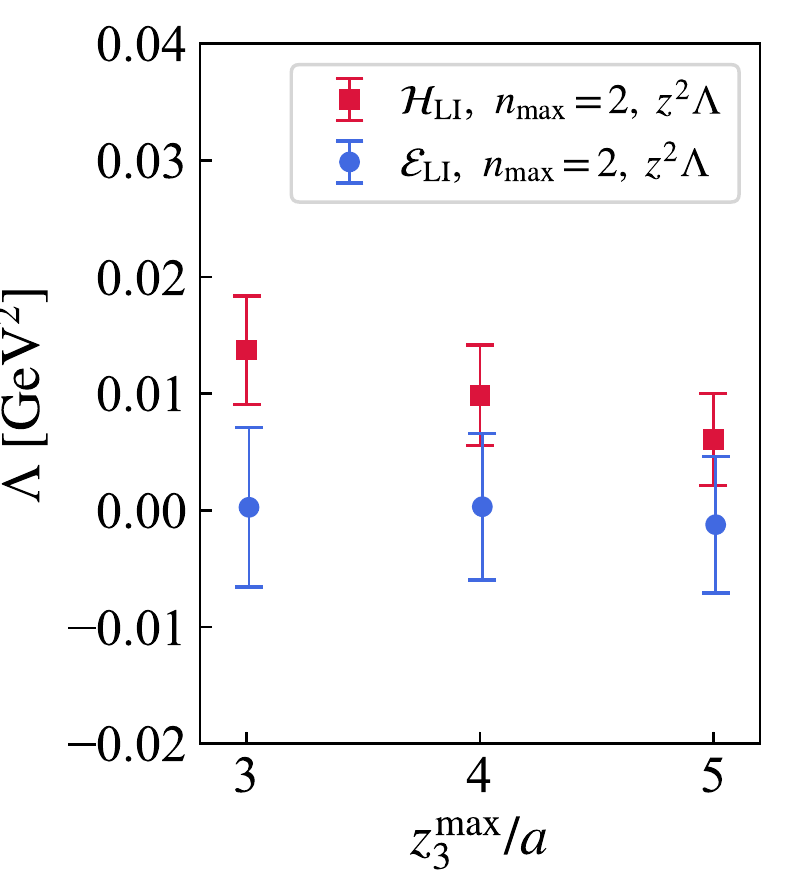}
	\caption{The $\Lambda$ term fitted from $\mathcal{H}_{\rm{LI}}$ and $\mathcal{E}_{\rm{LI}}$ renormalized matrix elements using \Eq{ratioOPEhtw}.\label{fig:momshtwLambda}}
\end{figure}

\begin{figure}[h!]
    \centering
    \includegraphics[width=0.3\textwidth]{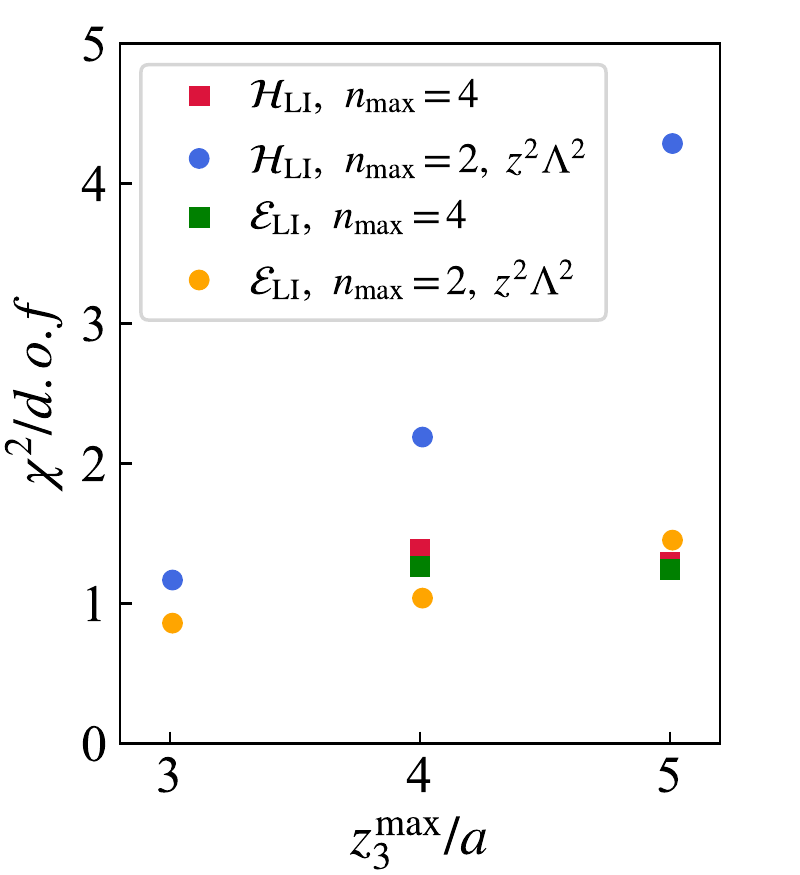}
	\caption{The $\chi^2/{d.o.f}$ of the fit from $n_{\rm max}=4$ SDF and $n_{\rm max}=2$ with leading higher-twist correction.\label{fig:momshtwshiq}}
\end{figure}

As has been mentioned in \sec{OPE}, the short distance factorization suffers from power correction $\mathcal{O}(z^2\Lambda_{\rm QCD}^2)$. 
Thus, one should keep the $z^2$ at a short distance. 
The ratio-scheme renormalization has the potential to reduce the power correction because of the possible cancellation between the numerator and denominator. In \sec{momsKernel}, we found the leading-twist factorization formula truncated at $n_{\rm max}$ = 4 can describe the data well, given the fact that higher moments are factorially suppressed. Nevertheless, one may question whether the higher-twist effects are small for the range of $z$ under consideration. Therefore, in this section, we will fit the data by including the leading power correction using the formula,
\begin{align}\label{eq:ratioOPEhtw}
	\begin{split}
		\mathcal{M}(z,P,\Delta)=\sum_{n=0}\frac{(-izP)^n}{n!}\frac{C^{\overline{\rm{MS}}}_n(\mu^2 z^2)\langle x^n\rangle}{C^{\overline{\rm{MS}}}_0(\mu^2 z^2)}+\Lambda z^2\,.
	\end{split}
\end{align}
Since the fourth and fifth moments are mostly consistent with zero and the fit becomes unstable with additional parameters, we simply truncate the $n_{\rm max}$ = 2 after adding $\Lambda z^2$. 
For the case of $-t=0.69~\rm{GeV}^2$, we have three different momenta $P_3=0.83,1.25$ and 1.67 GeV as shown in \fig{isoVfitzmax}. 
In \fig{momshtw}, we show the first three moments extracted from leading-twist $n_{\rm max}=4$ SDF formula and $n_{\rm max}=2$ formula with the $\Lambda z^2$ term. The fitted $\Lambda$ term and $\chi^2_{d.o.f}$ are shown in \fig{momshtwLambda} and \fig{momshtwshiq}. 
We find that in the short distance under consideration, the $\Lambda$ term is consistent with zero, and moments from different strategies agree with each other, suggesting that higher-twist corrections are negligible. 
Meanwhile, it can be seen that as the increasing of $z_3$, the $\chi^2_{d.o.f}$ of $n_{\rm max}$ = 2 with the $\Lambda z^2$ term becomes large while the ones of $n_{\rm max}$ = 4 are still good. 
That means the fits including higher moments can better describe the data instead of the leading power correction $\Lambda z^2$ term.

\bibliography{ref}

\end{document}